\documentclass[amsmath,amssymb,twocolumn,prd,showpacs,nofootinbib,floatfix]{revtex4}
\usepackage{graphics}
\usepackage{dcolumn}
\usepackage{bm}

\begin{document}
\newpage
\preprint{HEP-0405262}

\title{Sub-millimeter Tests of the Gravitational Inverse-square Law}

\author{C.D. Hoyle}
\author{D.J. Kapner}
\author{B.R. Heckel}
\author{E.G. Adelberger} 
\author{J.H. Gundlach}
\author{U. Schmidt}
 \altaffiliation[Currently at ]{Physikalisches Institut, Heidelberg, Germany}
\author{H.E. Swanson}
\affiliation{Department of Physics, Box 351560, University of Washington, Seattle, Washington 98195-1560}
\date{\today}
\begin{abstract}
Motivated by a variety of theories that predict new effects, 
we tested the gravitational $1/r^2$ law at separations between 10.77 mm and 
137 $\mu$m using two different 10-fold azimuthally symmetric torsion pendulums and 
rotating 10-fold symmetric attractors. Our work improves upon other
experiments by up to a factor of about $100$.
We found no deviation from Newtonian physics at the 95\% confidence level and interpret these results as constraints on extensions of the Standard Model that predict Yukawa or power-law forces. We set a constraint
on the largest single extra dimension (assuming toroidal compactification and that one extra dimension is significantly 
larger than all the others) of $R_{\ast} \leq 160~\mu$m, and on two equal-sized large extra dimensions of $R_{\ast} \leq 130~\mu$m. Yukawa interactions with $|\alpha| \geq 1$ are ruled out at 95\% confidence for
$\lambda \geq 197~\mu$m. Extra-dimensions scenarios stabilized by radions are restricted to unification masses
$M_{\ast} \geq 3.0$~TeV/c$^2$, regardless of the number of large extra dimensions. We also provide new constraints on power-law potentials $V(r) \propto r^{-k}$ with $k$ between 2 and 5 and on the $\gamma_5$ couplings of pseudoscalars with $m \leq 10$~meV/c$^2$.
\end{abstract}
\maketitle
\section{\label{sec:intro}Introduction}
\subsection{Background}
Until a few years ago, it was widely assumed that the Newtonian Inverse Square Law (ISL)
should be valid for length scales from infinity to roughly the Planck length 
$R_P=\sqrt{G \hbar/c^3}=1.6 \times 10^{-35}$ m, at which scale quantum effects
must become important. After all, the usual argument went, the exponent 2 in the force law simply reflects the fact that we live in a 3-dimensional world. A wide variety of recent theoretical speculations, 
motivated in part by string-theory considerations, have raised the possibility that fundamentally new 
phenomena could occur at length scales below 1 mm.
Many of these speculations are driven by the two so-called hierarchy problems of gravity:
\begin{itemize}
\item
{\em The gauge hierarchy problem.} Gravity is extraordinarily weak compared to the other fundamental forces. The Planck mass 
$M_P =\sqrt{\hbar c/G}= 1.2\times 10^{16}$~TeV/c$^2$
is huge compared to the electroweak scale $M_{EW} \sim 1$~TeV/c$^2$. It has been argued\cite{ar:98} that the true Planck mass, $M_{\ast}$, could be as low as 1 TeV/c$^2$ if some of the ``extra'' space dimensions demanded by string-theory are ``large'' compared to the Planck length. It is possible that the size of some of the ``large extra dimensions'' could be large enough to alter the gravitational Gauss law, so that gravity would become anomalously strong in an experimentally accessible regime\cite{ar:98}. 
\item 
{\em The cosmological constant problem.} The observed 
gravitating vacuum-energy density is vanishingly small compared to the predictions of quantum
mechanics. The gravitating energy density $\rho_{\rm vac}\sim 0.7\rho_{\rm c}$, inferred from a wide variety of astrophysical observations\cite{ca:01}, is at least $10^{60}$ times smaller
than the predicted zero-point energy for a cutoff of $M_P$. The observed energy density corresponds to a length scale
$R_{\rm vac}=\sqrt[4]{\hbar c / \rho_{\rm vac}} \approx 0.1$ mm and an energy of $\sqrt[4]{(\hbar c)^3\rho_{\rm vac}} \approx 2$ meV that may have fundamental significance\cite{be:97}. It has been suggested that the apparent inability of gravity to ``see'' the
vacuum energy could be explained if the effective theory of gravity had a cutoff of $\sim 1$~meV\cite{su:99,su:03},
so that gravity would effectively ``shut off'' at length scales less than $R_{\rm vac}$. \end{itemize}

Experimental tests of the gravitational ISL also probe certain speculations about {\em non-gravitational} physics. The standard model of particle physics cannot be complete and many ideas for extending it
predict very-low-mass scalar or vector bosons that could produce short-range
exchange forces that would appear as violations of the ISL (see, for example, an extensive summary in Ref.~\cite{ad:03}). 

The desire to test a basic law in a previously inaccessible, but very interesting,
regime motivated the work we report here. Some of the work we report
in this paper has already appeared in Letter form\cite{ho:01}. This paper includes
additional experimental work and an improved analysis; it supercedes Ref.~\cite{ho:01}. 
\subsection{\label{parameterization}Parameterizations}
It is now customary to interpret experimental tests of the ISL as setting bounds on a 
possible Yukawa addition to the familiar $1/r$ Newtonian potential
\begin{equation}
V(r)=-G\frac{m_1 m_2}{r}\left[ 1+\alpha e^{-r/\lambda}\right]~,
\label{eq:Yukawa}
\end{equation}
where $\alpha$ is a dimensionless strength parameter and $\lambda$ a length scale.
The Yukawa potential is, of course, the static limit of the interaction from exchange
of a boson of mass $m=\hbar/(c \lambda)$ in which case $\alpha$ is proportional to
the squared product of the appropriate coupling constants. This Yukawa form is obviously appropriate for the boson-exchange forces mentioned above.
It is also a good approximation to the effects one expects from large extra dimensions until the separation of the interacting bodies becomes
comparable to or smaller than the size of the large extra dimensions\cite{ke:99}. 

The ISL can also be violated by power-law potentials, which we parameterize as
\begin{equation}
V(r)=-G\frac{m_1 m_2}{r}\left( 1 + \beta_k \left[ \frac{1~{\rm mm}}{r} \right]^{k-1}\right)~;
\label{eq:power law definition}
\end{equation}
these can occur in processes where two massless particles are exchanged.
We therefore interpret our results as exclusion plots in the $\alpha-\lambda$
parameter space and as constraints on power-law parameters $\beta_k$ for $k=2$, 3, 4 and 5.
\subsection{\label{subsec:previous}Previous Experimental Limits}
The Yukawa constraints on ISL violation for length scales
$\lambda \leq 1$ cm, at the time this work was begun in late 1999, are shown in 
Fig.~\ref{fig:old constraints}. All the constraints in this Figure were obtained from torsion-balance
experiments: a null test using cylindrical geometry from the Irvine group\cite{ho:85}
set the most restrictive limits for length scales larger than about 2.5 mm, an absolute
comparison in Moscow of the force between spheres at two difference distances set the best
limit for $80~\mu\rm{m} \leq \lambda \leq 2.5$ mm\cite{mi:88} and a 
measurement of the Casimir force between a sphere and a plane\cite{la:97,la:98} set the tightest constraint for  $4~\mu\rm{m} \leq \lambda \leq 80~\mu$m. 
\section{\label{subsec:speculations}Theoretical Speculations}
A surprisingly rich array of theoretical speculations suggests that fundamentally new physics could show up at length scales between 1 mm and 100 $\mu$m. This new physics is clearly probed directly by ISL tests, but may also be revealed in collider experiments at the very highest energies or in astrophysical phenomena such as neutrino production in supernova explosions. The relative sensitivity of these very different probes depends strongly on the basic assumptions of specific models.
We give here only a brief summary of these considerations and refer the reader to two recent reviews\cite{ad:03,he:02} for  more comprehensive discussions.
\subsection{\label{subsec:extrad1}Large Extra Dimensions}
\begin{figure}
\hfil\scalebox{.58}{\includegraphics*[0.7in,0.4in][6.7in,5.7in]{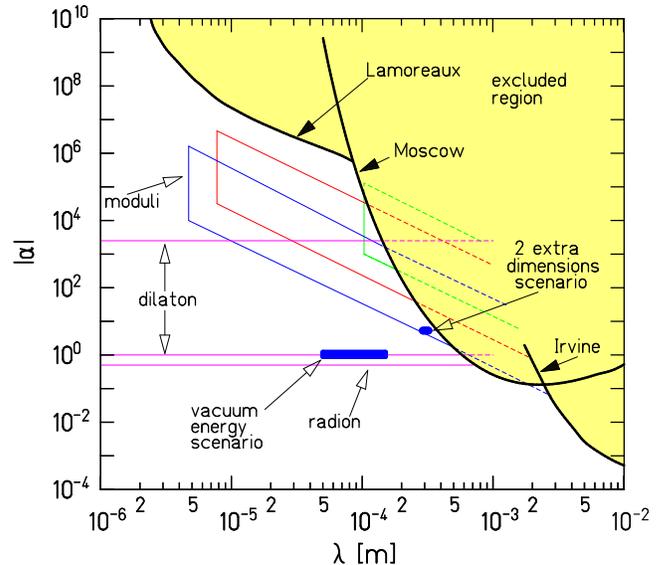}}\hfil
\caption{[color online] 95$\%$ confidence upper limits on a short-range Yukawa violation of the gravitational ISL as of 1999. The region excluded by previous 
torsion-balance experiments~\cite{la:97,la:98,mi:88,ho:85} lies above the heavy lines 
labeled Lamoreaux, Moscow, and Irvine, respectively. The constraints are taken from Ref.~\cite{lo:99}. The dilaton and moduli predictions are taken from Refs.~\cite{ka:00,di:96}, respectively. The extra dimensions and radion predictions are based on Eqs.~\ref{eq:R and alpha}, \ref{eq:true M_P} and \ref{eq:radion}, assuming $n=2$ and $M_{\ast}=1$~TeV.}
\label{fig:old constraints}
\end{figure}  
The discovery that string or M-theory necessarily contains objects called branes, where the number of spatial dimensions of the brane can be anywhere from 0 to 9, has opened new possibilities for explaining the gauge hierarchy problem. Arkani-Hamed, Dimopoulos and Dvali\cite{ar:98} proposed that the ``true'' Planck mass could be as low as 1 TeV; in their scenario all the Standard Model particles are represented by open strings whose ends are fixed to a 3+1 dimensional brane, while the graviton which is represented by a closed loop of string is free to propagate in the 10+1 dimensional ``bulk''. In their picture gravity appears weak not because it is inherently weak, but because most of its strength has leaked off into the extra (curled up) dimensions of the bulk which are inaccessible to physicists living on the brane. This notion can be tested very directly by probing the gravitational ISL at length scales comparable to the radius of the curled up dimensions $R_{\ast}$ which, it is suggested, could be much larger than the Planck length. If the effective number of large extra dimensions is $n$, the gravitational Gauss-law force will smoothly change from a $1/r^2$ form for $r>>R_{\ast}$ to a $1/r^{2+n}$ form for $r<<R_{\ast}$. For the experimentally relevant case where $r \agt R_{\ast}$,
the gravitational potential is expected\cite{ad:03,remark1}, assuming compactification on a torus, to have the form of Eq.~\ref{eq:Yukawa} with 
\begin{equation}
\lambda=R_{\ast}~~~~~{\rm and}~~~~~~~ \alpha=8n/3~,
\label{eq:R and alpha}
\end{equation}
so that gravity is predicted to get much stronger when probed at distances comparable to $R_{\ast}$. If all $n$ ``large'' extra dimensions have the same size $R_{\ast}$,
conservation of flux gives a relation between the ``true'' Planck mass $M_{\ast}$, $R_{\ast}$, $n$ and the Newtonian constant $G$. For compactification on a torus
\begin{equation}
R_{\ast}=\left[ \frac{M_P}{M_{\ast}} \right]^{2/n} \left[ \frac{\hbar}{2 \pi c M_{\ast}} \right]
\label{eq:true M_P}
\end{equation}
To lower $M_{\ast}$ to the 1 TeV scale of particle physics we require
\begin{equation}
R_{\ast} \approx \frac{1}{\pi}\;10^{-17+\frac{32}{n}}~{\rm cm}~.
\label{eq:range}
\end{equation}
Although the case with $n=1$, $R_{\ast}=3 \times10^{12}$~m is ruled out by solar-system observations, at the time this work was begun the case with $n=2$, $R_{\ast}=0.3$~mm, as well as scenarios with $n>2$, were consistent with ISL data.
\subsection{\label{subsec:newpart1}New Boson-Exchange Forces}
The character of the force produced by boson exchange depends on the spin and parity of the exchanged particle. Natural-parity ($0^+$ or $1^-$) bosons give forces between unpolarized bodies, while unnatural-parity ($0^-$ or $1^+$) bosons give purely spin-dependent forces. The forces from boson exchange violate the Weak Equivalence Principle and can be distinguished from forces due to extra dimensions by Equivalence Principle tests.
\subsubsection{\label{subsubsec:radions}Radion-mediated Forces}
In string theories, the geometry of spacetime is expected to be dynamical with the radii of new dimensions fluctuating independently at each point in our 4-dimensional spacetime. In an effective low-energy theory, the volume of the extra dimensions must be stabilized by {\em radions}, low-mass spin-0 fields with gravitational-strength couplings that determine the radius of the new dimensions. Radion exchange will produce a force with a strength\cite{ad:03,remark1}
\begin{equation}
\alpha=\frac{n}{n+2}
\label{eq:radion}
\end{equation}
and range of order
\begin{equation}
\lambda \sim \sqrt{\frac{\hbar^3}{c G M_{\ast}^4}} \approx 2.4 \left[ \frac{1~{\rm TeV}}{M_{\ast} c^2} \right]^2\;{\rm mm}~.
\label{eq:radion range}
\end{equation}
In many cases the radion-mediated force is the longest-range effect of new dimensions\cite{an:98}, and one which, unlike the direct effect (Eq.~\ref{eq:range}), does not diminish as the number of new dimensions increases.
\subsubsection{\label{subsubsec:string scalars}Exchange of String-Theory Scalars}
All known acceptable ground states of string theory are supersymmetric and contain large numbers of extremely weakly coupled, massless scalar fields called {\em moduli}. The expectation values of the moduli fields set the parameters of the effective theory.
The moduli necessarily couple weakly to the supersymmetry-breaking sector, and for a low supersymmetry-breaking scale are expected to acquire very small masses\cite{di:96}. Moduli couplings are computable in any given vacuum, so that ISL tests are the best way to search for these particles.
The best understood of these predicted scalar particles is the {\em dilaton} which determines the strength of gauge couplings. Its couplings to ordinary matter are nearly free of QCD uncertainties\cite{ka:00} so that its discovery could provide a smoking gun for string theory.
\subsubsection{\label{subsubsec:axion}Axion-Exchange Force}
The force due to exchange of axions, pseudoscalar particles invented to explain the very small upper limit on  $\Theta_{\rm QCD}$, is interesting. The first-order force from exchange of unnatural-parity bosons is purely spin-dependent and vanishes between unpolarized objects. However, if the axion does not drive $\Theta_{\rm QCD}$ all the way to zero, but merely to some very small value, the axion acquires a small, $CP$-violating, scalar admixture. This, in turn, leads to a spin-independent Yukawa potential between nucleons with\cite{mo:84}
\begin{eqnarray}
\alpha &=& \left[\Theta_{\rm QCD} \frac{m_a}{m_{\pi}} \frac{120~{\rm MeV}}{f_{\pi}}\right]^2 \frac{m_um_d}{(m_u+m_d)^2}\frac{\hbar c}{4 \pi u^2 G} \nonumber \\
 &\approx& \left[\left(\frac{\Theta_{\rm QCD}}{10^{-10}}\right) \left(\frac{m_a c^2}{1~{\rm meV}}\right)   \right]^2 1.3 \times 10^{-6} \label{eq:axion_alpha}\\
\lambda &=&\frac{\hbar}{m_a c} \approx \left( \frac{1~{\rm meV}}{m_a c^2} \right) 0.2 \:{\rm mm}~,
\end{eqnarray}
where $m_a$ is the axion mass which is constrained to lie between $\approx \!1\,\mu$eV/$c^2$ and $\approx \!10\,$meV/$c^2$~\cite{axion mass}, $m_{\pi}$ and $f_{\pi}$ are the pion mass and decay constant respectively, $m_u$ and $m_d$ are the current masses of the up and down quarks respectively, and $u$ is the atomic mass unit. The numerical value in Eq.~\ref{eq:axion_alpha} assumes $m_u/m_d=0.5$. The experimental upper limits on the neutron and mercury-atom electric dipole moments imply that $\Theta_{\rm QCD}\leq 6 \times 10^{-10}$\cite{po:99} and $\Theta_{\rm QCD}\leq 1.5 \times 10^{-10}$\cite{ro:01}, respectively.
\subsubsection{\label{subsubsec:2 pseudo}Multi-Particle Exchange Forces}
Power-law potentials such as Eq.~\ref{eq:power law definition} can arise from the 2nd-order process in which 2 massless particles are exchanged simultaneously. Such processes are particularly interesting when the exchanged particles are unnatural-parity bosons (for which the 1st-order force vanishes when averaged over unpolarized test bodies) or fermions (for which the 1st-order process is forbidden). The great sensitivity of ``gravitational'' experiments using unpolarized test bodies leads to useful constraints on the couplings in these cases.

Potentials with $k=2$ or $k=3$ may be generated by the simultaneous exchange of two massless scalar\cite{su:93}
or $\gamma_5$-coupled massless pseudoscalar\cite{fe:98} particles, respectively.
The exchange of two massive pseudoscalar bosons having 
$\gamma_5$ couplings to fermions $a$ and $b$ with coupling constants
$g_a$ and $g_b$, produces a long-range, spin-independent exchange potential\cite{fe:99,ad:03a},
\begin{equation}
V(r) = -\frac{\hbar}{c^3}\frac{g_a^2 g_b^2}{32 \pi^3 M_a M_b} \frac{K_1(2r/\lambda)}{\lambda r^2}~,
\label{eq:massive}
\end{equation}
where $K_1$ is a modified Bessel function, $\lambda=\hbar/(m c)$, and $M$ and $m$ refer to the fermion and pseudoscalar masses respectively.
In the limit where the pseudoscalar boson is massless Eq.~\ref{eq:massive} becomes
\begin{equation}
V(r) = -\frac{\hbar}{c^3}\frac{g_a^2 g_b^2}{64 \pi^3 M_a M_b} \frac{1}{r^3}~.
\label{eq:massless}
\end{equation}
Potentials with $k=5$ may be produced by the simultaneous exchange of two massless pseudoscalar particles with $\gamma_5\gamma_{\mu} \partial^{\mu}$ couplings such as axions or Goldstone bosons\cite{fe:98}, or by a massless neutrino-antineutrino pair\cite{fi:96}. 
\subsection{\label{subsec:cosmo1}Vacuum-Energy Scenarios}
The cosmological constant problem and theoretical attempts to solve it have been extensively reviewed (see for example
Refs.~\cite{ca:01,wi:00}). These efforts to solve the problem fall into 2 broad classes: attempts to find a mechanism for radically reducing the quantum mechanical prediction for the vacuum energy density, or attempts to find a mechanism for reducing the gravitational coupling to the standard vacuum energy. The latter attempts are particularly 
interesting from the standpoint of this work. Beane\cite{be:97} argued that in any local effective quantum field theory, naturalness implies new gravitational physics at length scales of about a millimeter that would cut off shorter distance contributions to the vacuum energy. Sundrum\cite{su:99}
proposed that the graviton is a ``fat'' object with a size of about 
$R_{\rm vac}$ and has been exploring how this might reduce its coupling to the vacuum energy\cite{su:03}, although it is not yet clear how self-consistent this is.  This scenario makes a definite prediction that gravity ``shuts off'' at length scales below about 100 $\mu$m.  
In the framework of a Yukawa ISL-violation, this corresponds to $\alpha = -1$ with $\lambda \sim 0.1$ mm.
%
%
\section{\label{sec:app}Apparatus}
\subsection{\label{subsec:princip}General Principles}
\begin{figure}[!]
\hfil\scalebox{.53}{\includegraphics*[1.2in,1.5in][7.3in,7.2in]{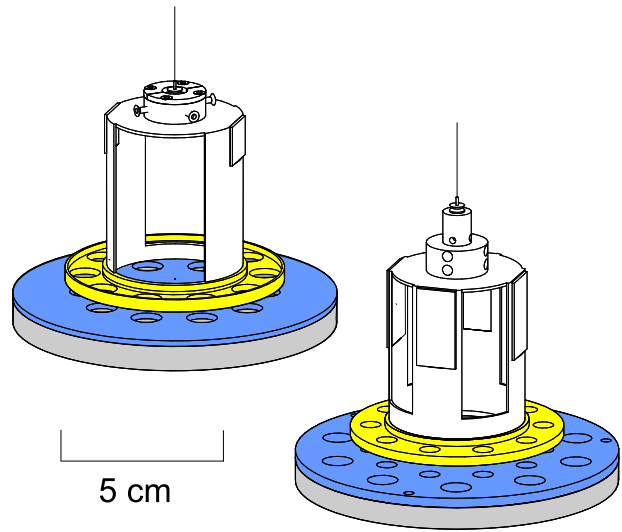}}\hfil
\caption{[color online] Torsion pendulums and rotating attractors used in Experiments I (left) and II (right). The active components are shaded. For clarity, we show an unrealistically large vertical separation between the pendulums and attractors, and omit the conducting membranes and attractor drive mechanisms.}
\label{fig:pendulums}
\end{figure}
To test the ISL at smaller length scales than had been studied before, 
we developed a new torsion-balance instrument,
shown schematically in Fig.~\ref{fig:pendulums}, that used planar test bodies rather than cylindrical\cite{ho:85} or spherical\cite{mi:88}
bodies that had been employed previously. Our test bodies were the ``missing masses'' 
of holes bored into cylindrically symmetrical plates. In each of the two
first-generation experiments reported here, the active
component of the torsion pendulum was a thin ring containing 10 cylindrical holes equally spaced around the azimuth, and 
the pendulum was suspended above a uniformly rotating, circular attractor 
disk containing
10 similar holes. In the absence of the holes, the disk's gravity
would pull directly down on the ring and could not
twist it. But because of the holes, the ring was twisted by a torque 
$N(\phi)=-\partial V(\phi) /\partial \phi$ 
where $V(\phi)$ is the potential energy of the ring in the field of the
disk when the disk's holes are displaced by an angle $\phi$ with respect to those in the pendulum. This torque 
oscillated 10 times for every revolution of the disk. $V(\phi)$ was not a simple sinusoidal function of $\phi$ so that
rotating the attractor at frequency $\omega$ produced torques at frequencies of $10\omega$ and its integer multiples. 
The $10\omega$, $20\omega$, 
and $30\omega$ torques, $N_{10}$, $N_{20}$, $N_{30}$, were measured as functions of the vertical separation between the bottom of the pendulum
and the top of the attractor (higher harmonic twists were greatly attenuated by the 
inertia of the pendulum and did not provide useful signals).
By placing the signals at high multiples of the
rotation frequency, $\omega$, we reduced many
potential systematic errors. 
We minimized electrostatic interactions between the attractor and pendulum
by interposing a stiff conducting membrane between the attractor and the pendulum
and surrounding the pendulum with an
almost complete Faraday cage.

The experiments were turned into approximate null measurements by attaching a second, 
thicker, disk to the bottom of each attractor. This disk
also had 10 equally-spaced holes bored into it, but the holes were rotated by 18 
degrees compared to those in the upper disk. The dimensions of these thicker and larger-diameter holes
were chosen so that the $10\omega$ Newtonian torque on the pendulum from the upper attractor holes was essentially canceled by the $10\omega$ Newtonian torque from the lower holes. On the other hand, torques
from a short-range interaction with a length-scale less than the thickness of the upper attractor disk could not be canceled because the lower attractor
was too far from the pendulum ring. The $20\omega$ and $30\omega$
Newtonian torques were reduced much less substantially than the Newtonian $10\omega$ torque.

\begin{figure}
\hfil\scalebox{.53}{\includegraphics*[0.8in,2.1in][5.5in,9.8in]{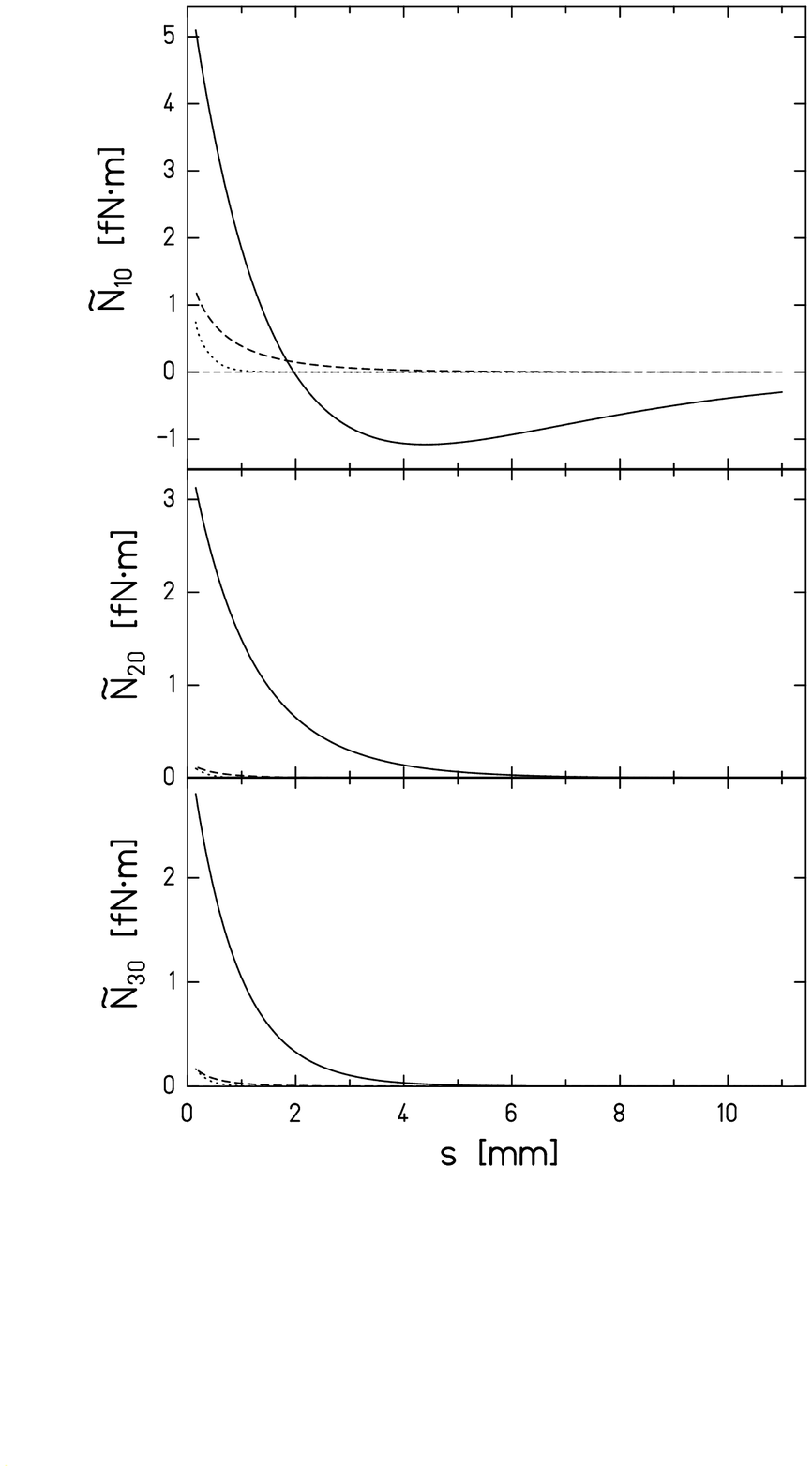}}\hfil
\caption{Predicted torques for Experiment I as a function of separation $s$. Solid curves are Newtonian gravity; dotted curves are for a Yukawa interaction with $\alpha=3$, $\lambda=0.25$ mm; dashed curves are for a power-law interaction with $\beta_3=0.05$.}
\label{fig:predicted torques}
\end{figure}

Data were taken at separations ranging from $s=10.77$ mm to $s=137~\mu$m, where $s$ is the distance from the top of
the attractor to the bottom of the pendulum. The signatures distinguishing conventional gravity from new short-range physics were
\begin{enumerate}
\item a characteristic shape of the Newtonian $10\omega$ torque $N^G_{10}$.
The cancelation of $N^G_{10}$ was a strong
function of $s$. In Experiment I, $N^G_{10}$ was exactly canceled at $s\approx 2$ mm,
undercanceled for $s<2$ mm
and overcanceled for $s>2$ mm. On the other hand, for ranges of interest the Yukawa torque $N^Y_{10}$
was a monotonically decreasing function of $s$. As a result, the exact location of the 
zero-crossing was very sensitive to any violation of the ISL. In Experiment II, the cancelation
occured at $s \approx 3.3$~mm.
\item a relatively high harmonic content of the Newtonian torque. $N^G_{20}$ and $N^G_{30}$ were comparable 
to $N^G_{10}$ because $N^G_{10}$ was highly canceled while $N^G_{20}$ and
$N^G_{30}$ torques were not.
On the other hand, $N^Y_{20}$ and $N^Y_{30}$ will be much less important than $N^Y_{10}$.
\end{enumerate}
The predicted Newtonian, Yukawa and power-law torques are shown as functions of $s$ in Fig.~\ref{fig:predicted torques}.

We inferred the harmonic components of the torque from the pendulum twist which we measured by reflecting
a laser beam from a mirror attached to the pendulum body. 
These measured torques 
were then compared to calculations of 
the expected Newtonian and possible Yukawa and power-law effects, allowing for uncertainties in key experimental parameters, such as the actual
masses of the holes. Agreement of the best-fit values of these
parameters with values determined  independently added to our confidence in the results.  
\subsection{\label{subsec:pends}Torsion Pendulum}
The torsion pendulums used in the two experiments, shown in Fig.~\ref{fig:pendulums}, were quite similar.
The active components of the pendulums were rings made from 7075~aluminum with 10 cylindrical
holes evenly spaced about the azimuth. The relevant properties of the rings---the hole radius $a_{\rm p}$ and height $h_{\rm p}$, the distance $r_{\rm p}$ from the hole centers to the pendulum's symmetry axis, and the ``missing mass''
$M_{\rm p}$ (the mass removed by drilling the holes)---are shown in Table~\ref{tab:pends}.  The mass measurements are discussed in Section~\ref{subsubsec:missmass}.
\begin{table}
\caption{Properties of the pendulum holes. All quantities are averages over the 10 holes. The mass refers to the missing mass of a single hole.}
\label{tab:pends}
\begin{ruledtabular}
\begin{tabular}{lcc}
Property       &   Experiment I     &   Experiment II \\
\hline
$h_{\rm p}$ (mm) & $2.002 \pm 0.002$  &  $2.979 \pm 0.002$ \\
$a_{\rm p}$ (mm)  & $4.7725 \pm 0.0015$  &  $3.1873 \pm 0.0015$ \\
$M_{\rm p}/10$ (g)       & $0.3972 \pm 0.0004$ & $0.266177 \pm 0.000014$ \\
$r_{\rm p}$ (mm) & $27.665 \pm 0.005$   & $26.667 \pm 0.005$  \\
\end{tabular}
\end{ruledtabular}
\end{table}
All quoted masses include small corrections for buoyancy in air and for the thin layer
of gold that covered the entire surface. This gold layer provided an uniform electrostatic surface that minimized variations in the surface electrical potential;
gold does not oxidize and has unusually small differences in the work functions of its
different crystal faces (see Ref.~\cite{ad:03}).
The entire pendulums, including the mirrors, as well as the tops of the upper attractor plates were gold coated for this same reason. In addition,
the pendulums were nested inside a gold-coated copper shroud that contained small holes for the torsion
fiber and the laser beam.

The pendulum rings were mounted on gold-coated aluminum frames 
that held the mirrors used by the optical system which
measured the pendulum twist. The frame for Experiment I had two-fold azimuthal symmetry with two
mirrors. In Experiment II we reduced the gravity-gradient noise by adopting a four-fold
symmetric frame with four mirrors. 
The upper portions of the frames contained screws for leveling the pendulum rings to 
the horizontal.
In Experiment I, these screws translated the fiber attachment point relative to the rest of the pendulum body, while in Experiment II, 
the screws adjusted the pendulum's center-of-mass so that it lay directly under the fiber. 
The total mass suspended from the fiber was $28.2527\pm 0.0002$~g in Experiment I and $39.6463\pm 0.0002$~g in Experiment II.
\subsection{\label{subsec-att}Rotating Attractor}
\begin{figure}
\hfil\scalebox{.75}{\includegraphics*[1.3in,0.9in][5.5in,3.2in]{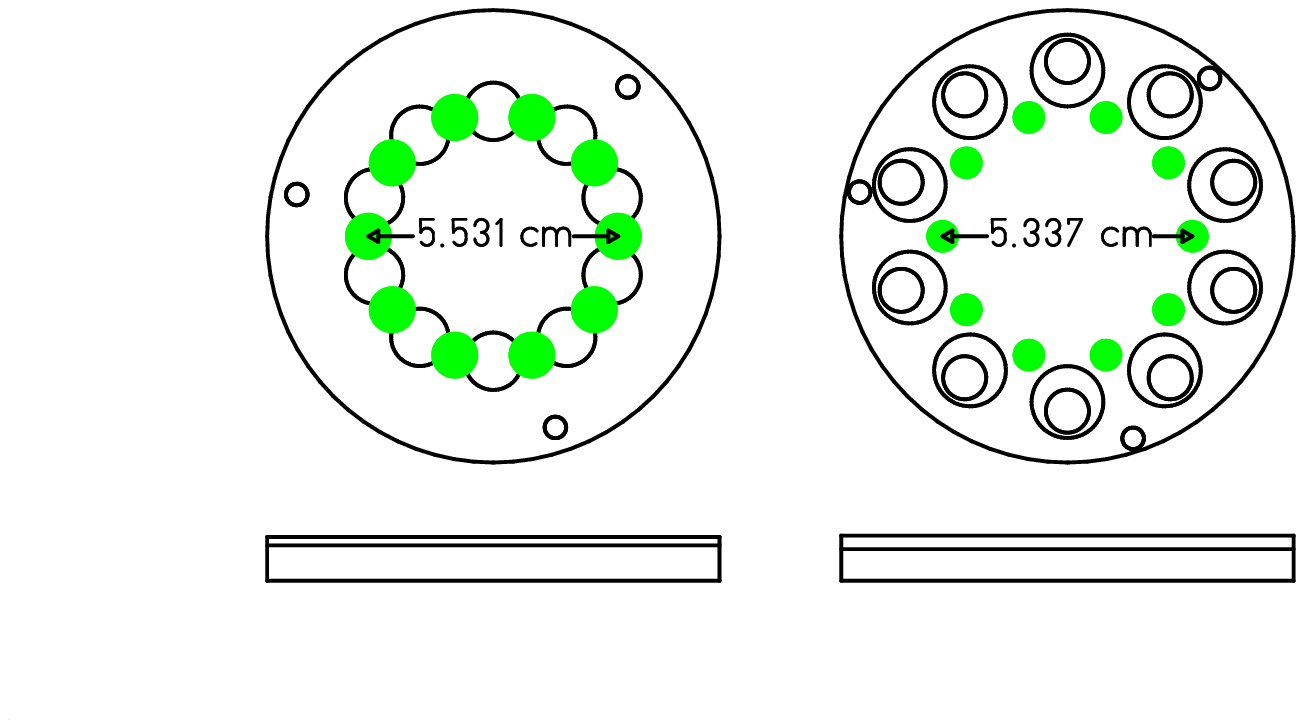}}\hfil
\caption{[color online] Top and side views of the attractor disks. The disks from Experiment I are on the left. The solid circles represent the in-phase upper-plate holes. The open circles show the out-of-phase holes in the upper and lower plates. The three small holes held pins that aligned the upper and lower disks.} 
\label{fig:attractors}
\end{figure}

The attractors for Experiments I and II, shown in Fig.~\ref{fig:attractors}, 
were pairs of flat, concentric, high-purity
copper disks. The upper disks were thinner and contained holes which, when properly rotated, aligned with those on the pendulum. The torque produced by these ``in-phase'' holes on the pendulum was opposed by the 10 ``out-of-phase'' holes in the lower disks that were rotated by $\pi/10$ relative to those in the upper plate (the upper attractor disk in Experiment~II also contained out-of-phase holes).  
The disks for both experiments were lapped to a flatness of $\pm0.003~$mm and the 
two sides of each disk were parallel to better than $30~\mu$rad. The top surface of the attractor assembly was coated with gold. Tables~\ref{tab:attI} and~\ref{tab:attII} show the relevant properties of the disks. 
\begin{table}
\caption{\label{tab:attI}Properties of the Experiment~I attractor holes. All dimensions are averages over the ten holes; $r_{\rm a}$ is the distance of the hole centers from the rotation axis, and the masses are given for a single hole.} 
\begin{ruledtabular}
\begin{tabular}{lcc}
Property & Upper Disk & Lower Disk\\
\hline
$h_{\rm a}$ (mm) & $1.847\,\pm\,0.005$ & $7.828\,\pm\,0.005$\\
$a_{\rm a}$ (mm) & $4.7690\,\pm\,0.0015$ & $6.3449\,\pm\,0.0015$\\
$M_{\rm a}/10$ (g) & $1.1770\,\pm\,0.0004$ & $8.8666\,\pm\,0.0019$\\
$r_{\rm a}$ (mm) & $27.655\,\pm\,0.005$ & $27.664\,\pm\,0.005$\\
\end{tabular}
\end{ruledtabular}
\end{table}

\begin{table}
\caption{\label{tab:attII}Properties of the Experiment~II attractor holes.
Uncertainties in the least significant figures are shown in parentheses.} 
\begin{ruledtabular}
\begin{tabular}{lccc}
Property 		 & \multicolumn{2}{c}{Upper Disk}  & Lower Disk \\
   &  In-phase  & Out-of Phase & Out-of-Phase \\
\hline
$h_{\rm a}$ (mm) & 3.005(2)       & 3.005(2)	  & 6.998(2) \\
$a_{\rm a}$ (mm)  & 3.1930(15)       & 4.7720(15)       & 7.9693(15)\\
$M_{\rm a}/10$ (g)   & 0.86228(4)     & 1.92592(5)     & 12.50984(6)\\
$r_{\rm a}$ (mm) 		   & 26.685(5)      & 38.697(5)      & 36.666(5) \\
\end{tabular}
\end{ruledtabular}
\end{table}
\subsection{\label{subsec:membrane}Electrostatic Shield}
The main component of the electrostatic shield was a tightly stretched 20-$\mu$m-thick, 11.4-cm-diameter BeCu foil clamped between two lapped aluminum rings. The foil and rings were assembled in a liquid nitrogen bath. As the unit warmed to room temperature, the differential thermal expansion between the BeCu foil and the aluminum rings created the high tension necessary to hold the foil flat. Inspection with a microscope showed that the membrane was flat to within $\pm$15~$\mu$m. 
The tension was sufficient to give the foil a fundamental ($m$=0, $n$=1) 
drum-head frequency of $f_{0,1}\approx$850~Hz. We found this resonance by raising the frequency
of an acoustical source until small particles placed on the membrane began to jiggle. 
\subsection{\label{subsec:susp}Pendulum Suspension System}
The pendulum suspension system consisted of a main torsion fiber hung from an eddy-current damper that,
in turn, hung from a short ``pre-hanger'' fiber that was attached to a
computer-controlled $x$-$y$-$z$-$\theta$ stage. The system was electrically insulated so that we could measure
the pendulum-electrostatic shield capacitance (see Sec.~\ref{subsec:vert}). The pendulum was grounded externally to 
the electrostatic shield in normal operation.
\subsubsection{Torsion Fiber}
The main torsion fiber was an 80-cm-long, 
20-$\mu$m-diameter, gold-coated tungsten wire. The fiber was attached to the pendulum frame with a 
crimp tube embedded in a copper screw. The fiber could support a 100~g load, so the pendulums, which had masses less than 40~g, 
did not stress the fiber near its elastic limit. 
The fiber's torsional spring constant, $\kappa$, may be determined from the relationship $\kappa=\Omega^2 I$, where $\Omega$ is the pendulum's resonant angular frequency and
$I$ is its rotational inertia which we computed from a detailed model of the pendulum based on the measured dimensions and masses of its components.

In Experiment I, $\Omega$ was $15.720{\pm}0.002$~mrad/s and we calculated $I=128\pm1$~g-cm$^2$, yielding $\kappa\,=0.0317\pm0.0002$~erg/rad. In Experiment II we found $\Omega=13.454\pm0.002$~mrad/s and computed $I=175 \pm1$~g-cm$^2$, yielding $\kappa\,=0.0317\pm0.0002$~erg/rad. As expected, the inferred $\kappa$'s were essentially identical because the same torsion fiber was used in both experiments. 
The fiber hung from
a passive damper consisting of an aluminum shaft and disk suspended between two 
ceramic ring magnets. The damper's axial symmetry minimized any damping of the torsional motion (which by the
fluctuation-dissipation theorem would have contributed to our noise), but 
the damper ``killed'' the swing, wobble and guitar-string modes of
the pendulum; the typical exponential damping time of such oscillations being approximately 
1000~s. 
The top of the damper shaft, in turn, hung from a 3-cm-long, 150-$\mu$m-diameter ``pre-hanger'' 
fiber. 
This torsionally stiffer prehanger acted as a gimbal, reducing 
any tilt-twist coupling to the torsion fiber. 

The fiber was prepared by heating it under load in vacuum to $\sim 70^{\circ}$C for
12-24 hours. After  a day of cooling, the fiber's drift rate--the slow change in the fiber's equilibrium twist
angle--was typically reduced to $\leq 1~\mu$rad/h.
\subsubsection{\label{subsubsec-phitop}Fiber Positioning Mechanism}
The pendulum was positioned relative to the attractor by hanging the suspension fiber from an
$x$-$y$-$z$-$\theta$ stage
located outside the vacuum chamber at the top of the instrument. 
The shaft holding the pendulum suspension system was fed through the vacuum vessel 
via a rotary O-ring seal, 
and a welded bellows allowed for horizontal and vertical translation. The full range of vertical translation was 25 mm, while the horizontal travel was limited to 8 mm.  

Three electronically controlled actuators (Newport Corporation model CMA-25PP with an ESP-300 controller) drove the
$x$, $y$, and $z$ stages. The rotation stage (Newport Corporation model URM80APE) controlled the equilibrium angle of the torsion fiber and was also used to remove torsional energy from the pendulum before starting a run. 
\subsection{\label{subsec-attdrive}Attractor Rotation Drive}
The attractor was held in a copper cup with three-fold azimuthal symmetry. The symmetry was designed to allow for gravitational centering (see, for example, Ref.~\cite{sm:00}), but was ultimately not used as a more effective method was found (see Sect.~\ref{subsec:cent}). The cup was mounted on an electrically insulated and largely
non-magnetic bearing consisting of an 8~mm-diameter stainless-steel shaft held by two
bearings with ceramic balls riding in Stellite races. 
The entire bearing assembly was mounted on 3 leveling screws that were used to make the
attractor precisely parallel to the electrostatic shield as discussed in Section~\ref{subsubsec:att-lev}.   

The attractor drive motor (made by SAIA-Burgess) was a 48-pole stepper with a 5000:1 reduction gear giving 240,000~steps per attractor revolution. Drive pulses were provided by a Stanford Research Systems DS345 Function generator and
the attractor angle was determined by counting the drive pulses. The absolute angle was obtained from a LED/photodiode pair that was interrupted once per revolution by a tab on the attractor drive mechanism,
producing an index pulse that was recorded by our data-taking computer. The index was constant to within 0.1~degrees throughout a run ($\sim$10~attractor revolutions).
The motor was mounted outside the vacuum vessel and drove the attractor
through an O-ring seal.
\subsection{\label{subsec-auto}Autocollimator}
\begin{figure}[b]
\hfil\scalebox{.72}{\includegraphics*[1.9in,0.4in][6.0in,4.3in]{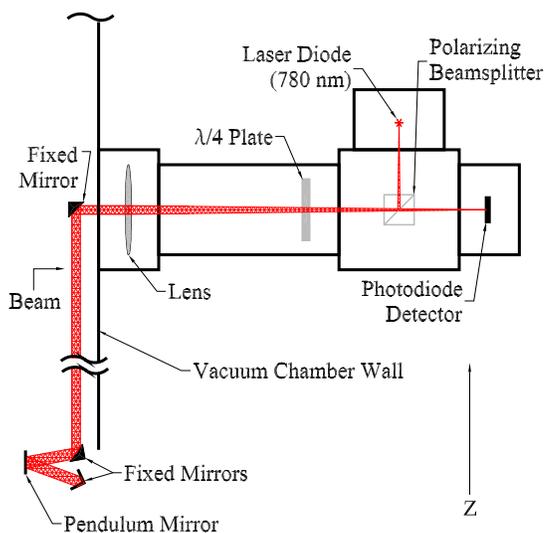}}\hfil
\caption{[color online] Schematic diagram of the optical system that measured the pendulum twist.}
\label{auto}
\end{figure}
The pendulum twist was measured by the autocollimator shown in Figure~\ref{auto}. 
Collimated light from a 20~mW, 780~nm, actively stabilized laser diode (Sony SLD301), placed at the focus of a 30~cm lens, was reflected by a polarizing beam splitter before
passing through the lens to produce a 6-mm-diameter parallel beam. The beam entered the vacuum vessel through an anti-reflective window, was reflected twice 
by two fixed, flat mirrors, and then struck a flat mirror on the pendulum. The light reflected off the pendulum struck a fixed 90$^\circ$ mirror which directed the light back to the pendulum. The beam then retraced its path to the autocollimator. A quarter-wave plate ensured that very little light bound for the detector returned to the laser diode. The optical power in the light beam was $\sim 15~\mu$W. This optical system gave a detector signal that was 
nominally insensitive to the `swing', `wobble' and `guitar-string' motion of the torsion pendulum
but had twice the optical gain compared to our previous\cite{sm:00} autocollimators.

The position-sensitive detector (United Detector Technology 1x3~mm model) was a 3-terminal, analog, resistive-division device. The location of the light spot (focused to a diameter of about 0.1~mm 
on the detector) was inferred from the ratio of the difference $\Delta$ and sum $\Sigma$  currents received at the two ends of the detector.
The laser beam was chopped at 141.7~Hz, and Stanford Research Systems model SR830 lock-in amplifiers generated the $\Delta$ and $\Sigma$ signals that were digitized and recorded by the data-taking computer. The time-constants of the lock-in amplifiers were 
normally set to nominal values of 10 s.
\subsection{\label{subsec:vac}Vacuum Vessel}
The vacuum vessel was used in a previous short-range test of the equivalence principle 
\cite{sm:00}. A small ion pump kept the pressure inside the chamber near 10$^{-6}$ Torr. 
The vacuum vessel was placed on a vibration-dampened kinematic mount directly attached to an unused cyclotron magnet in our laboratory. The magnet was relatively insensitive to building vibrations and sway because it had an independent foundation and a mass of about 200 tonnes. The vessel could be tilted with leveling screws to align the attractor assembly with the free-hanging pendulum as described in Sec.~\ref{subsubsec:att-lev}
\subsection{\label{subsec:supp}Mechanical Support and Shielding}
The entire apparatus and sensitive electronics were enclosed in a thermally insulating box made of 5-cm-thick foam. An external device (Neslab model RTE-221) circulated constant-temperature water through a radiator/fan assembly inside the thermal enclosure. The air temperature inside the box varied by $\leq$\,30~mK/day and variations at the $10\omega$ signal frequency
were less than about 1 mK. 

A single layer of 0.76-mm-thick mu-metal surrounded the lower half of the vacuum chamber. A second mu-metal plate was mounted between the motor and the vacuum vessel. This shielding reduced the ambient field by a factor of $\sim$\,80. The constant residual field inside the chamber was $<$5~mG.
\subsection{\label{subsec-env}Environmental Monitors}
We continuously monitored the in-phase
and quadrature outputs of the $\Delta$ and $\Sigma$ lock-ins as well as several important environmental parameters:
\begin{enumerate}
\item
two Applied Geomechanics Instruments model 755 electronic tilt sensors, attached to the top of the vacuum vessel, monitored apparatus tilt. 
\item
vertical and horizontal Teledyne Geotech model S-500 seismometers, mounted on the vacuum vessel support frame, monitored seismic disturbances during Experiment II. 
\item
eight Analog Devices model AD590 temperature sensors  monitored the temperature variations of the apparatus and the air around it: 3 were mounted on the
vacuum vessel, 4 measured air temperatures inside and outside the thermal shield, and 1 measured the temperature of the radiator water bath. All sensors were encased in small copper blocks to increase the short-term stability of the readings. 
\item the ion-pump current served as a vacuum gauge.
\end{enumerate}
\subsection{\label{subsec-calibtt}Calibration Turntable}
The gravitational calibration, described in Section \ref{subsubsec:abscal}, employed an external turntable that
rotated 2 brass spheres uniformly around the vacuum vessel. 
The turntable was made from aluminum and driven by a motor identical to that which drove the attractor and had a similar angle index marker. The turntable platter had 6 evenly spaced holes sized to seat the two 5.08~cm-diameter brass spheres used in the gravitational calibration. 
\subsection{\label{subsec:daq}Data Acquisition System}
We recorded data with a 80286 computer containing an Analogic LSDAS-16 interface card. The card had 16 analog input channels with 16-bit analog-to-digital converters, one of which was multiplexed to record up to 16 temperatures. Digital input registers counted the attractor drive pulses, giving us the instantaneous relative angle of the 
attractor, and the index pulse that gave us absolute angle information.
Digital outputs controlled the attractor rotation direction and turned on and off the rotation. 

Data were taken at regular intervals determined by the computer's clock, typically between 5 and 15~s. 
The data strobing interval was set to $\tau_{\circ}/4 n$ where $\tau_{\circ}$ was the nominal period of free
torsional oscillations and $n$ an integer that ranged between 8 and 17.
%
%
\section{\label{sec:alignpos}Alignment and Positioning}
\subsection{\label{subsubsec-attlev}Positioning the Attractor Relative to the Electrostatic Shield}
We aligned the attractor with respect to the electrostatic shield 
by temporarily replacing the Be/Cu foil and holder with a plate containing a hole through which we passed a depth micrometer probe
to measure the distance to the attractor. This plate had the same thickness as one of the aluminum rings that held the electrostatic shield membrane as they were lapped simultaneously.
By measuring the distance above each attractor leveling screw and making iterative adjustments, we made the attractor parallel to the lapped plate 
(and therefore to the electrostatic shield) to within 0.1~mrad. The distance $d_{\rm am}$ between the membrane and the upper surface of the attractor, 47$\pm$1~$\mu$m~(54$\pm$3~$\mu$m) in Experiment~I~(II), was determined by averaging
the results of the depth-micrometer measurements, 48$\pm$3~$\mu$m~(57$\pm$3~$\mu$m), with values inferred from 
membrane-attractor capacitance measurements, 47$\pm$1~$\mu$m~(51$\pm$1~$\mu$m). 
\subsubsection{\label{subsubsec:lev}Leveling the Pendulum}
\begin{figure}
\hfil\scalebox{.88}{\includegraphics*[0.6in,0.5in][4.3in,2.1in]{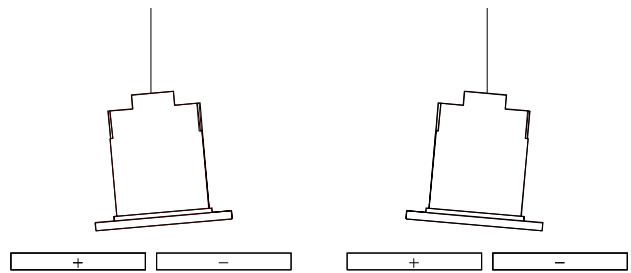}}\hfil
\caption{Schematic diagram of the leveling measurement. Suppose the pendulum were tilted so that the ``$+$'' plate has a higher capacitance as shown in the left diagram. The right diagram shows the pendulum after the upper fiber attachment has been rotated by 180$^\circ$; now the ``$-$'' plate has the higher capacitance.}
\label{cap_diff}
\end{figure}

\begin{figure}
\hfil\scalebox{.8}{\includegraphics*[0.6in,0.5in][4.8in,3.4in]{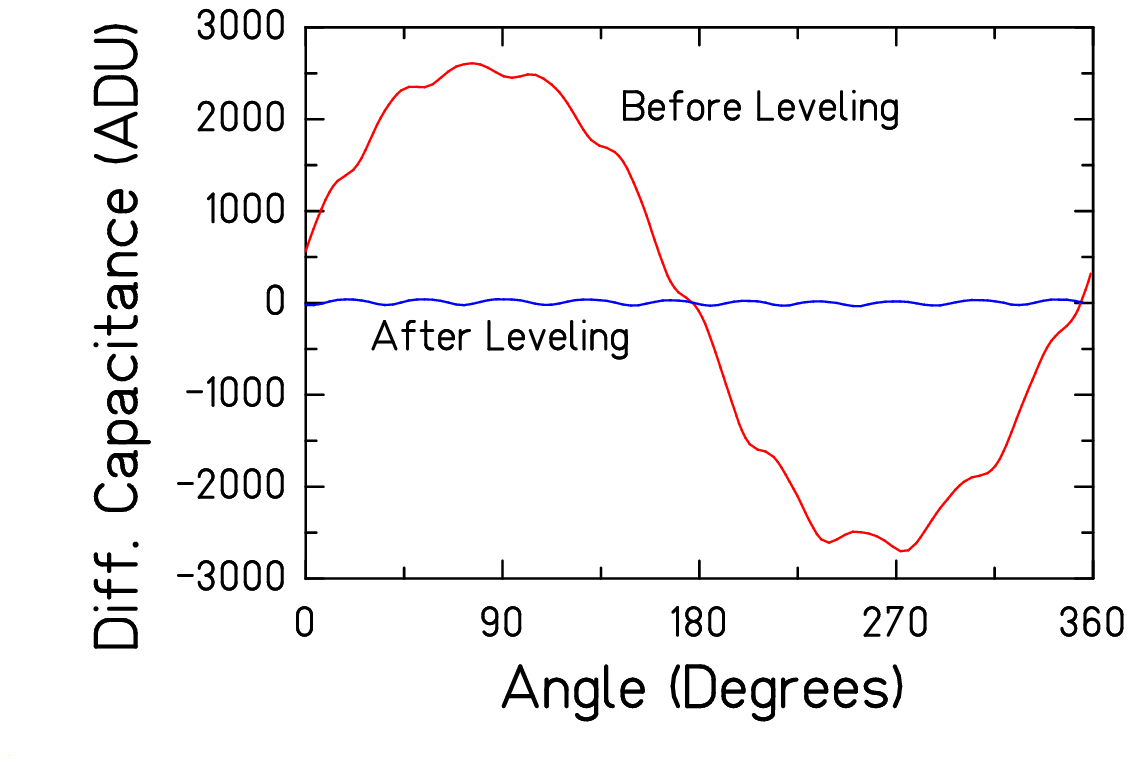}}\hfil
\caption{[color online] Differential capacitance for one revolution of the pendulum before and after leveling. The tilt signal occurs at 1$\omega$; the 10$\omega$ ripple results from the ten holes passing over the gap between the plates of the capacitor. The 1$\omega$ amplitudes correspond to pendulum tilts of 7.52$\pm$0.06~mrad and 0.03$\pm$0.01~mrad before and after leveling, respectively. }
\label{caprev}
\end{figure}

The pendulum ring was leveled (made perpendicular to the fiber) by using it as one plate of a differential capacitor as shown in Fig.~\ref{cap_diff}. The other two plates were semi-circular, horizontal copper pieces, separated by a thin gap, that were installed in place of the electrostatic shield. The pendulum was suspended $\sim$\,1~mm above the 
plates and rotated uniformly using the fiber positioner $\theta$ stage. If the pendulum disk were perfectly horizontal, the differential capacitance between the two sides would not change as the pendulum was rotated. However, any pendulum tilt would cause the differential capacitance 
to vary at the rotation frequency. This 1$\omega$ signal was independent of any static tilt of the copper plates but was a strong signature of a tilted pendulum.  Figure~\ref{caprev} shows the differential capacitance signal for one revolution of the stage before and after leveling. 
Initially, the differential capacitance was dominated by the 1$\omega$ signal. However, as the pendulum approached level, the 1$\omega$ signal became so small that the modulation caused by the 10 holes passing over gap between the plates became the dominant harmonic component (this 10$\omega$ signal indicates that the gap between the semi-circular plates was not
precisely aligned under the torsion fiber axis). 
We iterated the leveling process until the 1$\omega$ differential capacitance signal corresponded to a misalignment of only 0.03$\pm0.01$~mrad (0.12$\pm0.01$~mrad). 

A false leveling signal could occur if the fiber did not lie directly along the rotation axis of the stage. 
In this case the center of the pendulum ring would have moved in a circle with respect to the copper plates and the ``run-out'' would have modulated the differential capacitance at the rotation frequency. 
The fiber run-out was measured with a theodolite
and, after adjustments of the upper fiber attachment, was reduced to 74$\pm$23~$\mu$m (24$\pm$8~$\mu$m) for Experiment~I~(II). 
The effect of any stray signal induced by the residual fiber run-out (or stage and shaft motion) was checked as follows. First, we ``parked'' the pendulum  on the capacitor plates and turned the $\theta$ stage through $\sim$3~revolutions. Then, the pendulum was raised to the original height and left to unwind freely. With nothing but the pendulum's tilt to influence the differential capacitance, any difference in the peak-to-peak amplitude between the freely unwinding case and the
case when the stage was rotating must have been due to a false effect. The observed difference introduced an additional uncertainty of $\pm$0.11~mrad ($\pm$0.01~mrad). The resulting value
for the pendulum tilt in Experiment~I~(II) was 0.03$\pm$0.11~mrad~(0.12$\pm$0.01~mrad).
\subsubsection{\label{subsubsec:att-lev}Aligning the Attractor}
\begin{figure}[b]
\hfil\scalebox{.80}{\includegraphics*[0.35in,0.4in][4.4in,1.4in]{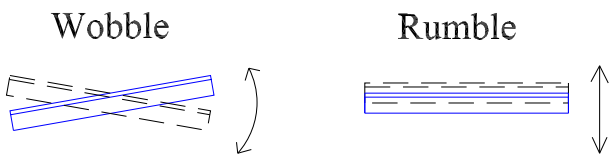}}\hfil
\caption{[color online] Possible attractor rotation imperfections. Vertical rumble was too small to measure with our instruments.} 
\label{rot-imp}
\end{figure}

Because our calculated Newtonian, Yukawa and power-law torques ignored most imperfections of the attractor's rotary motion, we had to reduce these effects to negligible levels. 
We measured the attractor's wobble and vertical and horizontal runouts (see~Figure~\ref{rot-imp}) with high-quality mechanical indicators. 
By adding thin shims between the attractor and holder, we reduced the wobble amplitude to $<\,$0.2~mrad~(0.05~mrad) in Experiment~I~(II). We found no measurable vertical runout within the $\pm3~\mu$m measuring error. The horizontal runout was $\pm 5~\mu$m. 

\begin{figure}
\hfil\scalebox{.56}{\includegraphics*[0.6in,0.5in][6.4in,4.6in]{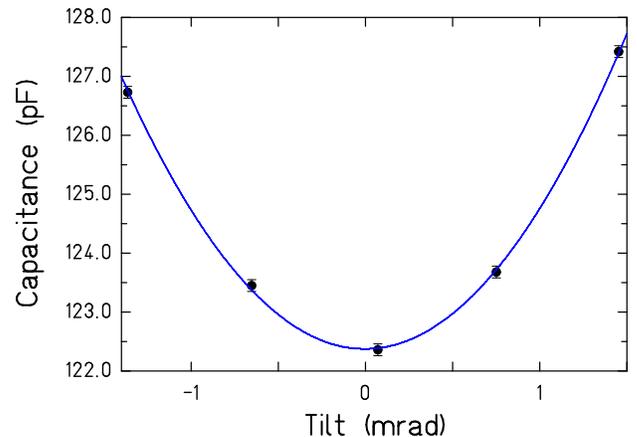}}\hfil
\caption{Data and parabolic fit for the pendulum-to-membrane alignment along one tilt axis. The capacitance is minimum when the pendulum and membrane are parallel. The error on the leveling, $\pm$18~$\mu$rad, was determined by the error on the fitted minimum position and the accuracy to which the tilt could be adjusted.} 
\label{pend-lev}
\end{figure}

We made the pendulum ring parallel to the conducting membrane using a capacitive technique. 
We tilted the entire apparatus and measured the pendulum-to-membrane capacitance as a function of the tilt using a low-noise, high-sensitivity Stanford Research Systems model SR720 LCR meter; because the capacitance is inversely proportional to the gap between flat plates, the minimum capacitance occurs when the pendulum and membrane are parallel. The capacitance at fixed z-actuator setting was measured for several values of the tilt along two axes and fitted with a parabolic function of tilt angle. The results for one scan are shown in Figure~\ref{pend-lev}. The tilt was then adjusted to the capacitance minimum along each axis. The error on this procedure was $\pm$18~$\mu$rad, determined by the accuracy of the tilt adjustment ($\pm$6~$\mu$rad) and the error on the fitted value of the minimum ($\pm$17~$\mu$rad).
\subsection{Centering the Pendulum on the Attractor}
The displacement between the pendulum and attractor was a key parameter in our experiments. The electronic actuators were sufficiently accurate (after accounting for the scale factors described in Section~\ref{subsubsec-stagealign}) and reproducible to give the {\em relative} $x$, $y$, and $z$ positions of the pendulum. It was more difficult
to determine $x_{\circ}$,  $y_{\circ}$ and $z_{\circ}$; the $x$, $y$, and $z$ values at which the pendulum 
would be resting on the attractor with its holes centered on the corresponding holes in the attractor.

\subsubsection{\label{subsec:cent}Horizontal Centering}
Our horizontal alignment relied upon
the gravitational torque on the pendulum being symmetric about $x_{\circ}$ and $y_{\circ}$. 
We took data at points offset along the $\hat{x}$ and $\hat{y}$ axes and used these points as part of our complete data set, the 
uncertainties in the $x$ and $y$ values were taken to be $\pm 2~\mu$m, the specified
reproducibility of the stages. The analysis left the alignment points, $x_0$ and $y_0$, as unconstrained free parameters. The fit was able to resolve the true alignment to $\pm$10~$\mu$m. Because the torque variation over an offset range of $\pm$100~$\mu$m was much less than our statistical uncertainty, this alignment was 
more than adequate.
\subsubsection{\label{subsec:vert}Vertical Separation}
We extracted the {\em absolute} vertical location of the membrane from ``z-scans'' where we measured the pendulum-to-electrostatic-shield capacitance, $C(z)$, as a function of z-actuator reading, $z$, using the
Stanford Research Instruments SR720 LCR meter. We checked that the AC fields used to measure
the capacitance did not appreciably deform the thin shield membrane by taking data at
drive voltages of 0.1 V and 1.0 V. 

We fitted the $C(z)$ data with
\begin{equation}
C(z)={\cal N} C_{\rm th}(z-z_{\ast})+C_{\rm st}~,
\label{eq:cap1}
\end{equation} 
where ${\cal N}$ was a normalization constant that accounted for the  
absolute calibration of the capacitance meter, $C_{\rm st}$ was the stray capacitance of
the suspension fiber, etc., and $z_{\ast}$ was the z-actuator reading
corresponding to zero vertical separation between the pendulum and the electrostatic shield.
The expected capacitance, $C_{\rm th}(\zeta)$, at a pendulum-shield separation $\zeta$ was calculated with an accurate finite-element model of the pendulum
and its electrostatic shield using ANSYS.

The torsion fiber was a linear spring as well, so that seismic vibrations excited 
the pendulum's vertical ``bounce'' mode. Ultimately, this motion limited the minimum achievable separation.
We also needed to account for bounce when determining the membrane position. The bounce amplitude was measured by raising the pendulum until its mirror intersected only half of the autocollimator light beam so that any vertical motion of the pendulum would
alter the amount of light returned to the detector. By measuring the change in the amount of returned light for a known
change in $z$ we determined that
$\Delta L / \Delta z\,=\,220~{\rm mV/mm}$,
where ${\Delta}L$ and ${\Delta}z$ are the changes in returned light and height, respectively. We monitored the 11.5~Hz (9.7~Hz) bounce oscillation on a spectrum analyzer and found the average bounce amplitude to be 13~$\mu$m, with a maximum value of 36~$\mu$m over the course of one day. 

The act of moving the vertical stage during a z-scan excited the bounce mode. We modeled the bounce as a sinusoidal modulation of the pendulum-to-membrane separation so that
Equation~\ref{eq:cap1} was modified to 
\begin{equation}
\label{eq:cap2}
C(z)={\cal N} \left[C_{\rm th}(\bar{z}) + \frac{\delta_{\rm b}^2}{2}\frac{\partial^2 C_{\rm th}(\bar{z})}
{\partial \bar{z}^2}\right]_{\bar{z}=z-z_{\ast}}\!\!\!\!\!\!\!\!\!\!\!\!\!+ C_{\rm st}~,
\end{equation}
where $\delta_{\rm b}^2$ is the mean square bounce amplitude.

\begin{figure}
\hfil\scalebox{.50}{\includegraphics*[0.6in,0.5in][7.5in,8.9in]{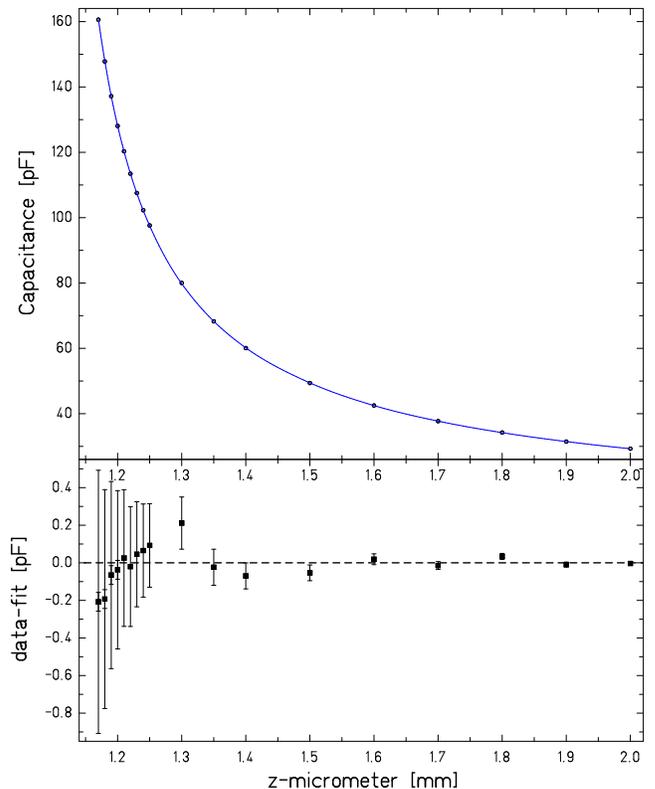}}\hfil
\caption{z-scan data and residuals from Experiment~I. The fit indicated the pendulum would touch the membrane at a z-actuator reading of 1.067$\pm$0.002~mm. The lower panel displays the fit residuals. The smaller set of error bars on the residuals represents the uncertainty in the capacitance values alone, while the larger ones include the effect of the $\pm$0.5~$\mu$m uncertainty in the relative stage positions.} 
\label{fig:fitcap}
\end{figure}

One z-scan typically contained about 25 different pendulum-shield separations between 
$\sim0.1-3$~mm. The data points and fit for one z-scan are shown in Figure~\ref{fig:fitcap}. The fitted values of the parameters in Equation~\ref{eq:cap2} are compared to their best estimates in Table~\ref{cappar}. The uncertainty in $z_{\ast}$ was typically $\pm$3~$\mu$m.
\begin{table}
\caption{\label{cappar}Comparison of the expected and fitted parameters of Equation~\ref{eq:cap2} for the z-scan of Figure~\ref{fig:fitcap}.} 
\begin{ruledtabular}
\begin{tabular}{cll}
Parameter & Expected Value & Fitted Value\\
\hline
${\cal N}$ & $1.00\pm0.01$ & $0.976\pm 0.005$ \\
$z_{\ast}$ & N/A & $1.067\pm 0.002$~mm \\
$C_{\rm st}$ & N/A & $0.78 \pm 0.11$~pF\\
$\delta_{\rm b}$ & $\sim 10~\mu$m & $5 \pm 14~\mu$m\\
\end{tabular}
\end{ruledtabular}
\end{table}
The z-actuator reading, $z_{\circ}$ at which the pendulum would just touch the attractor
was
\begin{equation}
z_{\circ}=z_{\ast} - t - d_{\rm am}~,
\end{equation}
where $t$ is the membrane thickness and $d_{\rm am}$ is defined in Section~\ref{subsubsec-attlev}.
The separation between the bottom surface of the pendulum and the upper surface of the attractor is then
\begin{equation}
s=z-z_{\circ}~.
\end{equation}
The uncertainty $\delta z_{\circ}= \pm 5~\mu$m was found by combining in quadrature
the uncertainties in $z_{\ast}$ ($\pm 3~\mu$m), $t$ ($\pm 3~\mu$m) and $d_{\rm am}$ ($\pm 2~\mu$m); the uncertainty in the $z$ readings was obtained by combining capacitance and z-micrometer data and varied between 3 (1.6)~$\mu$m and 6 (4)~$\mu$m in Experiment I (II).
For the upper-attractor-disk-only 
runs in Experiment~I the vertical position at vanishing pendulum-to-upper-disk separation, $z_{\circ}^{\prime}$, was
more uncertain, $\delta z_{\circ}^{\prime}=\pm0.025~$mm, because $d_{\rm am}$ was measured with a less precise micrometer.
\subsection{\label{subsubsec-stagealign}Alignment and Calibration of the Translation Stages}
\begin{figure}
\hfil\scalebox{.56}{\includegraphics*[0.6in,0.5in][6.4in,4.6in]{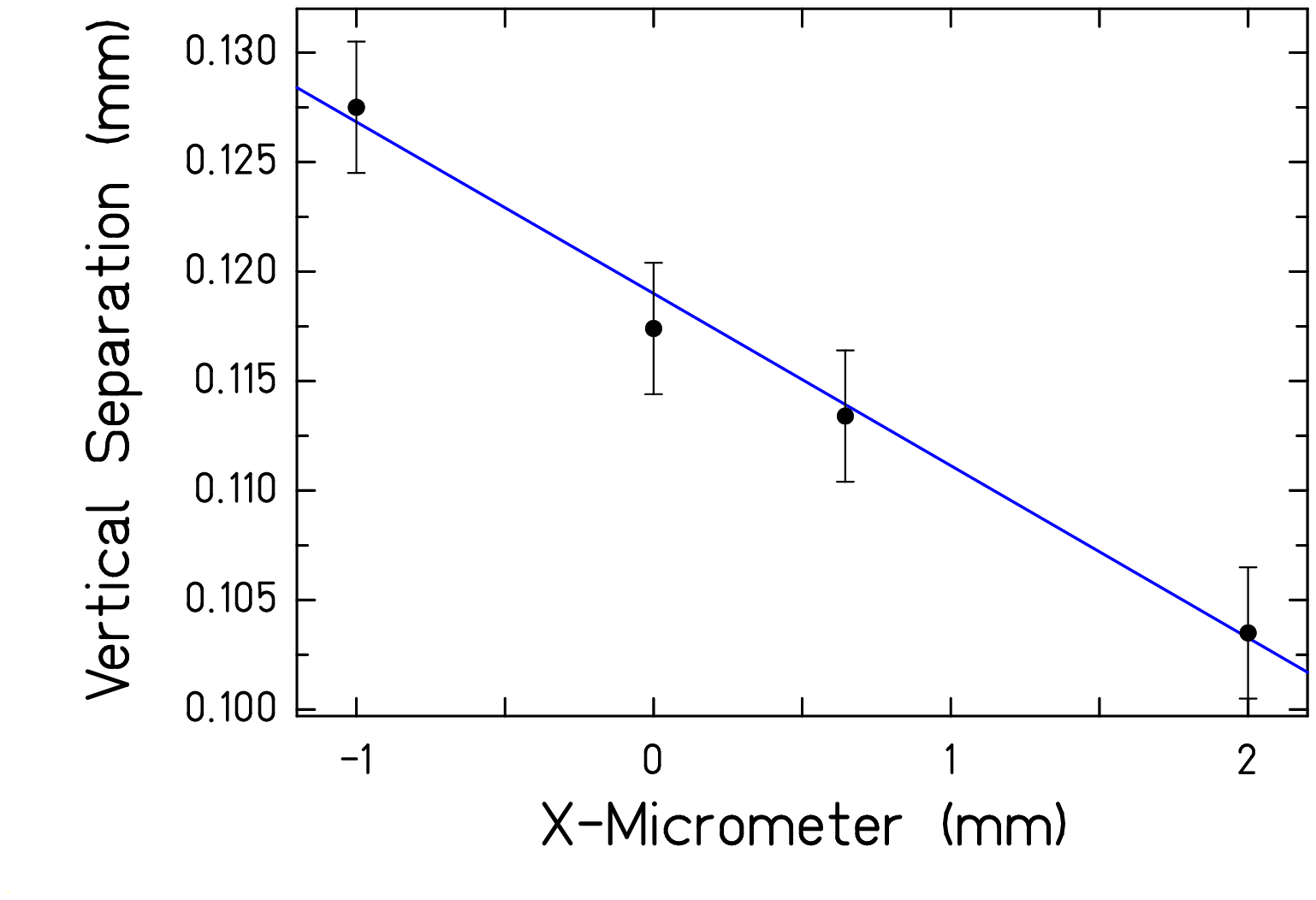}}\hfil
\caption{Dependence of the vertical separation on the horizontal position; the slope shows that the $x$-stage was tilted by 8~mrad from the horizontal.}
\label{stages}
\end{figure}

The alignment of the horizontal stages with respect to the plane of the electrostatic shield was checked by measuring
the change in the pendulum-shield capacitance as the $x$ and $y$-stages were moved (see Figure~\ref{stages}). 
We found a small misalignment ($x$-stage:~8~mrad, $y$-stage:~2~mrad) that was taken into account when 
computing the vertical separation values used in the data analysis. 

\begin{figure}
\hfil\scalebox{.56}{\includegraphics*[0.6in,0.5in][6.4in,4.6in]{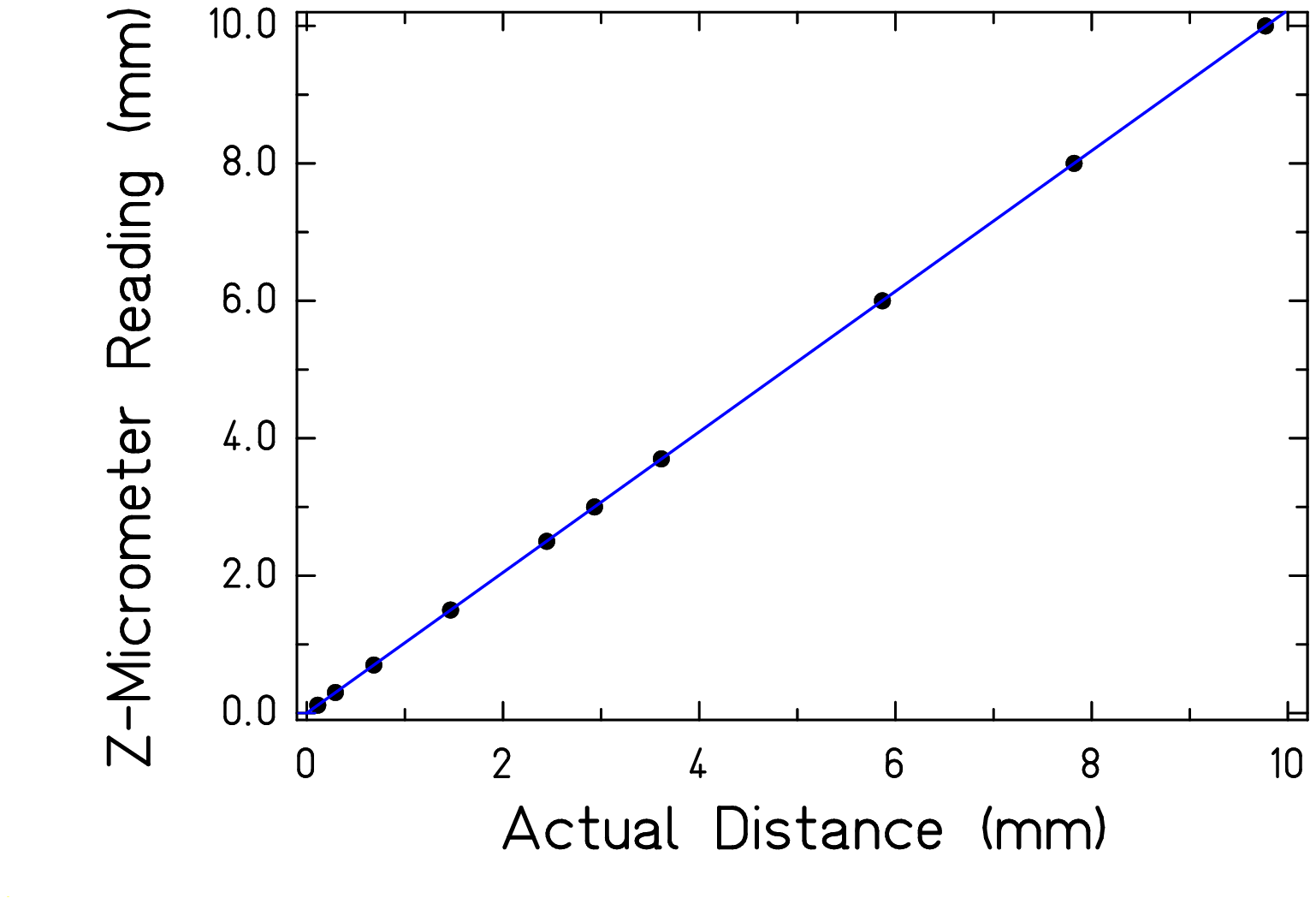}}\hfil
\caption{Calibration of the electronic z-actuator with a precision mechanical micrometer. The slope of the best fit line is 1.0230$\pm0.0002$, not the expected slope of 1.0000. The error bars are smaller than the size of the points.}
\label{zmic}
\end{figure}

The absolute calibration of the 3 electronically-controlled stage actuators caused us much trouble. Initially we assumed that the actuator readings were accurate to within 
their specifications. 
However, the Experiment I data did not agree with Newtonian predictions. Despite much effort,
we could not find any systematic error that could account for the results. Experiment II was designed solely to confirm this discrepancy, and showed a similar, but not identical, discrepancy. Finally, having exhausted all other possibilities, we
decided to check if the digital actuators were correct by
measuring the actual distance moved by each digital actuator with precision manual micrometers. 
The results are summarized in Figure~\ref{zmic} and Table~\ref{tab:mics}. We found highly linear relationships but the slopes were not unity! The position values used in our analysis were corrected with the appropriate calibration.
\begin{table}
\caption{\label{tab:mics}Ratio of apparent displacement of the computer controlled actuators
to the actual displacement determined using a precision manual counterpart. 
The ratios reveal that the stages did not move as far as expected.}
\begin{ruledtabular}
\begin{tabular}{cc}
Axis & Ratio\\
\hline
$x$ & 1.0239$\,\pm\,0.0002$\\ 
$y$ & 1.0248$\,\pm\,0.0002$\\
$z$ & 1.0230$\,\pm\,0.0002$\\
\end{tabular}
\end{ruledtabular}
\end{table}
%
%
\section{\label{sec:torque measurements}Torque Measurements}
\subsection{Protocols}
Data runs, which lasted about one day, were characterized by definite values of the vertical and horizontal positions of the pendulum. 
The data-sampling interval and the attractor rotation rate were both
synchronized to the pendulum's nominal free-oscillation period which was $\tau_0\,=\,399.7$~s (467.0~s)
in Experiment I (II).  
All sensors were recorded precisely 32 (68) times per free torsional oscillation period.
The attractor was rotated with a period of $\tau_{\rm att}= 17\tau_0$ (15$\tau_0$), so that the fundamental 10$\omega$ signal repeated every 1.7 (1.5) torsion cycles;
this placed the fundamental $10\omega$ signal at a sufficiently high frequency to suppress the $1/f$ noise 
(see Appendix~\ref{sec:noise}), but kept the $20\omega$ and higher harmonics from being too strongly attenuated by pendulum inertia. 
We analyzed signals up to 30$\omega$ (20$\omega$); higher harmonics were attenuated below the autocollimator noise level.
\subsection{Extracting the harmonic twist amplitudes}
\subsubsection{Converting the Autocollimator Output to a Twist Angle}
We converted the autocollimator difference $\Delta$ and sum $\Sigma$ signals into twist angles using 
\begin{equation} 
\theta\,=\,c_1\left(\frac{\Delta}{\Sigma}\right)\left[ 1\,+\,c_2\left(\frac{\Delta}{\Sigma}\right) \right],
\label{eq:theta def}
\end{equation}
where $c_1$ and $c_2$ are linear and quadratic calibration coefficients. The quadratic term arose
from small differences in the electronic gains of the signals from the two ends of the photo-detector as well as from 
non-linearities in the autocollimator's position-sensitive detector. We determined $c_2= -0.0416$ by
sweeping a light spot across the sensitive axis of the detector and
minimized our sensitivity to $c_2$ by periodically
adjusting the $\theta$-stage to compensate for fiber drift, keeping the mean position of light spot roughly constant. Our determination of $c_1$ is discussed in Sec.~\ref{subsubsec:abscal}.
\subsubsection{\label{subsubsec:dynamic-cal}Dynamic Calibration of the Autocollimator}
\begin{figure}
\hfil\scalebox{.56}{\includegraphics*[0.75in,0.5in][6.4in,4.6in]{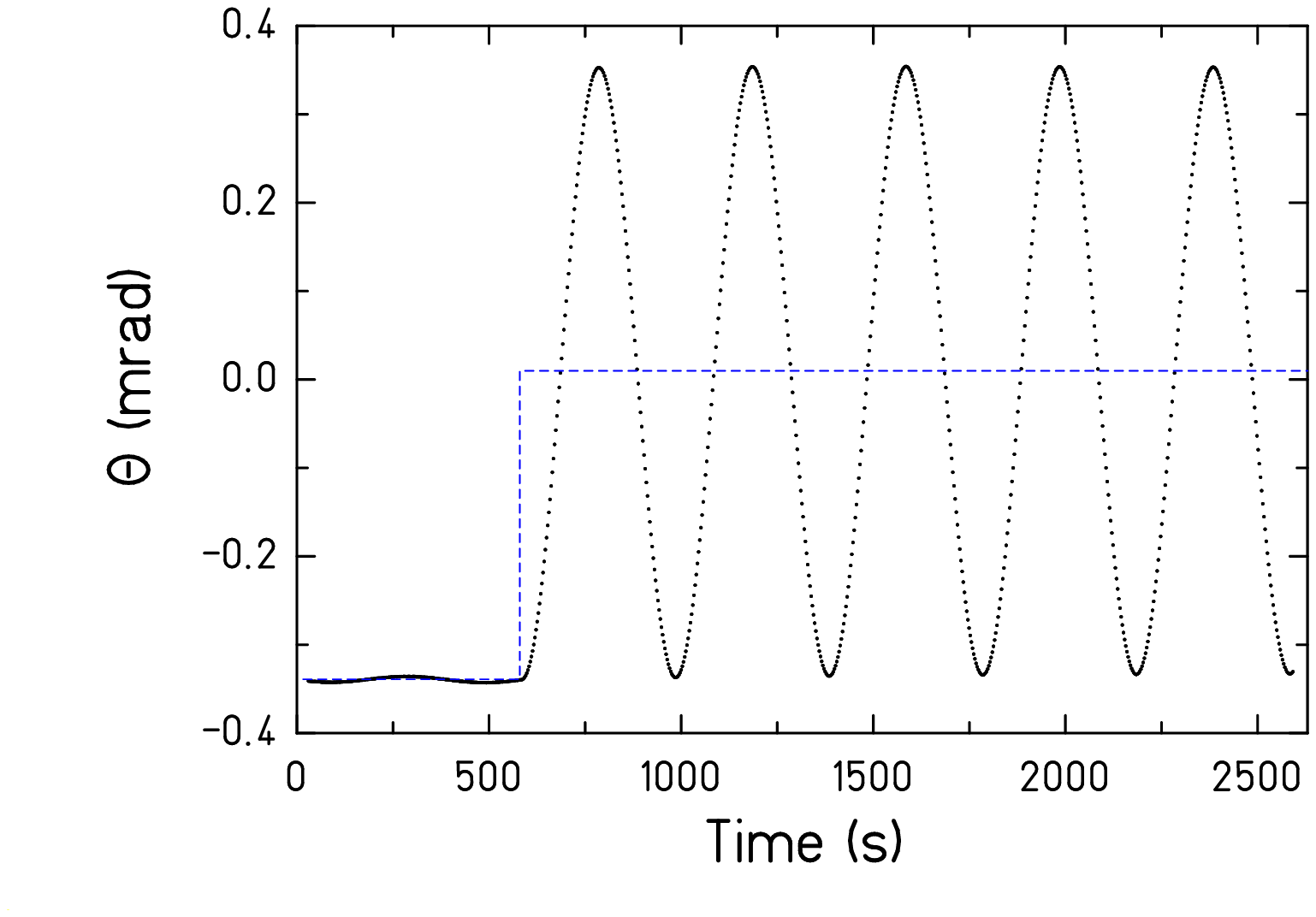}}\hfil
\caption{Dynamic calibration data. At t=580~s, the upper rotation stage changed the equilibrium position by 0.349~mrad. The dashed line depicts the equilibrium position as a function of time. The fit (not shown) passes so close to the data points that it cannot be distinguished from them in this plot.} 
\label{calsrt}
\end{figure}

We repeatedly checked the linearity of the deflection scale and measured the free-oscillation and decay times of the torsion oscillator with dynamic calibrations. Energy was first taken out of the pendulum's torsional mode by manipulating the $\theta$ stage until the pendulum had a very small twist amplitude, typically $A\sim\,5~\mu$rad. Then data taking was started. At a time $t_{\circ}$ after beginning the run the $\theta$ stage was abruptly rotated 
by a known amount, typically $\Delta \vartheta = 349~\mu$rad (0.02 degrees). The 
calibration coefficients were checked by fitting the pendulum's observed response $\theta(t)$ (which was an implicit function of $c_1$ and $c_2$)
to the expected change
\begin{eqnarray}
\theta_{\rm th}^{\rm dc}(t)&=& \alpha + \beta t + [A \sin(\Omega t + \gamma) +  \\
&&\Theta(t-t_{\circ}) \Delta
\vartheta (1-\cos \Omega (t-t_{\circ}))]e^{-(t-t_{\circ})/\tau_{\rm d}}~, \nonumber
\end{eqnarray}
where $\alpha$ and $\beta$ account for fiber ``drift'', $A$ and $\gamma$ account for the initial oscillation of the pendulum, $\Omega$ is the
free-oscillation frequency,
$\Theta$ is a unit step function, and $\tau_{\rm d}$ is the decay time of the pendulum amplitude.
Figure~\ref{calsrt} shows the data from one such calibration. 
The dynamic calibration uncertainty was dominated by the $\pm$8.5~$\mu$rad (2.5$\%$) 
uncertainty in the $\theta$-stage position (half the minimum angular step size and resolution of the encoder). Dynamic calibrations determined the nominal free-oscillation period, 
$\tau_{\circ}=2 \pi / \Omega$, and exponential amplitude-damping time, $\tau_{\rm d}$ (needed for the inertial corrections described in Section~\ref{subsec:atten}) and provided an excellent test of the linearity of the gross angular deflection scale. However, because the amplitude of the dynamic calibration data was necessarily much larger than the that of our ISL data ($\sim 5 \mu$rad) we did not use these data for the final calibration of the deflection (and therefore the torque) scale, but instead applied a known gravitational torque as described in Sec.~\ref{subsec: calibrate}.
\subsubsection{\label{subsubsec:filter}Filtering out the Free Oscillations}
The top panel of Figure~\ref{fig:dat} shows typical raw $\Delta$ autocollimator data.
The free torsional oscillation was the dominant component of the pendulum's twist
(we deliberately kept this amplitude large to facilitate precise measurements of $\Omega$). 
Because this motion provided no information about an ISL violation, we applied a digital filter to suppress the free oscillations before proceeding with the analysis. 
This ``notch'' filter averaged two data points separated by one half of a torsional period, and assigned the average value to their midpoint. 
\begin{equation}
\label{eq-tor}
\theta_{\rm f}(\phi)\,=\,\frac{1}{2}\left[\theta\left(t-\frac{\tau_{\circ}}{4}\right)\,+\,\theta\left(t+\frac{\tau_{\circ}}{4}\right)\right],
\end{equation}
where $t$ is the time when the attractor was at angle $\phi$, $\tau_{\circ}=2 \pi/\Omega$ is the torsional period, and  $\theta$ is the interpolated value of the twist. 
The middle panel of Figure~\ref{fig:dat} shows the filtered results.
\subsubsection{Harmonic Analysis of the Twist Data}
Each data run was subdivided into separate ``cuts'' containing exactly two oscillations of the 10$\omega$ signal (72$^{\circ}$ of attractor rotation). For each cut, $\theta_{\rm f}$ was fitted as a function of attractor angle $\phi$ with
\begin{equation}
\label{eq:fit}
\theta_{\rm th}(\phi)\,=\,\sum_{n}[b_n\,\sin(n\phi)\,+\,c_n\,\cos(n\phi)]\,+\,\sum_{m=0}^{2}d_mP_m~.
\end{equation}
\begin{figure}[ht]
\hfil\scalebox{0.85}{\includegraphics*[0.5in,0.4in][4.5in,7.6in]{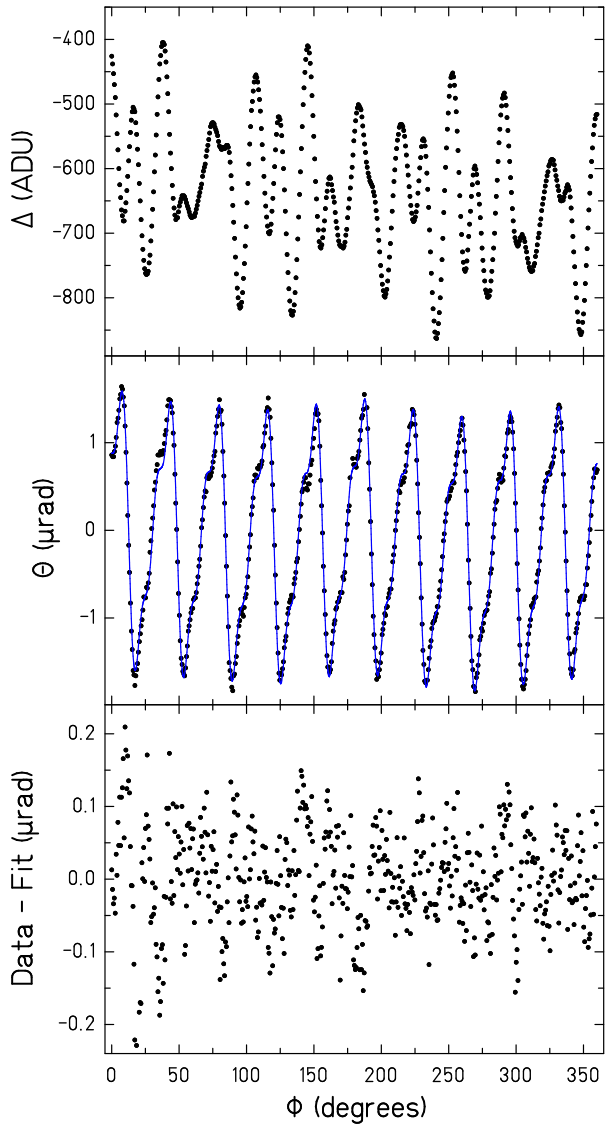}}\hfil
\caption{Top panel: raw autocollimator data for one attractor revolution (five cuts) at a vertical separation of 234~$\mu$m in Experiment~I. Middle panel: the same data after it has been filtered and fitted with Eq.~\ref{eq:fit}. The ten oscillations reflect the tenfold rotational symmetry of the instrument. The asymmetric shapes are due to phase shifts in the twist-to-torque conversion; the 10$\omega$ frequency is less than $\Omega$ while the $20\omega$ and $30\omega$ frequencies are greater than $\Omega$. Bottom panel: the fit residuals.} 
\label{fig:dat}
\end{figure}
All points in a given cut were treated equally and arbitrarily assigned an uncertainty $\pm 1$~ADU. 
The $d_m$ coefficients of the Legendre polynomials $P_m$ accounted for fiber drift.
The first sum in Eq.~\ref{eq:fit} ran over $n\,=\,$1, 3, 10, 20, and 30 (the $n\,=\,$30 term was not needed in Experiment~II). 
The floating $b_n$ and $c_n$ coefficients for the $n=10, 20,$~and~30 terms yielded the ``gravitational'' signals. 
The $n=3$ coefficients accounted for the three-fold symmetry of the attractor drive system while the $n=1$ coefficients allowed for any effect periodic at the attractor rotation frequency. 
(When fitting the individual cuts, the small $n=1$ and $n=3$ terms were fixed to values obtained from fitting data covering one complete 
attractor revolution.)  The smooth curve in the middle panel of Figure~\ref{fig:dat} shows the fit and the bottom panel shows the fit residuals.
For each cut we computed a ``chi-square'' of the $N_{\rm d}$ data points
\begin{equation}
\tilde{\chi}^2=\sum_{i=1}^{N_{\rm d}} \left( \theta_{\rm f}(\phi_i)-\theta_{\rm th}(\phi_i) \right)^2~,
\end{equation}
that reflected the goodness of the fit.
\subsection{Rejecting Bad Data}
We excluded cuts from our final analysis that were polluted by external disturbances with nongaussian distributions. These ``bad cuts'' were identified by a spike in the vacuum chamber pressure, an anomalously large $\tilde{\chi}^2$, or
an abrupt change in the amplitude or phase of the free torsional oscillation. These criteria rejected 228 (27) out of a total of 1599 (676) cuts In Experiment~I~(II).
Tests described below showed that the noise
in the remaining cuts was normally distributed. 
\subsubsection{\label{subsubsec:ioncut}Pressure Spikes}
\begin{figure}
\hfil\scalebox{.57}{\includegraphics*[0.6in,0.5in][6.4in,7.9in]{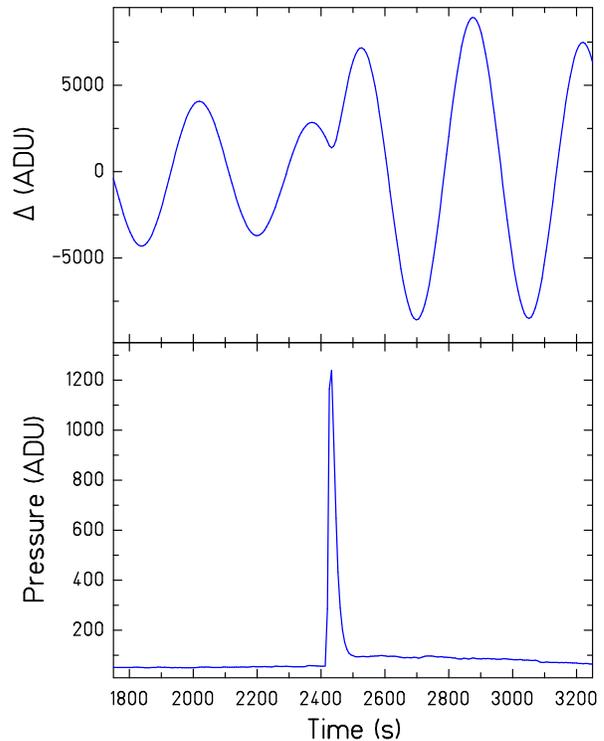}}\hfil
\caption{Upper panel: pendulum twist signal; lower panel: vacuum chamber pressure.}
\label{ion}
\end{figure}

The first rejection criterion was an abrupt disturbance of a sensor. The only observable irregularities were spikes in the vacuum chamber pressure which often visibly excited the pendulum (see Figure~\ref{ion}). We rejected all cuts with pressure spikes greater than ten standard deviations above the mean pressure. This criterion rejected 108 (5) cuts in Experiment I (II).
\subsubsection{\label{subsubsec:chi}Bad $\tilde{\chi}^2$ of the Harmonic Fit}
\begin{figure}
\hfil\scalebox{.56}{\includegraphics*[0.80in,0.4in][6.7in,3.6in]{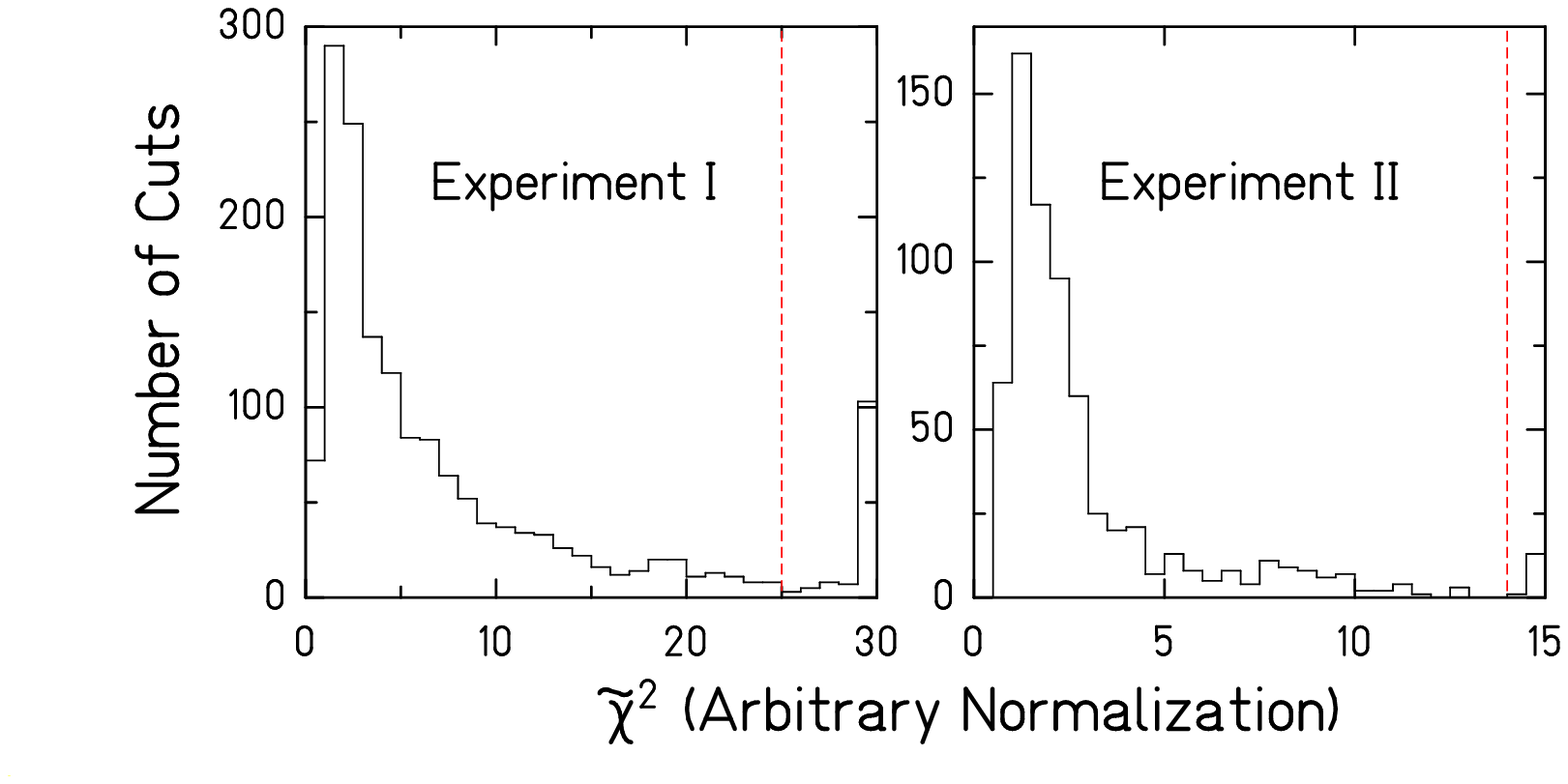}}\hfil
\caption{Distributions of $\tilde\chi^2$ values for all 1599 (676) cuts in Experiment I (II). The dashed vertical lines show the cut rejection levels.
The rightmost bins contain all cuts with $\tilde\chi^2$'s extending past the ends of the plots.} 
\label{chis}
\end{figure}

An unusually large value of $\tilde\chi^2$ indicated that the pendulum had been disturbed in some way. 
The $\tilde\chi^2$ distributions and rejection levels for Experiments~I and II are shown in Figure~\ref{chis}. This criterion rejected
111 (14) of the cuts passing the first criterion in Experiment I (II).
\subsubsection{\label{subsubsec:jump}Jumps in the Free Torsional Oscillation Amplitude and Phase}
\begin{figure}
\hfil\scalebox{.55}{\includegraphics*[0.7in,0.4in][6.8in,3.6in]{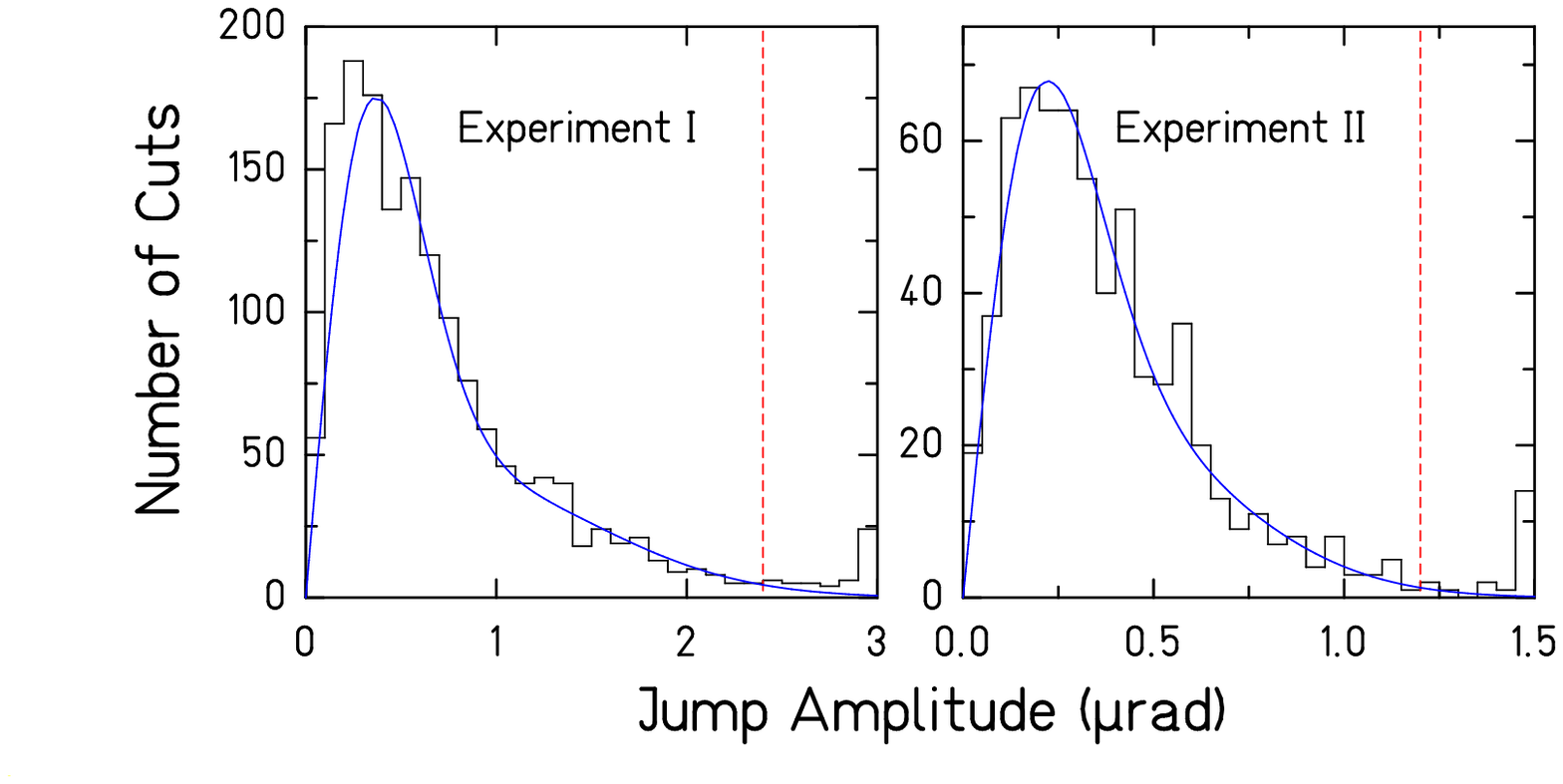}
}\hfil
\caption{Distribution of jumps in the 2 experiments. The curves, based on Eq.~\ref{eq-jump} with $M=2$ provide good fits to the data to the left of the vertical dashed lines, indicating that these jumps are normally distributed with two independent sources with $\sigma_1=0.32\pm 0.02~\mu$rad 
($0.20\pm 0.02~\mu$rad) and $\sigma_2 = 0.78\pm 0.05~\mu$rad ($0.41\pm 0.03~\mu$rad) in Experiment I (II). The vertical dashed lines are the rejection levels; larger jumps are due to spurious events such as small earthquakes, pressure spikes, or spontaneous unwinding of the torsion fiber. Note that in thermal equilibrium the mean free oscillation amplitude is $\theta_{\rm rms}\,=\,\sqrt{k_B\,T/\kappa}\,=\,1.1~\mu$rad, where $k_B$ is Boltzmann's constant.} 
\label{jumpcut}
\end{figure}

The final rejection criterion was an abrupt change in the amplitude or phase of the free torsional oscillation.
We determined the two quadrature amplitudes of the free oscillation in each cut and found the differences between adjacent cuts $\Delta a_{\sin}$ and $\Delta a_{\cos}$. The difference $J=\sqrt{\Delta a_{\sin}^2+\Delta a_{\cos}^2}$ is the ``jump'' in the free oscillation amplitude. If 
$\Delta a_{\sin}$ and $\Delta a_{\cos}$ are distributed about zero with a common spread, and there are $M$ different, normally-distributed sources  producing jumps,
each with its own average amplitude $\sigma$, the
jump distribution has the form
\begin{equation}
\label{eq-jump}
dN(J)\,=\,\sum_{i=1}^{M}\frac{N_i}{\sigma_i^2} \,J\,e^{(-J^2/\sigma_i^2)} dJ~,
\end{equation}
where $N_i$ are the numbers of events of each type. Figure~\ref{jumpcut} shows the jump data from both Experiments. Equation~\ref{eq-jump}, with $M=2$, and $N_i$ and $\sigma_i$ as free parameters, provides an excellent fit to the observed jump distributions, giving $\chi^2/\nu$ of 0.98 (0.68) in Experiment~I~(II); on the other hand, fits with $M=1$ were poor. The two stochastic sources may, for example, be vertical and horizontal seismic vibrations to which the apparatus responds quite differently. 
We eliminated cuts with jumps that were not distributed normally by setting the rejection level at the point where the fits began to deviate significantly from the data: 2.4~$\mu$rad (1.2~$\mu$rad) in Experiments I (II). The rejection level was lower in Experiment II because the magnetic damper, which did not function properly in Experiment I, was fixed before beginning Experiment II. This criterion rejected 9 (8) of the surviving cuts
in Experiment I (II).
\subsection{\label{subsec:good}Combining Good Data}
\begin{figure}[b]
\hfil\scalebox{0.75}{\includegraphics*[0.7in,0.5in][4.3in,4.5in]{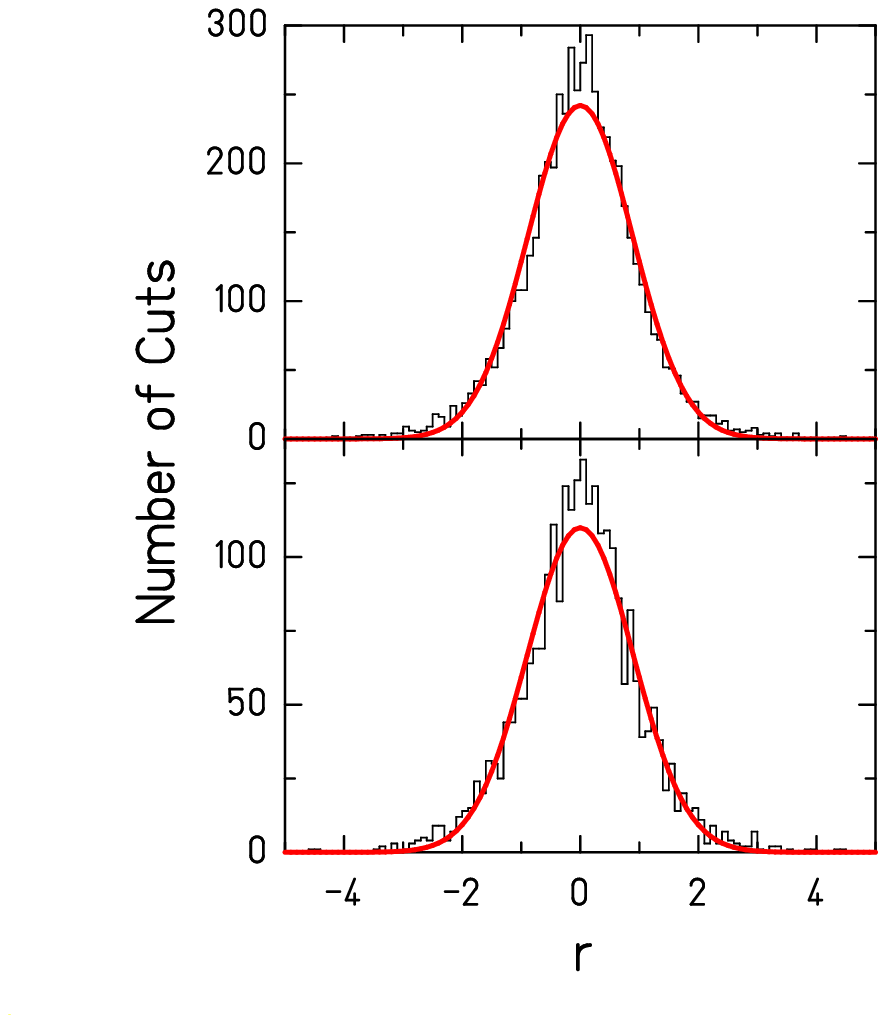}}\hfil
\caption{Normalized residuals $r_i=(b_i-\bar{b}_i)/(\sqrt{N_i} \sigma_i)$ for the accepted data, where $\sigma_i$ is given by Eq.~\ref{dev}. Top panel: all 5574 $10\omega$, $20\omega$ and $30\omega$ $b_n$ and $c_n$ coefficients of the 2-disk attractor data from Experiment I. Bottom panel: all 2596 $10\omega$ and $20\omega$ coefficients from Experiment II. The smooth curves are Gaussian fits that give $\sigma = 0.90\pm 0.01$ and $\sigma = 0.91 \pm 0.01$ respectively, indicating that our assigned errors
(Eq.~\ref{dev}) were slightly conservative.} 
\label{cuts}
\end{figure}

All $N_{\rm c}$ cuts within a given run that survived the rejection criteria were treated as equivalent so that the mean and its standard deviation
of the Eq.~\ref{eq:fit} coefficients 
from an individual run were 
\begin{equation}
\label{mean}
\bar{b}_n\,=\,\frac{1}{N_{\rm c}}\,\sum_{i=1}^{N_{\rm c}} b_n(i)
\end{equation}
\begin{equation}
\label{dev}
\sigma_{\bar{b}_n}\,=\,\sqrt{\frac{1}{N_{\rm c}(N_{\rm c}-1)}\,\sum_{i=1}^{N_{\rm c}}(b_n(i)-\bar{b}_n)^2}~;
\end{equation}
similar equations apply to the $c_n$ coefficients. Figure~\ref{cuts} shows that the $b_n$ and $c_n$ coefficients of the individual 
accepted cuts were essentially normally distributed.
\subsection{\label{subsec:atten}Converting Twists to Torques}
The applied torque on the pendulum, $N$, at frequency $\omega_{\rm s}$ was inferred from the twist angle $\theta(\omega_{\rm s})$ using
\begin{equation} 
N(\omega_{\rm s})=I \Omega^2 \theta(\omega_{\rm s}) f(\omega_{\rm s},\Omega)~,
\label{eq:torque from theta}
\end{equation}
where $I$ is the rotational inertia of the pendulum, $N$ and $\theta$ are complex numbers whose real and imaginary parts represent the
components varying as $\cos \omega_{\rm s}t$ and $\sin \omega_{\rm s}t$, respectively;
$f=f_{\rm i} f_{\rm tc} f_{\rm f}$ is a complex quantity that accounts for the effects of pendulum inertia, electronic time constants and digital filtering---its imaginary part corresponds to a phase shift.
We extracted the torques using Eq.~\ref{eq:torque from theta} rather than $N=\kappa \theta$ (where $\kappa$ is the torsional spring constant of the fiber) because electrostatic couplings 
(see Section~\ref{subsubsec:elec}) can cause the effective
$\kappa$ to vary slightly with $s$, whereas $I$ is truly constant.

\begin{figure}
\hfil\scalebox{.45}{\includegraphics*[0.9in,0.6in][8.0in,6.2in]{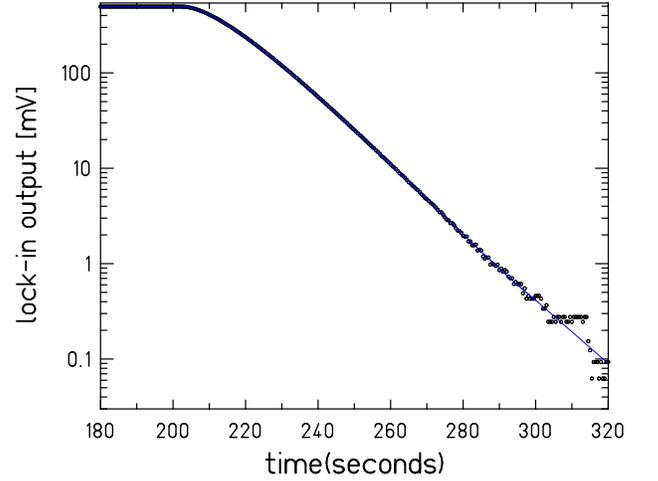}}\hfil
\caption{Response of the lock-in amplifier after a steady input was abruptly removed at time $t_{\circ}$. The fitted curve (largely indistinguishable from the points in this plot) assumes an ideal 2-pole filter response with time constant $t_{\rm tc}$, where $t_{\circ}$ and $t_{\rm tc}$ are free parameters.}
\label{fig:lockin tc}
\end{figure}

The differential equation describing the twist of a pendulum subjected to a driving torque $N$ and to anelastic damping (which acts like an imaginary component of the torsion constant\cite{sa:90}), is
\begin{equation}
\label{eq-osc}
I\ddot{\theta}+\kappa\left(1+ \frac{i}{Q}\right)\theta\,=\,N~,
\end{equation}    
where $Q=\tau_{\rm d} \Omega /2$ is the quality factor of the torsion oscillator.
A periodic driving torque, $N\,=\,N_0{\rm e}^{i{\omega}t}$, produces a steady-state twist
\begin{equation}
\label{eq-res}
\theta(t)\,=\,\left(\frac{1}{1-(\omega/\Omega)^2+i/Q}\right)\frac{N_0}{I \Omega^2}\,{\rm e}^{i{\omega}t},
\end{equation}
where $\Omega=\sqrt{\kappa/I}$. Equation~\ref{eq-res} shows that $f_{\rm i}=A_{\rm i} \exp{i \Phi_{\rm i}}$, where
\begin{equation}
A_{\rm i}=\sqrt{[1\!-\!(\omega/\Omega)^2]^2 + (1/Q)^2} ~;~~
\Phi_{\rm i}=\tan^{-1}\frac{\Omega^2}{Q(\Omega^2\!-\!\omega^2)}~.
\end{equation}
The torsion filter described in Section~\ref{subsubsec:filter} has a response $f_{\rm f}=A_{\rm f} \exp{i\Phi_{\rm f}}$, where
\begin{equation}
A_{\rm f}=\sec\left(\frac{\pi\omega}{2\Omega}\right)~;~~~  \Phi_{\rm f}\,=\,0~.
\end{equation}
In computing the effect on $\theta$ of the digitally synthesized 2-pole filters of the $\Delta$ and $\Sigma$ lock-in amplifiers, we could neglect
the time constant of the $\Sigma$ amplifier because its signal was nearly constant in time. Then
$f_{\rm tc}=A_{\rm tc} \exp{i \Phi_{\rm tc}}$, where
\begin{equation}
A_{\rm tc}=(1+\omega^2 t_{\rm tc}^2)~;~~~\Phi_{\rm tc}=2 \tan^{-1}(\omega t_{\rm tc})~.
\end{equation}
The lock-in amplifiers had nominal time constants of 10 s;
we found actual value of $t_{\rm tc}$ by using a function generator to drive the amplifier's analog and reference inputs with sine-wave and square-wave signals, respectively. The time dependence of the output after the input was abruptly disconnected, shown in Fig.~\ref{fig:lockin tc},
revealed that $t_{\rm tc}=10.082\pm0.011$~s . 
\subsection{Calibrating the Torque Scale}
\label{subsec: calibrate}
\subsubsection{Absolute gravitational calibration of Experiment II}
\label{subsubsec:abscal}
\begin{figure}
\hfil\scalebox{.56}{\includegraphics*[0.6in,0.5in][6.4in,4.6in]{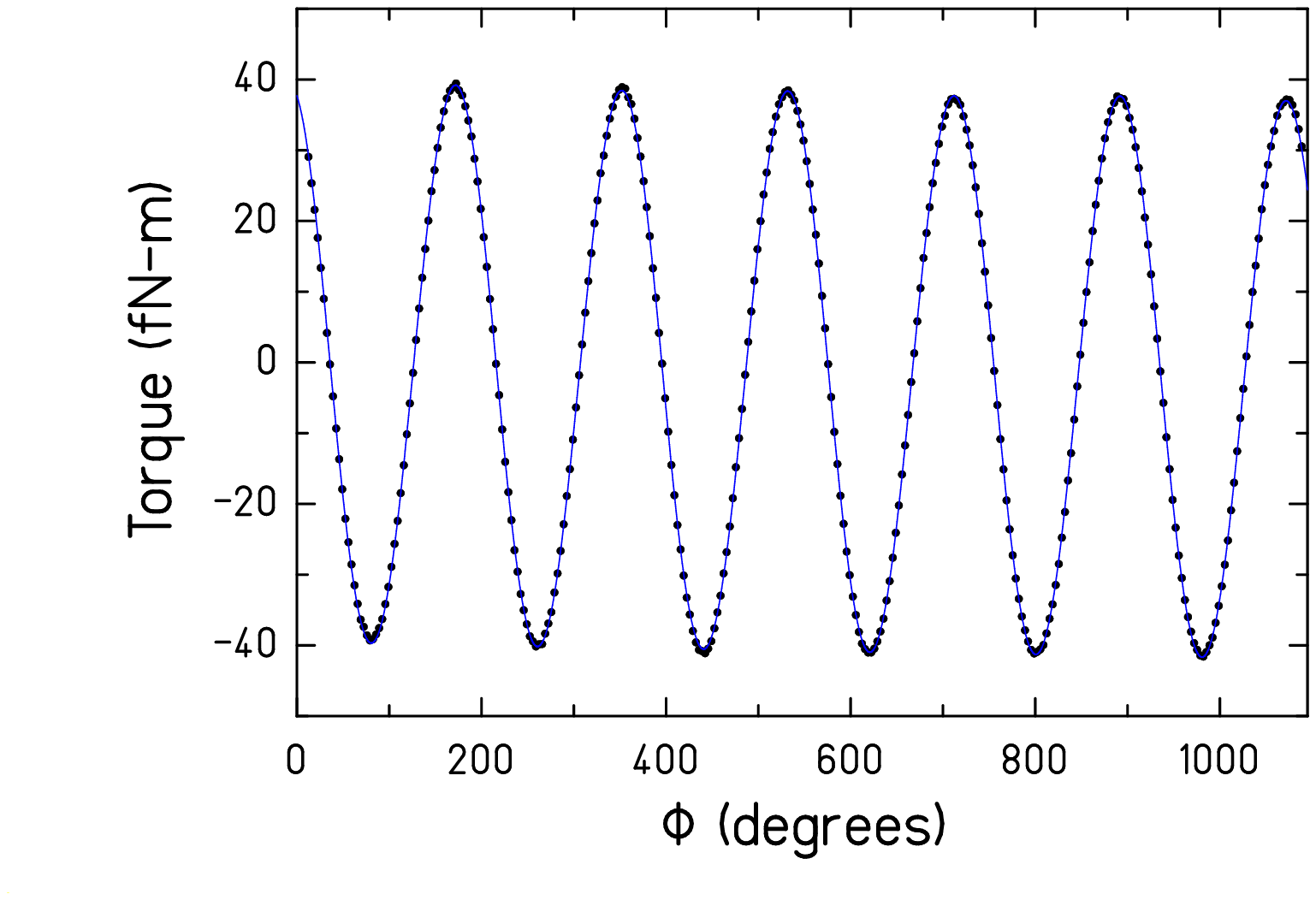}}\hfil
\caption{Data and fit for three revolutions of the calibration turntable. The 2$\omega_{\rm c}$ signal is dominant, as expected.} 
\label{fig:2omega}
\end{figure}

The torque scale was calibrated in Experiment II by applying a known gravitational torque on the pendulum. Two  7.938$\pm0.003$ mm-diameter aluminum spheres with average masses (corrected for buoyancy in air) of $M_{\rm s}=0.7348 \pm 0.0001$~g were placed in diametrically opposed holes on the pendulum ring, producing known $q_{lm}$ gravitational inner multipole moments, where $l$ and $m$ define the spherical multipole order. At the same time, two brass spheres with average masses (in air) of $515.1809\pm 0.0016$~g\cite{calib} were placed in diametrically opposite holes in the calibration turntable platter  
$13.983 \pm 0.004$ cm from the
rotation axis. These spheres produced known $Q_{lm}$ outer multipole moments so that as the calibration turntable rotated at a frequency $\omega_{\rm c}$, it produced a torque on the pendulum given by\cite{sm:00} 
\begin{equation}
\label{t22} 
\tilde{N}_{\rm cal}=8 \pi G \sum_{l=2}^8 \frac{1}{2l+1}\,\sum_{m=2}^l {\rm Im}
(m q_{lm}\,Q_{lm}\,e^{im\omega_{\rm c}t})  ~.
\end{equation} 
The sum was terminated at $l=8$ because higher-order torques were negligible. We considered only the $m=2$ component of the torque, which was calculated from the masses and locations of the spheres to be 
$\tilde{N}_{\rm cal}^{m=2}=40.042\pm 0.039$~fN-m. 
Figure~\ref{fig:2omega} shows typical data. Because the 13.98 cm length scale of our calibration was close
to the 16.76 cm scale used in measuring\cite{gu:00} $G$, any ISL violation (see Fig.~\ref{fig:old constraints}) could not have significantly effected the accuracy of our calibration.

\begin{figure}
\hfil\scalebox{.56}{\includegraphics*[0.6in,0.5in][6.4in,4.6in]{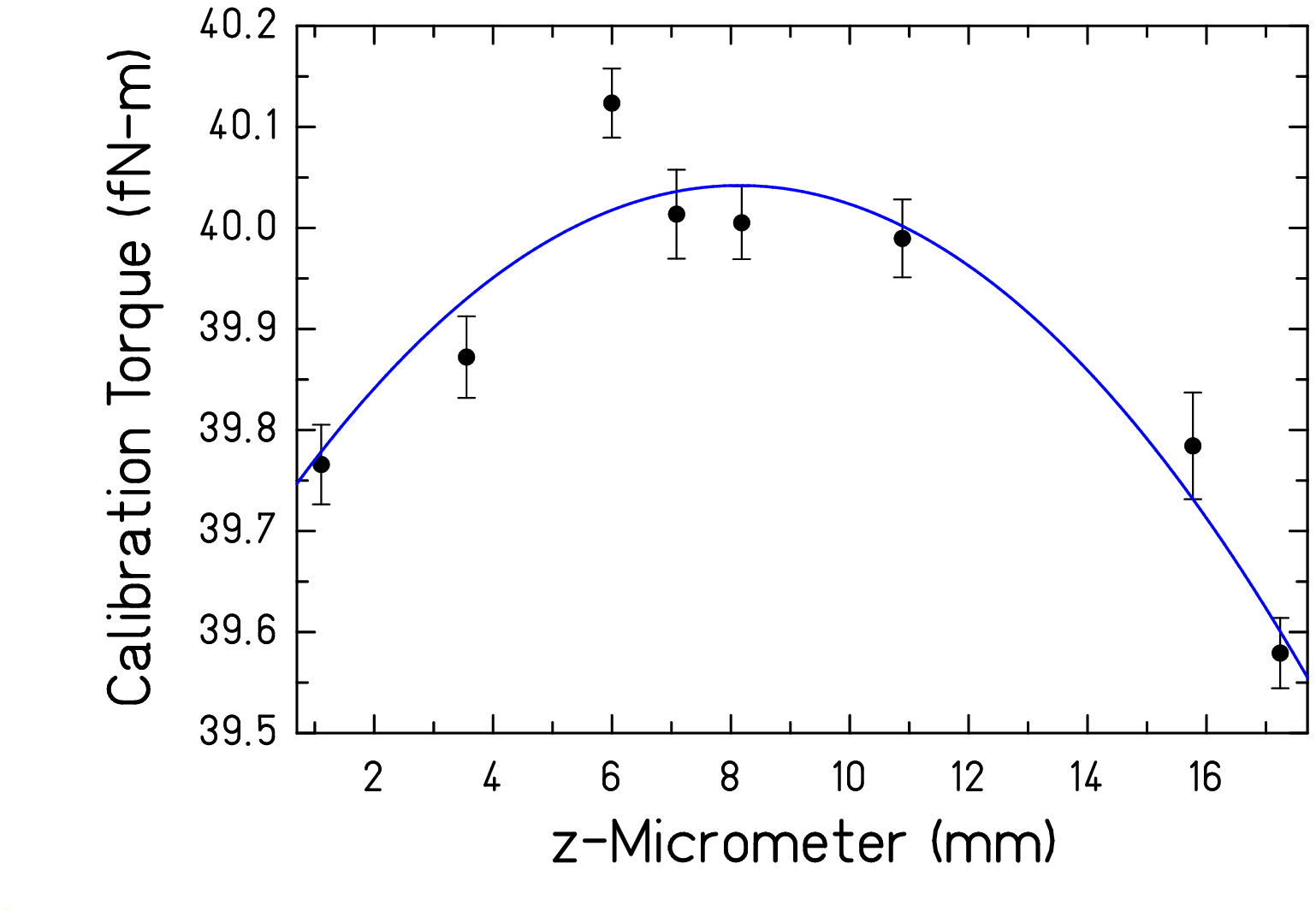}}\hfil
\caption{Absolute calibration data as a function of vertical separation. The curve is a polynomial fit whose curvature 
agrees well with the multipole prediction.
\label{caldat}}
\end{figure}

We fitted data with the calibration turntable displaced from its nominal center position by a series of horizontal and vertical points and corrected for 
the small measured backgrounds from mass asymmetries in the pendulum body and the calibration turntable; the maximum twist was $\theta_{\rm cal}^{m=2}= 12.6330 \pm 0.0048~\mu$rad,  Figure~\ref{caldat} shows the observed signal as a function of vertical position and the best fit based on the multipole prediction (the effects of displacements on the $q_{lm}$ and $Q_{lm}$ moments were calculated using multipole translation theorems\cite{sm:00}).

The torque corresponding to a twist signal $\theta_{\rm s}$ was computed as
\begin{equation}
N(\omega_{\rm s})=
\tilde{N}_{\rm cal}^{m=2} \frac{I}{I_{\rm cal}}\frac{\Omega^2}{\Omega^2_{\rm cal}} \frac{f(\omega_{\rm s},\Omega)}{f(2\omega_{\rm c},\Omega_{\rm cal})} \frac{\theta_{\rm s}}{\theta^{m=2}_{\rm cal}}~,
\label{eq:calII}
\end{equation}
where $I=175 \pm 1$ g cm$^2$ is the normal pendulum's rotational inertia, $I_{\rm cal}=I+I_{\rm sph}$ where $I_{\rm sph}=10.544 \pm 0.004$ g cm$^2$ is the rotational inertia of the small calibration spheres used on the pendulum.
The uncertainties in $I$,
$I_{\rm sph}$, $\tilde{N}^{m=2}_{\rm cal}$, $\Omega_{\rm cal}$ and $\theta^{m=2}_{\rm cal}$ produce a torque scale-factor error of 0.12\%.  We folded in an additional 0.5\% error to account for autocollimator non-linearities, measured in data taken subsequent to Experiment II, to give an overall torque scale factor $C_{\rm II}=1.000 \pm 0.005$; this was incorporated into our analysis as described in Section~\ref{subsec:fitting functions} below.

Note that the left hand side of Eq.~\ref{eq:calII} is independent of the autocollimator coefficient $c_1$ defined in Eq.~\ref{eq:theta def}. However we can use the gravitational calibration data to determine 
\begin{equation}
c_1=1.2758 \pm 0.0073~{\rm mrad}~.
\label{eq:c_1}
\end{equation}
\subsubsection{\label{subsubsec:relcal}Relative calibrations in Experiment II}
\begin{figure}
\hfil\scalebox{.57}{\includegraphics*[0.65in,0.5in][6.4in,4.6in]{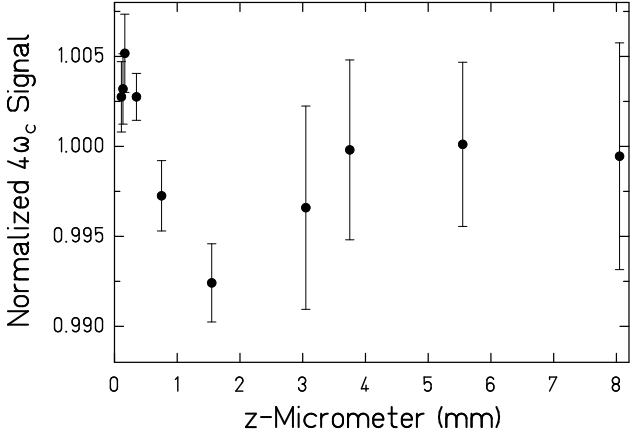}}\hfil
\caption{Calibration torque signal (relative units) as a function of the pendulum's vertical position in Experiment~II.}
\label{relcals}
\end{figure}

We made relative gravitational calibrations of each Experiment II data run to test for and correct any $z$-dependence of the torque scale. The calibration turntable with the two brass spheres was rotated continuously
at frequency $\omega_{\rm c}$. 
The $q_{l4}$ moments of the pendulum frame and $Q_{l4}$ moments of the brass spheres on the calibration
turntable interacted to produce a 4$\omega_{\rm c}$ torque. The calibration turntable was raised and lowered along with the pendulum to ensure the calibration torque was independent of $z$. Figure~\ref{relcals} shows
the observed $4\omega$
twist as a function of vertical position. The small $z$-dependence of the signal could have been caused by irregularities in the reflectivity or curvature of the pendulum mirror. The torques for each point were corrected for this variation of the calibration.
\subsubsection{Calibration of Experiment I}
Because Experiments I and II used the same autocollimator and suspension fiber, we could infer the calibration for Experiment I from the absolute calibration of Experiment II. We adopted the $c_1$ value determined by the Experiment II gravitational calibration (Eq.~\ref{eq:c_1}) but increased the uncertainty to $\pm 1.5$\;\% to account for the uncertainty in $I$ and autocollimator non-linearities. Experiment I torques were then computed using Eq.~\ref{eq:torque from theta}; we adopted a scale-factor uncertainty $C_{\rm I}=1.000 \pm 0.015$.
\subsubsection{Checking the calculated rotational inertias}
The torques measured using Eq.~\ref{eq:torque from theta} were proportional to the pendulum rotational inertias which were calculated from detailed models of the pendulums. We verified the calculation of $I$ for Experiment II using the known rotational inertia of the spheres $I_{\rm sph}$ and the measured free-oscillation frequencies from the gravitational calibration data 
\begin{eqnarray}
\Omega &=& 13.4584 \pm 0.00027~{\rm mrad/s} \\
\Omega_{\rm cal} &=& 13.0709 \pm 0.00173~{\rm mrad/s}~,
\end{eqnarray}
where $\Omega$ was extracted from data with $s \geq 370~\mu$m.
We accounted for the effect on the torsional constant $\kappa$ from fiber stretch due to the extra weight of the spheres
\begin{equation}
\frac{\Delta \kappa}{\kappa} = 4\frac{\Delta d}{d} - \frac{\Delta L}{L}
\end{equation}
where $d\approx 20~\mu$m  and $L\approx 800$~mm are the diameter and length of the fiber. We determined $\Delta L$ from the pendulums's observed bounce frequency $\nu_{\rm b}$ and total mass $M$ so that
\begin{equation}
\frac{\Delta L}{L}=\frac{2 M_{\rm s} g}{(2 \pi \nu_{\rm b})^2 M L} = -0.00012
\label{eq:length correction}
\end{equation}
The tabulated Poisson's ratio for tungsten
gives $\Delta d/d=-0.28 \Delta L/L$, so that $\Delta \kappa/\kappa=-2.12\Delta L/L$, or
\begin{equation}
\frac{\kappa}{\kappa_{\rm cal}}=1.00026~.
\label{eq:kappa correction}
\end{equation}
Therefore
\begin{equation}
\frac{I_{\rm II}}{I_{\rm II}+I_{\rm sph}} = 
\frac{\kappa}{\kappa_{\rm cal}} \,\frac{\Omega^2_{\rm cal}}{\Omega^2}~,
\end{equation}
which can be solved to give $I_{\rm II}=175.3 \pm 0.8$ g-cm$^2$, which agrees well with the value $175 \pm 1$ g-cm$^2$ computed using our detailed model of the pendulum.

We checked the calculated $I$ for Experiment I in a similar fashion, using the measured free-oscillation frequencies of the Experiment I and II pendulums and correcting for the effect on $\kappa$ of the differing weights of the two pendulums using equations similar to Eqs.~\ref{eq:length correction} and \ref{eq:kappa correction}. 
We found $I_{\rm I}=128.5 \pm 0.7$~g-cm$^2$ which agreed well with the detailed model calculation which gave $128 \pm 1$ g-cm$^2$.
\subsection{Torques from Experiment I}
Experiment~I contained 27~runs, each of which consisted of at least 36~cuts. The data were collected over a period of $\sim\,$45~days. Several runs were taken with the upper attractor disk alone at vertical separations $s>\,$8~mm. 

The full range of vertical separations was 216~$\mu$m to 10.77~mm, including the upper-disk-only runs. The measured harmonic components of the torque as a function of $s$ are shown in Table~\ref{tab:oncend1}. $N_{10}$ changes sign at a separation of $\sim\,$2~mm where the torque from the lower attractor plate overcomes that from the upper plate.
\begin{table}
\caption{\label{tab:oncend1}Harmonic torque components from the on-center data in Experiment~I. All torques have units of fN-m. The errors do not include the overall 1.5\% scale-factor uncertainty.}
\begin{ruledtabular}
\begin{tabular}{lrrr}
$s$ (mm) & \multicolumn{1}{c}{$N_{10}$} & \multicolumn{1}{c}{$N_{20}$} & \multicolumn{1}{c}{$N_{30}$} \\
\hline
0.216 & $5.327  \pm 0.051$ & $2.358  \pm 0.040$ & $2.866  \pm 0.031$\\
0.234 & $5.283  \pm 0.022$ & $2.287  \pm 0.015$ & $2.794  \pm 0.024$\\
0.382 & $4.471  \pm 0.032$ & $1.998  \pm 0.026$ & $2.344  \pm 0.036$\\
0.577 & $3.729  \pm 0.044$ & $1.671  \pm 0.031$ & $1.835  \pm 0.026$\\
1.067 & $2.051  \pm 0.041$ & $1.046  \pm 0.031$ & $1.054  \pm 0.033$\\
1.556 & $0.974  \pm 0.052$ & $0.642  \pm 0.043$ & $0.608  \pm 0.046$\\
2.045 & $0.140  \pm 0.030$ & $0.406  \pm 0.021$ & $0.387  \pm 0.035$\\
3.022 & $-0.752 \pm 0.040$ & $0.228  \pm 0.023$ & $0.139  \pm 0.037$\\
3.999 & $-0.970 \pm 0.034$ & $0.059  \pm 0.031$ & $0.045  \pm 0.033$\\
4.977 & $-1.031 \pm 0.026$ & $0.066  \pm 0.028$ & $-0.027 \pm 0.038$\\
6.443 & $-0.909 \pm 0.042$ & $0.063  \pm 0.028$ & $0.008  \pm 0.031$\\
8.132\footnotemark[1]  & $5.005 \pm 0.038$ & $ 0.035 \pm 0.033 $ & $ 0.030 \pm 0.075 $\\
8.134\footnotemark[1]\footnotemark[2]  & $5.008 \pm 0.065$ & $ 0.077 \pm 0.396 $ & $ 0.056 \pm 0.228 $\\
8.134\footnotemark[1]  & $4.993 \pm 0.046$ & $ 0.067 \pm 0.028 $ & $ 0.036 \pm 0.042 $\\
8.798\footnotemark[1]  & $3.846 \pm 0.066$ & $ 0.075 \pm 0.048 $ & $ 0.044 \pm 0.076 $\\
8.817\footnotemark[1]  & $3.823 \pm 0.057$ & $ 0.035 \pm 0.043 $ & $ 0.044 \pm 0.037 $\\
10.773\footnotemark[1] & $1.780 \pm 0.045$ & $ 0.058 \pm 0.027 $ & $ 0.039 \pm 0.025 $\\
10.773\footnotemark[1] & $1.830 \pm 0.025$ & $ 0.050 \pm 0.022 $ & $ 0.063 \pm 0.033 $\\
\end{tabular}
\end{ruledtabular}
\footnotetext[1]{Upper-disk-only runs}
\footnotetext[2]{Run taken at twice the attractor rotation frequency of the other runs}
\end{table}
Additional runs were taken at varying horizontal offsets to determine the unknown alignment point $(x_{\circ},y_{\circ})$. 
\subsection{Torques from Experiment II}
Experiment~II contained 17~runs, each of which consisted of least 19~cuts.  The entire data set was collected over a period of~$\sim\,$30~days. 
The full range of vertical separations was 137~$\mu$m to 7.90~mm. The measured harmonic components of the torque as a function of separation are shown in Table~\ref{tab:oncend2}. In this measurement the $N_{10}$ phase reversal occured at a separation $s \sim 3.3$~mm.

\begin{table}
\caption{\label{tab:oncend2}Harmonic torque components for the on-center data runs of Experiment~II in units of fN-m.
The errors do not include the 0.5\% scale-factor uncertainty.}
\begin{ruledtabular}
\begin{tabular}{crr}
$s$ (mm)& \multicolumn{1}{c}{$N_{10}$} & \multicolumn{1}{c}{$N_{20}$} \\
\hline
0.137 & $31.287 \pm 0.081$ & $10.024 \pm 0.042$\\
0.137 & $31.240 \pm 0.077$ & $10.029 \pm 0.030$\\
0.162 & $30.750 \pm 0.094$ & $9.842  \pm 0.034$\\
0.191 & $30.079 \pm 0.072$ & $9.625  \pm 0.029$\\
0.372 & $26.382 \pm 0.047$ & $8.368  \pm 0.023$\\
0.763 & $19.671 \pm 0.052$ & $6.169  \pm 0.028$\\
1.545 & $10.030 \pm 0.037$ & $3.349  \pm 0.021$\\
3.011 & $1.008  \pm 0.023$ & $1.057  \pm 0.018$\\
3.700 & $-0.992 \pm 0.016$ & $0.655  \pm 0.018$\\
5.459 & $-2.828 \pm 0.015$ & $0.183  \pm 0.014$\\
7.903 & $-2.492 \pm 0.018$ & $0.056  \pm 0.012$\\
\end{tabular}
\end{ruledtabular}
\end{table}
Six of the 17 runs were taken for horizontal centering and were not used to constrain ISL violation because the relative calibration setup was not employed. These 6 runs were used only to determine the central values and errors of $x_{\circ}$ and $y_{\circ}$.
%
%
\section{\label{sec:sys}Systematic Errors}
\subsection{\label{subsec:false}False Effects}
False effect are any effects, other than the gravitational interaction between the pendulum ring and the attractor disks, that produced a real or apparent coherent torque on the 
pendulum at our signal frequency. Whenever possible, we tested the sensitivity of our apparatus to individual false effects by exaggerating the effects and report the systematic effect on the fundamental 10$\omega$ signal. 
Unless otherwise noted, the effect on the higher harmonics was negligible. We took explicit account of those systematic effects that were not negligible compared to the typical 8(4)~nrad statistical error per Experiment I(II) run.     
\subsubsection{\label{subsubsec:therm}Thermal}
We measured the temperature-twist coupling by subjecting the apparatus to a 1$\omega$ temperature modulation 
and measuring the pendulum's response (the large thermal mass of the apparatus prevented us from obtaining an adequate temperature variation at higher frequencies). 
Three sensors attached to the vacuum vessel quantified the temperature change. The radiator water bath had a 1$\omega$ temperature amplitude of 910$\pm$1~mK, while the average of the three sensors mounted on the vacuum vessel recorded a 123$\pm0.5$~mK 1$\omega$ variation. The induced twist amplitude was 1.17$\pm0.07~\mu$rad, corresponding to a 9.51$\pm0.57~\mu$rad/K temperature 
feed-through. The variations in the temperature at the fundamental 10$\omega$ frequency were $\leq\,$0.1~mK  during normal data taking so that the systematic error associated with temperature modulations was $\lesssim$1~nrad. 

A false signal from thermal effects could also have occurred because heat generated by the drive motor made the attractor slightly higher warmer than its surroundings. When the motor drive was energized, its temperature increased by 1 K.
Consider Experiment I. The holes in the upper attractor plate were essentially blocked off by the lower plate, so that information about the attractor motion could be transmitted thermally only to the extent that the
top surface of the upper attractor plate was at a different (lower) temperature than the 
top surface of the lower plate. This temperature difference $\Delta {\cal T}_{\rm ul}$ could have produced rotating ``hot spots'' on the electrostatic shield that might have transmitted information about the attractor rotation to the pendulum. 

Several time scales are relevant for assessing the size of such effects. The time scale for the foil temperature, ${\cal T}$ to relax by radiation is
\begin{equation}
\tau_{\rm rad}=\frac{C_{\rm p} \rho t}{8 \sigma \epsilon {\cal T}^3} \approx 55~{\rm s}~,
\end{equation} 
where $\rho\,\simeq\,8900$~kg/m$^3$ is the density of copper,  
$C_{\rm p}$and $\rho$ are the heat capacity at constant pressure and density of copper, $\sigma$ is the Stefan-Boltzmann constant, and $\epsilon \simeq 0.1$ and $t=20~\mu$m are the emissivity and thickness of the foil, respectively. The time scale for the foil temperature to relax by conduction through the copper is
\begin{equation}
\tau_{\rm con} = \frac{C_{\rm p} \rho L^2}{4 k} \approx 0.2~{\rm s}~,
\end{equation}
where $k$ is the thermal conductivity of copper and $L$, taken to be 1 cm above, is the distance
between 2 points on the foil.
The time scale for the upper attractor disk temperature to relax by residual gas conduction to objects at a temperature ${\cal T}$, separated by less than a mean-free-path, is 
\begin{equation}
\tau_{\rm gas} \approx \frac{C_{\rm p} \rho w}{7 p \sqrt{k_B/(m {\cal T})}} \sim 3 \times 10^5~{\rm s}~,
\end{equation}
where  $k_B$ is Boltzmann's constant, $m$ is the mass of a gas molecule (assumed to be H$_2$), $p \sim 10^{-3}$ Pa is the pressure, and $w=2$~mm is the thickness of the upper attractor plate. 

We estimated $\Delta {\cal T}_{\rm ul}$ by noting that
the lapped plates are in good thermal contact and that the time constant for the top plate to cool via radiation
($\simeq 10^5 ~{\rm s}$) is much longer than the time constant for thermal conduction between the plates so that we can
assume the bottom surface of the upper plate is at the same temperature as the top surface of the lower plate. The energy
lost by the upper plate through radiation to the foil is supplied by thermal conduction through the upper plate, giving
a temperature difference between the top and bottom surfaces of the upper attractor plate, which is essentially $\Delta {\cal T}_{\rm ul}$,
\begin{equation}
\Delta {\cal T}_{\rm ul} = \frac{4 \epsilon^{\prime} \sigma w {\cal T}^3 \Delta {\cal T}_{\rm uf}} {k}
\end{equation}
where $\Delta {\cal T}_{\rm uf}$ is temperature difference
between the upper attractor plate and the conducting foil, and $\epsilon^{\prime} = \epsilon/2 \simeq 0.05$ was obtained by
assuming that the upper plate and foil have the equal emissivities. Taking an upper limit $\Delta {\cal T}_{\rm uf} \simeq 1 ~{\rm K}$, we find that $\Delta {\cal T}_{\rm ul} \simeq 1.5 \times 10^{-6} ~{\rm K}$.

The temperature variations on the foil ($\Delta \cal T$ in Eqs.~\ref{eq:tpress}, \ref{eq:tforce} and \ref{eq:tortemp}) due to $\Delta {\cal T}_{\rm ul}$ can be estimated as follows. The rotation period of the attractor, $\simeq 7000 ~{\rm s}$, is much longer than the thermal equilibration time of the foil so that we 
can consider the case
of a stationary source. If we model the thermal variations in the azimuthal coordinate of the foil as a one-dimensional
thermal conductor subject to a spatially-sinusoidal heat source, then the steady-state heat equation becomes:
\begin{equation}
\frac {\partial^2 {\cal T}} {\partial x^2} = -\frac {S ~\cos\:(2\pi x/L)} {k}
\end{equation}
where $S$ is the rate of heat input per unit volume. The solution to this equation gives us
$\Delta {\cal T} = S L^2/(4 \pi^2 k) = \epsilon^{\prime} \sigma L^2 {\cal T}^3 \Delta {\cal T}_{\rm ul}/(2 \pi^2 k t)$
where $t = 20 ~\mu{\rm m}$ is the thickness of the foil. Taking $L \simeq 2$ cm as the distance between the
holes in the source plate, we find 
\begin{equation}
\Delta {\cal T} \simeq 6 \times 10^{-10} ~{\rm K}~.
\label{eq:delta T}
\end{equation}

Finally, we estimate the effect on the pendulum of hot spots on the foil as follows. Assume that moving
spots, which are hotter by an amount $\Delta \cal T$, are approximated as squares of side length $2a$ where $a$ is the radius of a hole. The outgassing rate $q({\cal T})$ of the foil depends strongly on the temperature $\cal T$,
\begin{equation}
q({\cal T})=Q \exp(-\Theta/{\cal T})~,
\label{eq:outgas rate}
\end{equation}
where $\Theta$ is an activation temperature that has a characteristic value in the neighborhood of $10^4$ K.
Assuming that the pressure, $p$, is dominated by the outgassing rate, Eq.~\ref{eq:outgas rate} implies that
a local temperature difference $\Delta {\cal T}$ will produce a local pressure difference
\begin{equation}
\frac{\Delta p}{p}\approx \frac{\Theta \Delta \cal T}{{\cal T}^2}~.
\label{eq:tpress}
\end{equation}
The molecules travel ballistically because the mean free path is long compared to the relevant dimensions. The maximum force on a pendulum hole
occurs when an attractor hole is displaced horizontally by $a$. Molecules emitted by a hot area $2 a^2$ hitting an area $ah$, where $h$ is the thickness of a pendulum hole, apply an extra horizontal force on a single hole
\begin{equation}
\Delta F \sim \frac{a h p}{4 \pi} \frac{\Theta \Delta \cal T}{{\cal T}^2}~. 
\label{eq:tforce}
\end{equation}
This leads to a torque
\begin{equation}
N \sim 10^{3} \frac{\Theta}{\cal T} \Delta \cal T~\text{fN-m}~,
\label{eq:tortemp}
\end{equation}
where we assumed $p \sim 10^{-3}$ Pa.
Equations~\ref{eq:tortemp} and \ref{eq:delta T} predict a
thermally induced torque is $4 \times 10^{-5} ~\text{fN-m}$ or a spurious signal of $\simeq 0.01 ~{\rm nrad}$.

We checked for a temperature gradient effect by deliberately heating the attractor with resistors attached to the attractor drive motor, raising the temperature of the motor mount by 4.8$\,^\circ$C---a five-fold increase over the temperature rise associated with turning on the motor. The 10$\omega$ twist signal in this configuration   
was consistent at the 1.5$\sigma$ level with the predicted negligible size of this effect.  

Thermal expansion from the heating associated with turning on the motor could have affected the attractor-to-membrane distance. The attractor-membrane capacitance was unchanged within $\pm0.01$~pF out of 1.33~pF when the motor was turned on, corresponding to a distance change of $<1~\mu$m for Experiment~I. We therefore assumed that systematic errors associated with the thermal gradient produced by the motor were negligible.     
\subsubsection{\label{subsubsec:tilt}Tilt}
Tilt can produce two separate false effects. The first arises if the apparatus tilt is modulated at the signal frequency. Tilting the fiber attachment is known to twist the pendulum, so that any tilt at the signal frequency would produce a false signal. 
We tilted the vacuum vessel by 1.02$\,\pm\,0.01$~mrad and found that the pendulum equilibrium position changed by 16$\,\pm\,2~\mu$rad.
We analyzed the tilts recorded by the AGI sensors in each normal run and an upper limit of 4.0~nrad on signal periodic at 10$\omega$; this 
corresponded to a tilt systematic at 10$\omega$ of $\leq\,$0.06~nrad.

The second tilt effect arises if there is a static tilt between the pendulum and attractor. Our calculations of the expected Newtonian and Yukawa torques assumed that the plates were perfectly parallel. 
In normal data taking, the tilt between the pendulum and attractor was $<0.2$~mrad (see Section~\ref{subsubsec:lev}). The effect of ignoring this tilt was calculated as follows. Let the calculated $10\omega$ gravitational signal of an untilted pendulum at a pendulum-to-attractor separation of $s$
be $\tilde{N}_{10}(s)$. A tilt of size $\psi << 1$ would change the effective separations between the centers of masses of the holes. This effect causes the torque on a tilted pendulum 
\begin{equation}
{\tilde N}_{10}^{\rm tilt}(s)\,\approx\,\sum_{j=1}^{10}\frac{{\tilde N}_{10}(s+r_{\rm p}\psi\sin(j\pi/5))}{10}, 
\end{equation}
to depend quadratically on $\psi$
(the sum runs over the 10 pendulum holes and $r_{\rm p}$ is the distance of the holes from the pendulum's symmetry axis). The maximum expected difference for any 
Experiment I(II) run was 
\begin{equation}
|{\tilde N}_{10} \, - \, {\tilde N}_{10}^{\rm tilt}| \,=\, 1.2 \,\times \, 10^{-4}~(1.6 \, \times \, 10^{-4})~\text{fN-m}~.
\label{ttilt}
\end{equation}
Tilt also causes the direction of the symmetry axes of pendulum holes to deviate from the assumed vertical. But this second-order effect is smaller than that given in Eq.~\ref{ttilt} by an amount $\sim (a_{\rm p}/r_{\rm p})^2$ and therefore can be neglected.

Before beginning Experiment I we checked the static tilt effect by deliberately tilting the apparatus by 1 mrad and measuring the change in $N_{10}$. We attempted to recenter the pendulum after the tilt but uncertainties in the centering prevented us from being able to verify the very small effect predicted by Eq.~\ref{ttilt}. We adopted the value in Eq.~\ref{ttilt} for the systematic error due to tilt. 
\subsubsection{\label{subsubsec:mag}Magnetic Background}
We measured the magnetic moment of the pendulum by applying a time-varying external magnetic field. A coil placed outside the thermal shielding produced a horizontal 6.3~mG field at the pendulum location. We modulated this field at 10$\omega$ and saw induced twists of 335$\,\pm\,79$~nrad (252$\,\pm\,25$~nrad). 
Under normal operating conditions, the amplitude of 10$\omega$ magnetic field variations at the site of the pendulum, measured by a flux-gate magnetometer, was $\leq0.6~\mu$G. Hence, the error associated with a magnetic coupling was $\pm$0.03~nrad ($\pm$0.02~nrad), much smaller than our statistical errors. 
\subsubsection{\label{subsubsec:elec}Electrical Background}
Suppose the attractor, membrane and pendulum were ideal conductors but at slightly different potentials
with voltage differences $V_{\rm am}$ and $V_{\rm mp}$. The attractive pressure between two flat conducting
plates with a potential difference $V$ and separated by a distance $d$ is $\epsilon_0 V^2/(2d^2)$.
This pressure is essentially absent over the holes, so the membrane feels a missing pressure above
each attractor hole $P_{\rm am}=\epsilon_0  V_{\rm am}^2/(2 d_{\rm am}^2)$, where $d_{\rm am} \approx .05$ mm (see Sec.~\ref{subsubsec-attlev}). This will excite the $m\!=\!10,~n\!=\!1$
drumhead mode of the membrane (the lowest mode that could produce a $10\omega$ electrostatic coupling
of the attractor to the pendulum).  The deflection in this mode is 
$\delta(r,\phi)=\alpha J_{10}(k_{10}r) \cos(10\phi)$, where $J_{10}$ is a Bessel function of the first kind and
\begin{equation}
\alpha=\frac{C}{4 \pi^2 \rho t}\frac{P_{\rm am}}{f_{10,1}^2}=  \left[ \frac{V_{\rm am}}{1~{\rm mV}}\right]^2 6\times 10^{-16}~{\rm mm}~,
\end{equation}
$C\approx 0.07$ is the projection of the attractor hole pattern on the shape of the
membrane mode, $\rho t$ is the areal density of the membrane, and its resonant
frequency is $f_{10,1}=6.02 f_{0,1}\approx 5.1$~kHz 
(see Sec.~\ref{subsec:membrane}). These rotating ``bumps'' will in turn change the electrostatic binding energy
of pendulum-membrane system by an amount $\Delta U \approx \alpha \epsilon_0 V_{\rm pm}^2 A/(2 d_{\rm pm}^2)$ where $A$
is the total area of the 10 pendulum holes and $d_{\rm pm}\approx 60 \mu$m is the minimum pendulum-to-membrane separation. The moving bumps therefore exert a torque on the pendulum
\begin{equation}
N=\Delta U/\Delta \phi \leq \left[ \frac{V_{\rm am}}{1~{\rm mV}} \right]^2  \left[ \frac{V_{\rm pm}}{1~{\rm mV}} \right]^2 ~2 \times 10^{-15}~\text{fN-m}
\end{equation}
which is negligible.
\subsubsection{\label{subsubsec:grav}Gravitational Background}
Our calculations of Newtonian and Yukawa torques assumed that the torque was due entirely to the holes in the pendulum ring and attractor disk. However, residual gravitational coupling between the pendulum frame and the attractor drive mechanism could have produced small spurious signals at 10$\omega$, 20$\omega$, etc. 
The apparatus was designed to suppress such couplings---the pendulum frame had two-fold (four-fold) rotational symmetry and the attractor drive was 3-fold symmetric---so we did not expect significant torques from these components at any harmonic of interest.
We tested for residual torques at the signal frequencies by taking data with the attractor disks 
removed from their holder and found no resolved effect. We then replaced the attractor in its original position and removed the pendulum ring from its frame and again found no signal.  
Because we did not expect or observe any gravitational background signals, we assigned a negligible systematic error for this effect. 
\subsubsection{Electrostatic Effects on $\kappa$}
\begin{figure}
\hfil\scalebox{.58}{\includegraphics*[1.42in,0.5in][7.2in,4.8in]{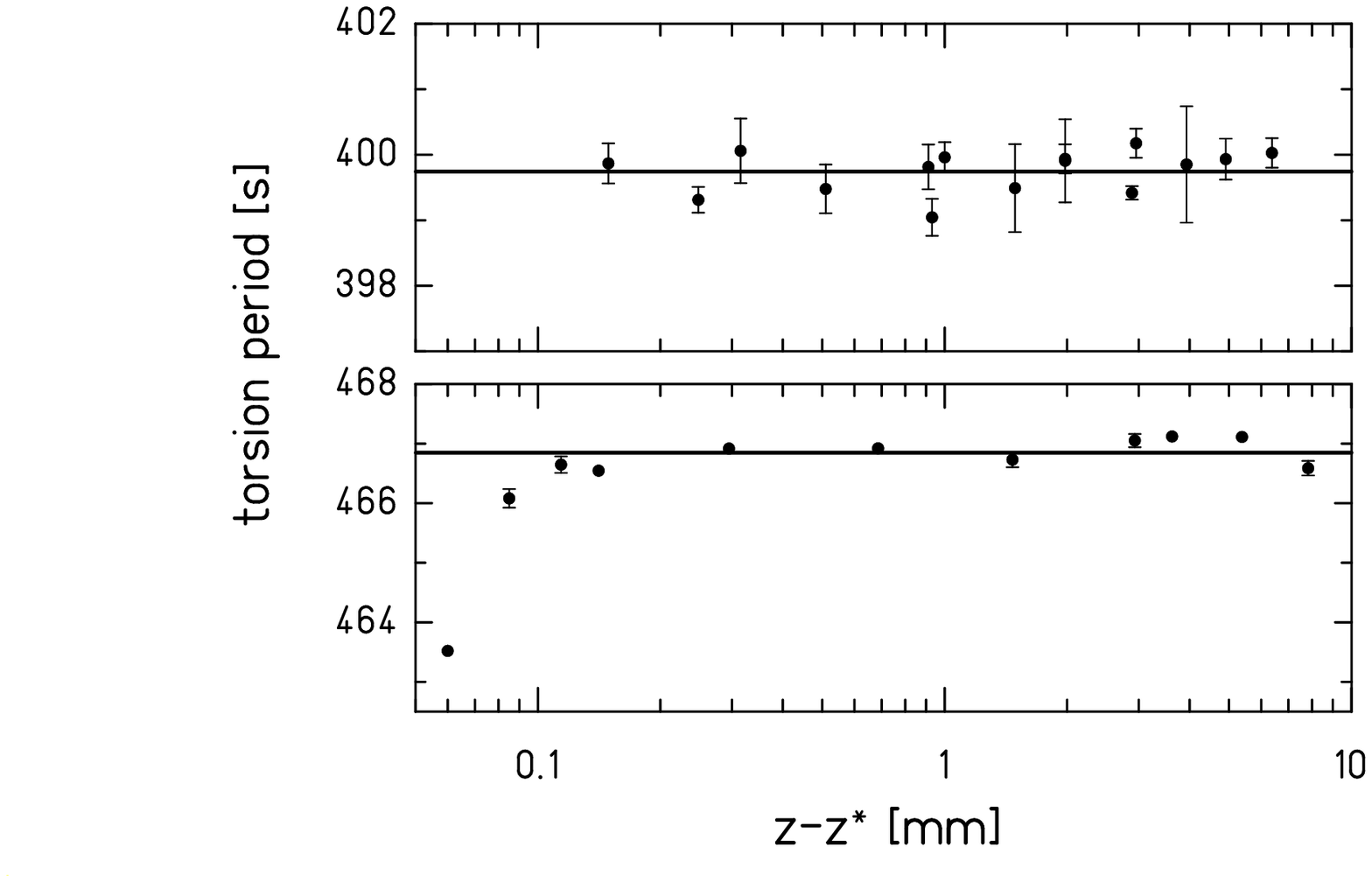}}\hfil
\caption{The torsion period as a function of the pendulum-to-membrane separation in Experiments~I (upper panel) and II (lower panel). The horizontal lines show the mean periods for separations $> 100 ~\mu$m).}
\label{ttors}
\end{figure}

Direct electrostatic interactions between the pendulum and attractor were essentially eliminated by the  conducting membrane. However, real conductors differ from ideal ones and electrostatic potential wells on the surface of the membrane could interact with patch charges on the pendulum to change the effective $\kappa$ of the
torsion fiber, making it a function of $z-z^{\ast}$. Figure~\ref{ttors} shows the torsion period, $2 \pi/\Omega$ as a function of separation for Experiments~I and II. A small systematic variation in the period occurred at very close separations.
We sidestepped this problem by measuring the free oscillation frequency
$\Omega$ for each run, and inferring the torques from the measured twist using Eq.~\ref{eq:torque from theta}
which does not explicitly depend on $\kappa$.
Note that {\em any} effect that changed the oscillator properties was eliminated by this
procedure. 
\subsubsection{\label{subsubsec:falsebugd}False Effect Error Budget}
Table~\ref{tab:false} summarizes the systematic twist error produced by false effects. The sum of these effects is far smaller than our statistical noise, which was $\sim8$ (4)~nrad per data point.  
\begin{table}
\caption{\label{tab:false}Contributions to the systematic twist error from dominant false effects.}
\begin{ruledtabular}
\begin{tabular}{lc}
Effect & 10$\omega$~Twist~Error~(nrad)\\
\hline
Temperature Modulation & $\lesssim$\,1\\
Apparatus Tilt & $\leq$\,0.06\\
Relative Tilt & $\leq$\,0.05\\
Magnetic & $\leq$\,0.03\\
\end{tabular}
\end{ruledtabular}
\end{table}
\subsection{\label{subsubsec:twist conversion}Errors in the Twist-to-Torque Conversion}
We checked the validity of the frequency-dependent twist-to-torque conversion described in Section~\ref{subsec:atten} by comparing data taken at different attractor rotation rates $\omega$. In Experiment~I, where the normal rotation period was $\tau_{\rm att}\,=\,17\tau_{\circ}$, we also took data with attractor periods of $7\tau_{\circ}$ and $43\tau_{\circ}$. The results are summarized in Table~\ref{tab:speeds}. As we found no measurable discrepancy and expected none, we assigned a negligible error to the conversion (except for that introduced by measurement uncertainties in $\Omega$, $t_{\rm d}$, $\omega$
and $t_{\rm tc}$).  
\begin{table}[b]
\caption{\label{tab:speeds} Comparison of the torques measured at two different attractor rotation speeds in Experiment~I.}
\begin{ruledtabular}
\begin{tabular}{lccc}
Attractor Period & \multicolumn{3}{c} {Measured Torques [fN-m]} \\
   & 10$\omega$ & 20$\omega$  & 30$\omega$ \\
\hline
$7\tau_{\circ}$ & 3.204$\,\pm\,0.040$ & 1.564$\,\pm\,0.116$ & 1.630$\,\pm\,0.144$\\
$43\tau_{\circ}$ & 3.227$\,\pm\,0.037$ & 1.590$\,\pm\,0.019$ & 1.702$\,\pm\,0.052$\\
\hline
Signal change & 0.023$\,\pm\,0.054$ & 0.026$\,\pm\,0.118$ & 0.072$\,\pm\,0.153$\\
\end{tabular}
\end{ruledtabular}
\end{table} 
\subsection{\label{subsec:inst}Uncertainties in the Instrumental Parameters}
Uncertainties in important instrumental parameters (such as the pendulum-to-attractor separation, torque scale calibration, and the missing masses of the holes) affect the predicted Newtonian and Yukawa torques. We measured these parameters and used their values and errors as constrained parameters when fitting the torque data, in effect letting gravity give us an independent measure of these experimental parameters. The parameters were $C$ (the overall torque-scale factor), $M_{\rm p}$ (the total missing mass of the pendulum holes), $M_{\rm a}^{\rm up}$ (the mass of the upper attractor holes), $M_{\rm a}^{\rm lo}$ (the mass of the lower attractor holes), $z_{\circ}$ (the vertical position reading at vanishing pendulum-to-attractor separation), $g$ (the gap between the upper and lower attractor disks), and $\varphi$ (the angular
misalignment of the two attractor disks). In addition for Experiment I we needed $z_{\circ}^{\prime}$ (the $z_{\circ}$ value for the upper-disk only runs) and for Experiment II we needed $M_{\rm a}^{\rm up,out}$ (the mass of the out-of-phase holes in the upper attractor) and $\varphi^{\prime}$ (the misalignment angle between the in-phase and out-of-phase holes in the upper attractor plate).
\subsubsection{\label{subsubsec:missmass}Missing Mass of the Holes}
The missing masses of the holes were key factors in our experiment. 
In Experiment~I, we inferred the missing masses from the densities of the materials and the hole dimensions. The densities were determined
after we had taken all the data for that experiment.
We measured the masses of the pendulum ring and the attractor disks, and then drilled 10 extra holes in the pendulum ring and 
8 extra holes in each of the attractor disks. These masses, together with the measured volumes  of the normal and extra holes gave us the missing masses of the normal holes.

A more direct technique was used to determine the missing masses of the holes in Experiment~II. 
Each disc was weighed before the active holes were drilled ($M_1$) and then again immediately after drilling the holes ($M_2$). Since the upper plate had 2 sets of holes, the mass was measured after drilling each set. The attractor disks were then lapped to the final thickness and weighed again ($M_3$). (The pendulum was not lapped so that, for it, $M_3\,=\,M_2$.)
Then the gold layer was applied and the piece weighed again ($M_4$). The resulting hole masses were 
\begin{equation}
M\,=\,(M_1-M_2)\,\frac{M_3}{M_2}\,+\,f_{\rm Au}\,(M_4-M_3),
\label{eq-missmassII}
\end{equation} 
where the computed factor $f_{\rm Au}$ accounted for the gold missing from the flat surfaces of the holes as well as that present
on the round edges of the holes.
\subsubsection{\label{subsubsec:holep}Hole Placement}
\begin{figure}
\hfil\scalebox{.55}{\includegraphics*[0.5in,0.5in][6.4in,4.6in]{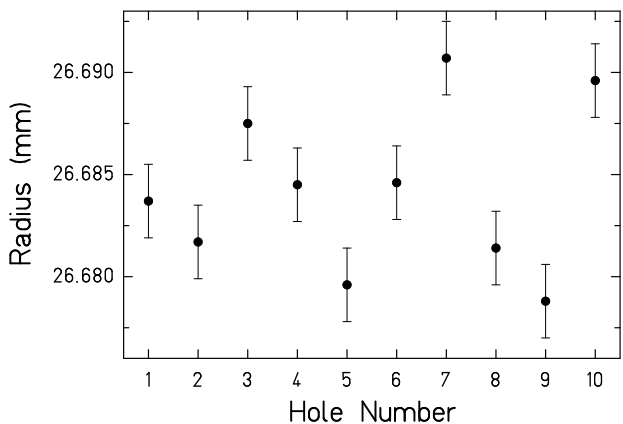}}\hfil
\caption{Radial positions of the individual inner holes of the Experiment~II upper attractor disk.} 
\label{rads}
\end{figure}

\begin{figure}
\hfil\scalebox{.55}{\includegraphics*[0.5in,0.5in][6.4in,4.6in]{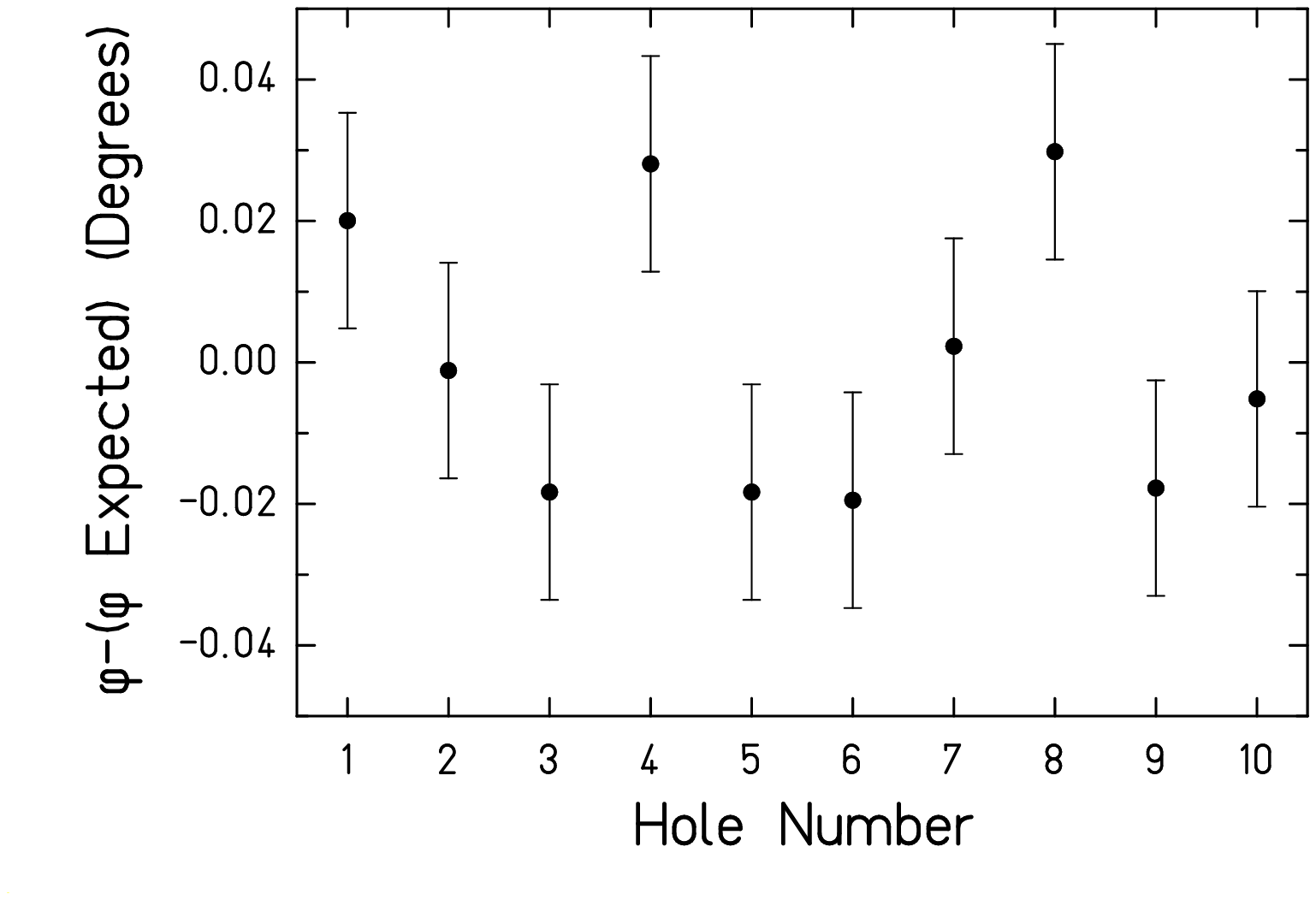}}\hfil
\caption{Residuals of the angle between the individual inner holes of the Experiment~II upper attractor disk.} 
\label{phis}
\end{figure}

The calculated torques were based on the average values of the distance of the holes from the rotation axis and the angle between them. It was important to know whether there were any systematic deviations from the average values (for example, the holes could have been on an ellipse instead of a circle). Figures~\ref{rads} and~\ref{phis} show the radii and angles of the inner holes of the upper attractor disk in Experiment~II, respectively. There are no systematic variations or outliers. The holes in the other disks (and pendulum rings) used in the two experiments displayed similar behavior. 
\subsubsection{\label{subsubsec:gap}Gap between the Attractor Disks}
Any gap $g$ between the attractor disks (for example, because they were not perfectly flat) 
would affect the cancelation of the $10\omega$
torques. Although $g$ was a parameter in fitting our data, we needed an independent constraint
on $g$ to minimize the effect of correlations between $g$ and other fitting parameters that would increase
the uncertainties in the extracted physics parameters $\alpha$ and $\lambda$.

In Experiment~I, the upper and lower disks had thicknesses of 1.846$\,\pm\,$0.005~mm and 7.825$\,\pm\,$0.005~mm, respectively. When the disks were stacked the combined thickness was 9.681$\,\pm\,$0.007~mm, yielding  $g\,=\,(0.010\,\pm\,0.010)$~mm.
We used the same technique in Experiment~II, but took many more thickness measurements and used a more precise micrometer to obtain $g\,=\,(0.0014\,\pm\,0.0010)$~mm.   
\subsubsection{Angular Misalignment of the Attractor Disks}
The angular alignment of the upper and lower attractor disks was maintained by three pins that passed through holes in the upper and lower disks. We
determined the angular misalignment, $\varphi$ between the two disks by examining the disk assembly with a measuring microscope. We found that $\varphi = 0.00\pm$0.01~degrees (0.13$\pm$0.01~degrees) for Experiment~1~(II).
\section{\label{sec:predicted torques}Predicting the Expected Newtonian, Yukawa, Power-Law and Massive Pseudoscalar Torques}
\subsection{Conventions}
We assume that the pendulum ring and attractor disks are perpendicular to the vertical torsion fiber, so that a torque about the fiber axis
can only be caused by a horizontal force $F_{\rm t}$, perpendicular to the fiber and tangential to the pendulum ring. The horizontal force between 
any two perfectly circular holes is directed along the horizontal projection, $t$, of the line between their centers, so that
the torque exerted by the pendulum on the fiber is 
\begin{equation}
N = \sum_{i,j} \vec{r}(i) \times \vec{F_{\rm t}}(i,j) + \sum_{i,k} \vec{r}(i)\times \vec{F_{\rm t}}(i,k) ~,
\end{equation}
where $i$ runs over the 10 pendulum holes, $j$ and $k$ run over the holes in upper and lower attractor disks, respectively, and
$F_{\rm t}$ has Newtonian, Yukawa, power-law and massive pseudoscalar components 
\begin{equation}
F_{\rm t}=F_{\rm t}^G + \alpha(\lambda) F_{\rm t}^Y(\lambda) + \beta_k F_{\rm t}^P(k) + \gamma(\lambda) F_{\rm t}^M(\lambda)~.
\end{equation}
where $\alpha$, $\beta_k$ and $\gamma$ are dimensionless parameters. The forces $F^G_{\rm t}$, $F^Y_{\rm t}$, $F^P_{\rm t}$ and $F^M_{\rm t}$ were computed as  
\begin{eqnarray}
\label{eq:force defs}
F_{\rm t}^G(t,s)\! &=& \!G \rho_1 \rho_2 \!\oint_{V_1} \!\!\oint_{V_2}\!\! dV_1 dV_2 \frac{\partial}{\partial t} \left[\frac{1}{R}\right]  \\ 
F_{\rm t}^Y(t,s,\lambda)\!&=&\! G \rho_1 \rho_2 \!\oint_{V_1}\!\! \oint_{V_2}\!\! dV_1 dV_2 \frac{\partial}{\partial t} \left[\frac{\exp(-R/\lambda)}{R}\right] \nonumber \\  
F_{\rm t}^P(t,s,k) \!&=&\! G \rho_1 \rho_2 \!\oint_{V_1}\! \!\oint_{V_2} \!\! dV_1 dV_2 \frac{\partial}{\partial t} \left[ \left( \frac{r_0}{R}\right)^{\! k} \frac{1}{R}\right]      \nonumber \\ 
F_{\rm t}^M(t,s,\lambda) \!&=& \!G \rho_1 \rho_2 \!\oint_{V_1}\! \!\oint_{V_2} \!\!dV_1 dV_2 \frac{\partial}{\partial t}\left[\left( \frac{r_0}{R} \right)^{\!2} \frac{K_1(2R/\lambda)}{\lambda} \right] \nonumber 
\end{eqnarray}
where $R=|R_2 - R_1|$ is the distance between infinitesimal volumes in the two cylinders, $r_0=1$~mm, and the integrals extend over the volumes $V_1$ and $V_2$ of the two cylinders.
\subsection{\label{subsec:calctorq}Calculating the Harmonic Torque Database}
\begin{figure}
\hfil\scalebox{.45}{\includegraphics*[0.3in,0.5in][7.5in,5.4in]{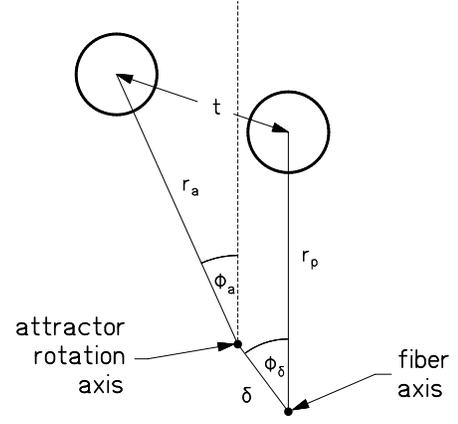}}\hfil
\caption{Geometry used to calculate the torque on the pendulum from the rotating attractor. The attractor hole is on the left.}
\label{geometry}
\end{figure}

For the geometry shown in Figure~\ref{geometry}, the torque about the torsion fiber axis from a single pendulum/attractor 
hole pair is 
\begin{eqnarray}
\tilde{N}_1(\phi_a,\,\delta,\,\phi_{\delta},\,s)&=&r_{\rm p}\,F_{\rm t}(t, s)\\
&&\times\left(\frac{r_{\rm a}\sin \phi_a+\delta\sin \phi_{\delta}}{t}\right)~, \nonumber
\end{eqnarray} 
where $F_{\rm t}(t, s)$ is the total horizontal force on the hole at a pendulum-to-attractor vertical separation $s$, and 
\begin{equation}
t=\sqrt{(r_{\rm a}\sin\phi_a+
\delta\sin\phi_{\delta})^2+(r_{\rm a}\cos\phi_a+\delta\cos\phi_{\delta}-r_{\rm p})^2}~. \nonumber
\end{equation}
The procedures for efficiently calculating $F_{\rm t}^G$, $F_{\rm t}^Y(\lambda)$, $F_{\rm t}^P(k)$ and $F_{\rm t}^M(\lambda)$ are described in Appendix~\ref{sec:calc}.

\begin{figure}
\hfil\scalebox{.54}{\includegraphics*[0.69in,0.5in][6.5in,5.3in]{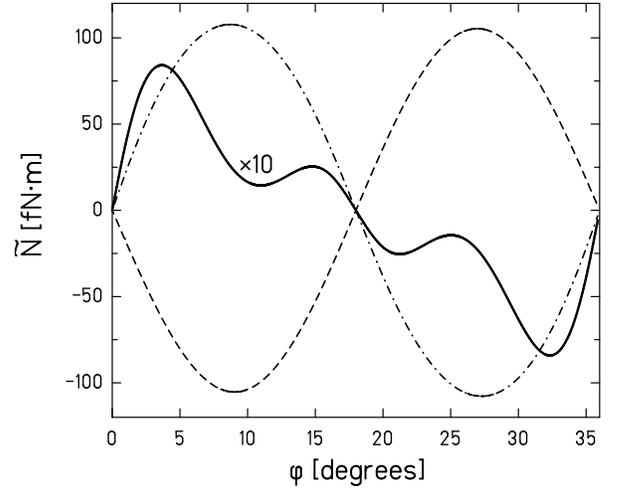}}\hfil
\caption{Calculated Newtonian torque on the Experiment~I pendulum at $s=0.2$~mm as a function of attractor rotation angle. The dash-dotted and dotted curves are the torques from the upper and lower attractor disks, respectively. The solid curve is the total torque magnified by a factor of ten.} 
\label{piover5}
\end{figure}

The torque on the pendulum from one complete set of attractor holes
contains 100 terms, 
\begin{eqnarray}
&~&\hat{N}(\phi_{\rm a}, \delta, \phi_{\delta}, s)= ~~~~~~~~~~~~~~~~~~\\
&~&~~~~~~~~~\sum_{j=0}^{9}\sum_{k=0}^{9}\tilde{N}_1(\phi+\phi_{\circ}+j\,\pi/5,\,\delta,\,\,\phi_{\delta}-k\,\pi/5,s)~, \nonumber
\end{eqnarray}
where the actual attractor angle is $\phi_a=\phi_{\circ}+\phi$, with $\phi$ being the attractor angle reading and $\phi_{\circ}$ is a $\phi$ value at which the torque vanishes.
We computed $\tilde{N}$ for $\phi_{\delta}=0$ and $\phi_{\delta}=\pi/2$; because the results were identical to within 10$^{-18}~$N-m, it was sufficient to fix $\phi_{\delta}=0$.
Figure~\ref{piover5} shows the calculated torques from the upper and lower disks of Experiment I for attractor angles between $-\pi/10$ and $\pi/10$ 
and a vertical separation of $s=0.2$~mm. 

The tenfold rotational symmetry of the attractor was exploited to compute a database of the 
harmonic torque amplitudes at frequencies $n\omega$,
\begin{equation}
\tilde{N}_n(\delta, s)\,=\,\frac{20}{\pi}\int_0^{\pi/10}\!\!\!\!\hat{N}(\phi, 0, \delta, 0, s)\sin(n\,\phi)\,d\phi~.\,
\end{equation}
for 20 different $\delta$'s between 0 and 2~mm  
and 100 values of $s$.
Numerical evaluation of this integral is exact as long as a sufficient number of equally spaced points are used. 
\subsection{\label{subsec:torqinterpolate}Generating Interpolated Torque Functions}
\begin{figure}
\hfil\scalebox{.58}{\includegraphics*[0.67in,0.5in][6.5in,4.8in]{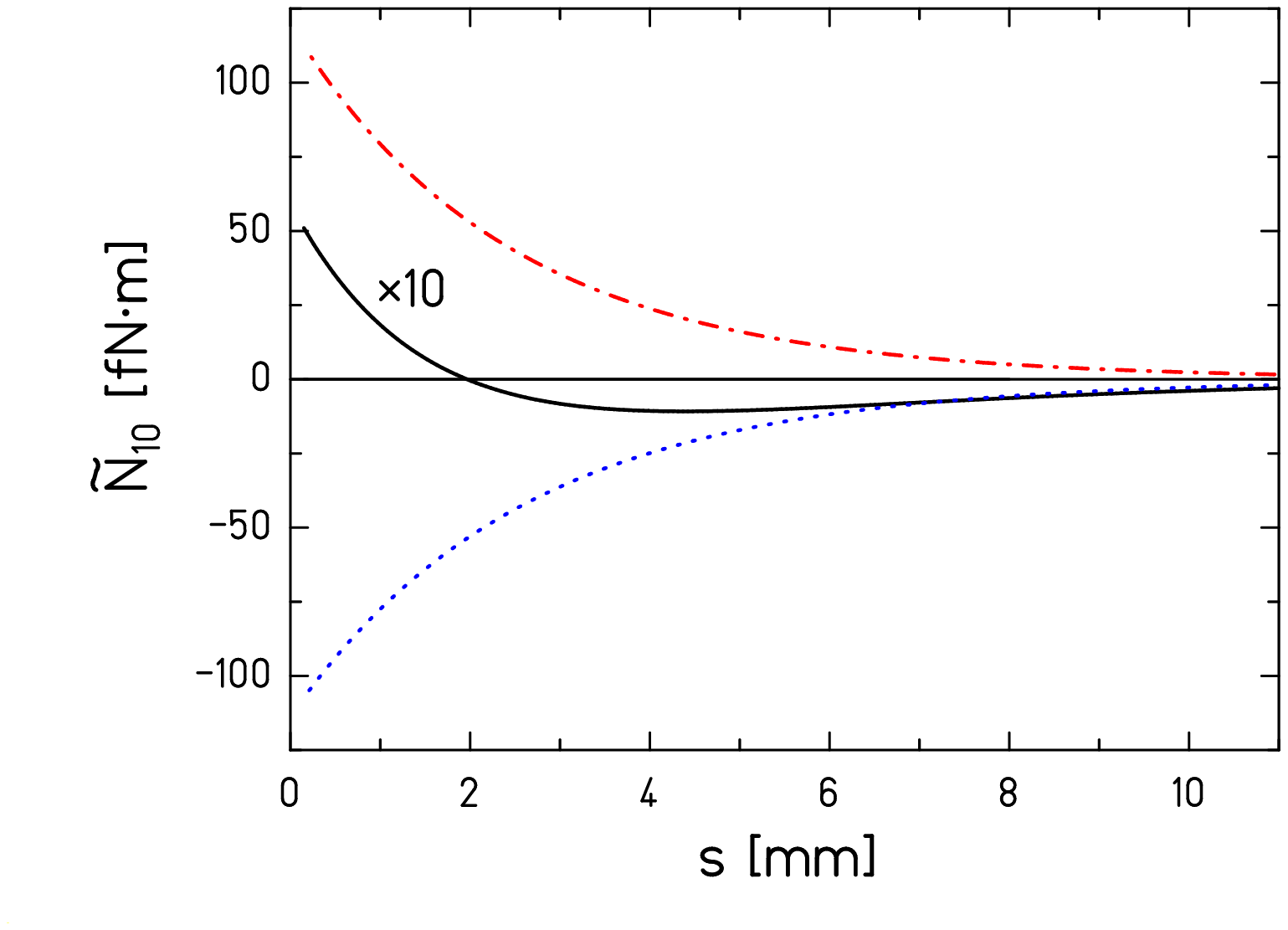}}\hfil
\caption{[color online] Predicted Newtonian $10\omega$ torque on the Experiment~I pendulum as a function of $s$. The solid curve shows the total torque multiplied by a factor of~10. The 
dash-dotted and dotted curves show the torques from the upper and lower disks, respectively.} 
\label{tint}
\end{figure}

We generated two-dimensional analytic  
harmonic torque amplitudes for each set of attractor holes at each signal frequency of interest by cubic-spline interpolation of the harmonic torque database.
Each potential required six interpolating functions: one for each of the 2 (3) sets of holes and the 3 (2) frequencies in Experiment I (II). Figure~\ref{tint} shows the interpolated $10\omega$ torque from both attractor disks of Experiment~I for $\delta=0$ as functions of vertical separation $s$.
\subsection{\label{subsec:accuracy}Precision Requirements}
We wanted the overall {\em absolute} precision of the torque calculations, $\delta \tilde{N}$, to be much better than the statistical uncertainty of the experimental values, which was as good as 
$\delta N \approx 1 \times 10^{-17}$~N-m (see Table~\ref{tab:oncend2}). Because of the large cancelation between the torques from the upper and lower attractor disks,  
$\delta \tilde{N} \approx \sqrt{2}\; \delta \tilde{N}_{\rm d}$,
where $\delta \tilde{N}_{\rm d}$ is the precision of the calculated torques on the pendulum from an individual upper or lower attractor disc. We therefore require 
\begin{equation}
\delta \tilde{N}_{\rm d} <<  7 \times 10^{-18}~{\rm N\,m}~.
\label{eq:torque precision}
\end{equation}
\begin{figure}
\hfil\scalebox{.6}{\includegraphics*[0.95in, 0.5in][6.5in,6.0in]{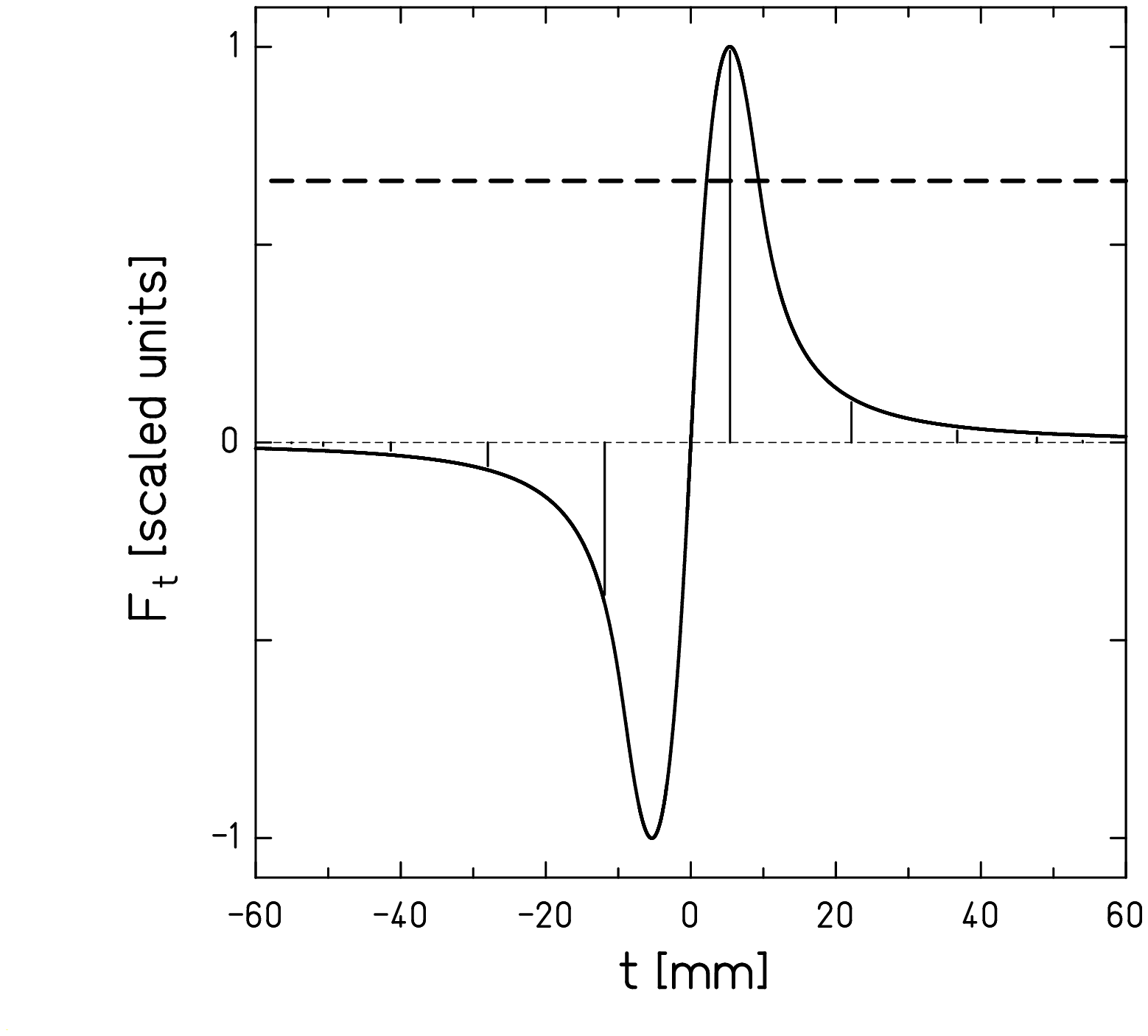}}\hfil
\caption{The solid line shows the horizontal Newtonian force, $F^G_{\rm t}$, on a single pendulum hole in Experiment I (scaled to a maximum value of unity) as a function of horizontal hole separation $t$. One attractor hole has been placed where the force
between cylinders is maximum ($t \sim 5.4$ mm). The horizontal distances of the
other 9 attractor holes are shown. The vertical lines show the contributions of each upper attractor hole to the force vector.  The
horizontal dashed line is the net force from all 10 attractor holes; it is largely determined by the two attractor holes closest to the pendulum hole.}
\label{fig:torque}
\end{figure}
\begin{figure}
\hfil\scalebox{.6}{\includegraphics*[0.7in,0.3in][6.7in,6.0in]{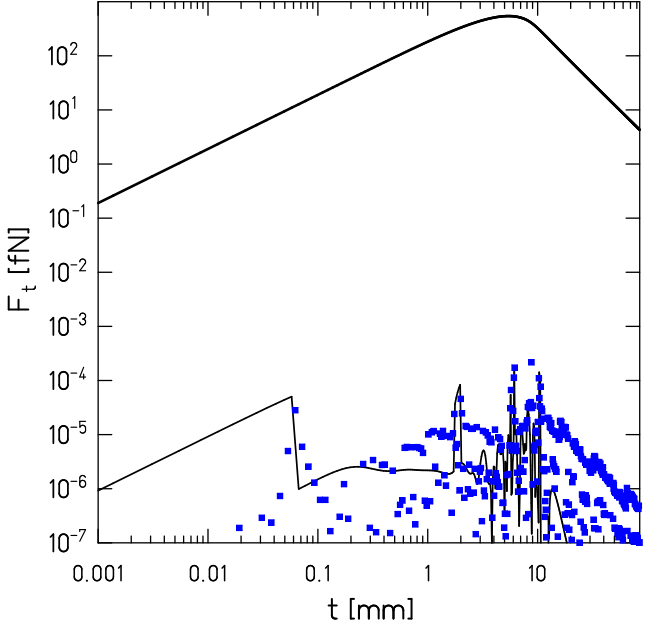}}\hfil
\caption{A typical Newtonian force calculation. The heavy line shows $F_t$ from the upper attractor disk of Experiment I at a separation of
$s=250~\mu$m. The light line shows the integration error, determined by doubling the number of steps in evaluating each of the two dimensions of the force integral. The points show the error associated with the spline interpolation
which we found by comparing the values interpolated halfway between the data-base points to a direct numerical evaluation of the force integrals.}
\label{fig:force accuracy}
\end{figure}
The torque $\tilde{N}_{\rm d}=10 \, r \, F_{\rm t}^{\rm d}$ where $r=.03$~m and the force on a single pendulum hole, $F_{\rm t}^{\rm d}$, consists of a sum over 10 attractor holes. Although
Fig.~\ref{fig:torque} shows that the force on a given pendulum hole is dominated by the forces from the two nearest attractor holes and that these two forces tend to cancel,  $\delta F_{\rm t}^{\rm d} \approx \sqrt{10}\; \delta F_{\rm t}$. We satisfy Eq.~\ref{eq:torque precision} if
\begin{equation}
\delta F^G_{\rm t} \approx \frac{\delta \tilde{N}_{\rm d}}{10 \sqrt{10}\,r} << 7 \times 10^{-18}~{\rm N}~.
\end{equation}
We adopted the requirement $\delta F^G_{\rm t} < 1 \times 10^{-19}~{\rm N}$. 
Two sources of error had to be considered. The desired precision determined the stopping point of the numerical integration procedure, as well as the spacing of the database points that were used to interpolate the force at arbitrary $s$ and $t$. Figure~\ref{fig:force accuracy} shows the calculated Newtonian force and the errors associated with the database and its interpolation.
The contributions to the error of the Newtonian torque calculations are listed in Table~\ref{tab:errcalcn}.
\begin{table}
\caption{Error budget for the Newtonian torque calculations.}
\begin{ruledtabular}
\begin{tabular}{lc}
Source & Error \\
\hline
Force Calculation & $\lesssim 1\times10^{-20}~$~N-m \\
Force Interpolation & $\lesssim 1\times10^{-20}~$~N-m \\
Torque Interpolation & $\lesssim 2\times10^{-20}~$~N-m \\ 
Total & $\sim2.5\times10^{-20}~$~N-m \\
\end{tabular}
\end{ruledtabular}
\label{tab:errcalcn}
\end{table}

Our calculations of the Yukawa, power-law and massive pseudoscalar forces could be less precise because, in essence, these forces were compared to the Newtonian residuals of our data that were always within about $2\sigma$ of zero. Because the fractional uncertainties in the best-fit values of $\alpha$, $\beta_k$ and $\gamma$ from uncertainties in the measured torques were greater than 0.4, the systematic uncertainties from the corresponding calculations of the forces between the two closest holes (see Fig.~\ref{fig:torque}) could be neglected as long as
\begin{equation}
\frac{\delta F_{\rm t}}{F_{\rm t}} << 0.4~.
\label{eq:non-Newtonian}
\end{equation}
We computed the non-Newtonian forces with an absolute accuracy good enough to satisfy
Eq.~\ref{eq:non-Newtonian} in all cases. For example, the fractional accuracy of the Yukawa force 
calculation was $\sim 4.5 \times 10^{-2}$ at $\lambda=10~\mu$m, $\sim 10^{-3}$ at $\lambda=100~\mu$m, and $2 \times 10^{-4}$ at $\lambda=1$~mm.				7
\section{\label{sec:results}Results}
\subsection{\label{subsec:fitting functions}Fitting Functions}
We constrained ISL violations by comparing the measured complex $n\omega$ torques, $N_n(\zeta_j) \pm \delta N_n(\zeta_j)$, at pendulum positions indicated symbolically as $\zeta_j=(x_j,y_j,z_j)$, to calculations of the expected Newtonian and possible Yukawa, power-law, or massive pseudoscalar torques. The calculated torques 
were explicit functions of $\zeta_j$ as well as of $P$ key instrumental parameters (the missing masses $M_{\rm p}$, $M_{\rm a}^{\rm up}$, $M_{\rm a}^{\rm lo}$ of the holes in the pendulum and the lower and upper attractors, the alignment parameters $x_{\circ}$, $y_{\circ}$, $z_{\circ}$, $\phi_{\circ}$, and the angular offset $\varphi$ and gap $g$ between the two attractor disks) denoted symbolically by the array $\eta_m$, and the ISL-violating parameters $\alpha$,  $\lambda$ and $\beta_k$ denoted symbolically as $\varsigma$. The complete data sets for both experiments were fitted by minimizing 
\begin{eqnarray}
\chi^2 &=& \sum_j \sum_n \left[ \frac{N_n(\zeta_j)-\tilde{N}_n(\zeta_j,\eta,\varsigma)}{\Delta N_n(\zeta_j)} \right]^2  \nonumber \\ 
&+& \sum_{m=1}^{P_{\rm max}} \left[ \frac{\eta_m - \eta_m^{\rm exp}}{\delta \eta_m^{\rm exp}} \right]^2~, 
\label{eq:chisqr}
\end{eqnarray}
where $j$ refers to the particular run, $n$ to the signal frequency ($10\omega$, $20\omega$, etc.), and $m$ to the individual experimental parameters ($P_{\rm max}$ of which were independently constrained) with $\eta_m^{\rm exp}$ and $\delta\eta_m^{\rm exp}$ being their independently measured values. For Experiment II we included an uncertainty in the phase of measured torques, $n \delta \phi_{\circ}$, where $\delta \phi_{\circ}=\pm 0.046$~degrees was deduced from the deviation of the index angle readings.
We accounted for the uncertainty in the independent variable $s$ by setting
\begin{equation}
\Delta N_n= \sqrt{(\delta N_n)^2+(\delta s~ \partial \tilde{N}_n/\partial s)^2}~.
\end{equation} 
Because $x_{\circ}$, $y_{\circ}$ and $\phi_{\circ}$ were not independently measured they were not included in the last sum in Eq.~\ref{eq:chisqr}.

The two-disk attractor data of Experiment I were fitted with
\begin{eqnarray}
\label{eq:experiment I fitting function}
&~&\tilde{N}_n(\zeta_j,\eta,\varsigma)=e^{i n \phi_{\circ}}M_{\rm p}\Bigl[M_{\rm a}^{\rm up}\hat{N}_{\rm up}(n,\delta_j,z_j\!-\!z_{\circ},\varsigma)~~~~~~~~~ \\
&~&~~~~~~~~~~~~~~+ {\rm e}^{i n \varphi} M_{\rm a}^{\rm lo}\hat{N}_{\rm lo}(n,\delta_j,z_j\!-\!z_{\circ}\!+\!h_1\!+\!g,\varsigma)\Bigr]~. \nonumber
\end{eqnarray} 
where $\hat{N}_{\rm up}$ and $\hat{N}_{\rm lo}$ (which were calculated assuming unit mass for the holes) have the forms
\begin{eqnarray}
\label{eq:torque defs}
\hat{N}(n,\delta,z,\varsigma)&=&\hat{N}^G(n,\delta,z)+\alpha \hat{N}^Y(n,\delta,z,\lambda) \\
&+&\beta_k \hat{N}^P(n,\delta,z,k) + \gamma \hat{N}^M(n,\delta,z,\lambda)~,\nonumber
\end{eqnarray}
and $\delta_j\,=\,\sqrt{(x_j-x_0)^2+(y_j-y_0)^2}$.
The upper-disk-only runs of Experiment I used only the first term on the right-hand side of Eq.~\ref{eq:experiment I fitting function}.

The parameter $M_{\rm p}$, which multiplies all torques, is degenerate with
the experimental torque scale factor $C$. We therefore assumed $C=1$ exactly and combined the scale factor uncertainty with the measurement uncertainty in $M_{\rm p}$ when specifying $\delta M_{\rm p}$.
The Experiment II data were fitted with a function similar to Eq.~\ref{eq:experiment I fitting function} but with additional terms that accounted for the out-of-phase holes in the upper attractor disk.
\subsection{\label{subsec:newtpred}Comparison of Data with Newtonian Predictions}
We first compared our data with the Newtonian prediction by fixing $\alpha$, the $\beta_k$'s and $\gamma$ in Eq.~\ref{eq:torque defs} to zero.
\subsubsection{\label{subsubsec:constI}Experiment I}
\begin{figure}
\hfil\scalebox{.55}{\includegraphics*[0.9in,0.4in][6.7in,5.1in]{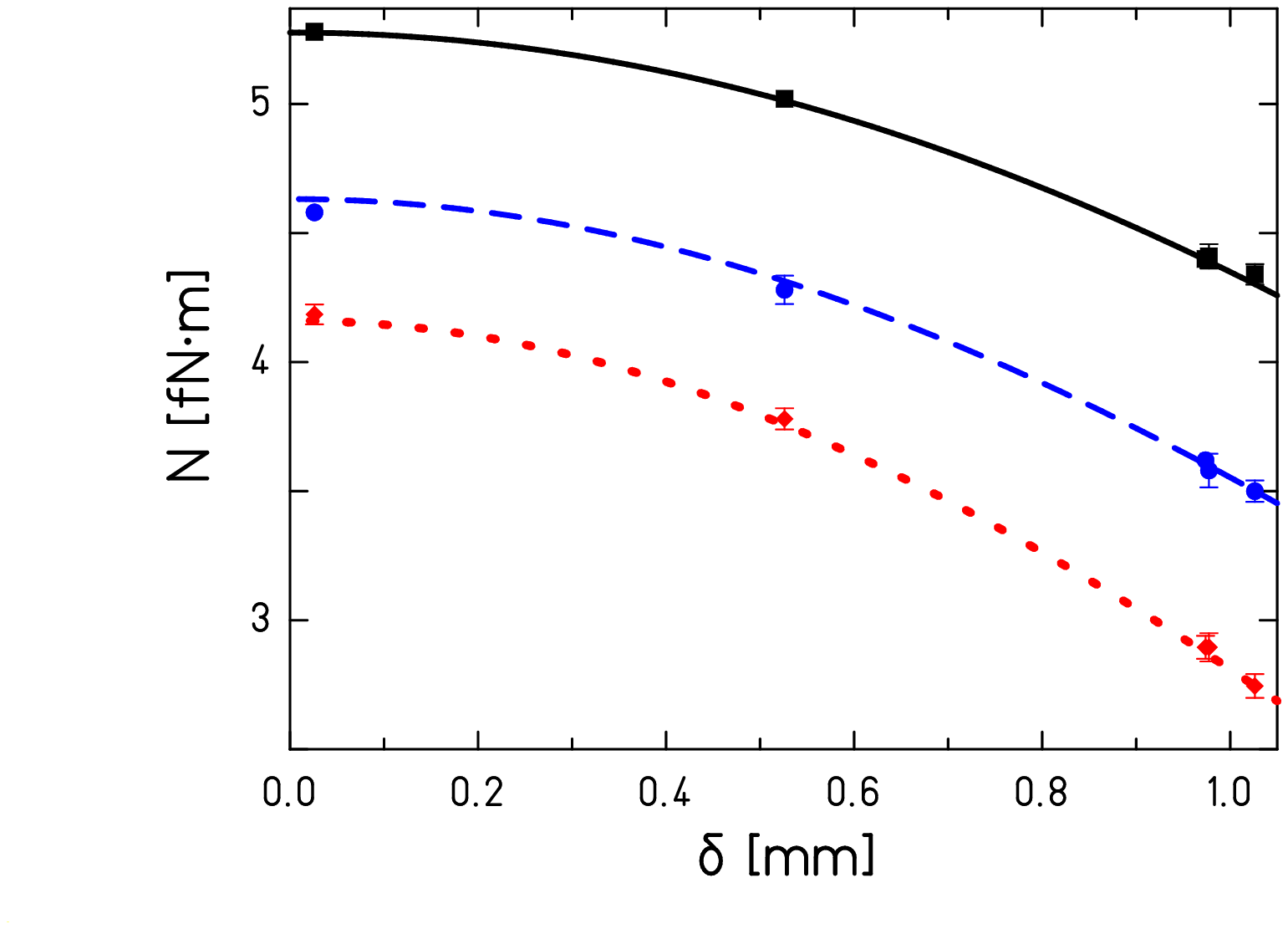}}\hfil
\caption{[color online] Centering data and Newtonian fits as a function of the magnitude, $\delta$, of the offset. This plot includes points displaced in both $x$ and $y$. Solid, dashed and dotted lines denote $N_{10}$, $2 \times N_{20}$ and $1.5 \times N_{30}$, respectively.} 
\label{fig:cent1}
\end{figure}

\begin{figure}[h]
\hfil\scalebox{.59}{\includegraphics*[0.64in,0.4in][6.5in,7.3in]{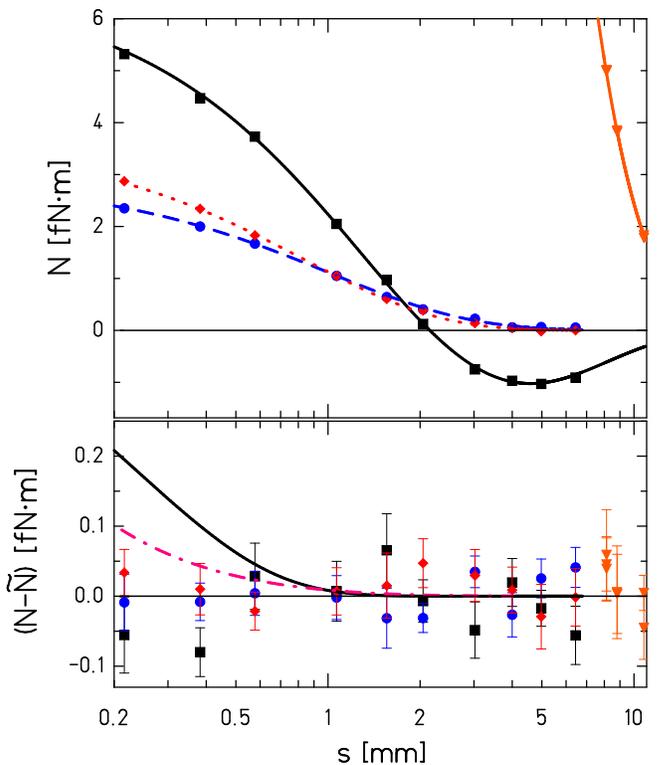}}\hfil
\caption{[color online] Upper panel: torques and Newtonian fit for on-center data from Experiment~I. Errors are smaller than the point size. Square, round, and diamond-shaped points show $N_{10}$, $N_{20}$, and $N_{30}$ torques. Triangular points are $N_{10}$ of the upper-disk-only runs; these show the degree to which our 2-disk attractor attenuates the Newtonian signal. Lower panel: residuals; the solid and dot-dashed curves show the expected $N_{10}$ residuals from an $\alpha=1$, $\lambda=250~\mu$m Yukawa and from a $\beta_5=0.005$ power-law interaction, respectively. The corresponding Yukawa curves in Figure 3 of Ref.~\cite{ho:01} and Figure 2 of Ref.~\cite{ad:03} were incorrectly designated as $\alpha=3$.} 
\label{fig:oncent1}
\end{figure}

\begin{table}[t]
\caption{Comparison of the measured and fitted experimental parameters for Experiment~I. The uncertainty in 
$M_{\rm p}$ 
reflects the uncertainty in the overall torque-scale factor as well as in the hole masses. Primed quantities refer to the upper-disk-only runs.}
\begin{ruledtabular}
\begin{tabular}{ccc}
Parameter & Measured Value & Fitted Value\\
\hline
$M_{\rm p}$ & $3.972\,\pm\,0.060$ g & $4.096\,\pm\,0.014$ g\\
$M_{\rm a}^{\rm up}$ & $11.770\,\pm\,0.004$ g & $11.7707\,\pm\,0.0039$ g\\
$M_{\rm a}^{\rm lo}$ & $88.666\,\pm\,0.019$ g & $88.665\,\pm\,0.019$ g\\
$x_0$ & N/A & $-0.413\pm 0.007$~mm\\
$y_0$ & N/A & $-0.977\pm 0.012$~mm\\
$z_{\circ}$ & 0.000$\,\pm\,0.005$~mm & 0.001$\,\pm\,0.002$~mm\\
$\phi_{\circ}$ & N/A & $0.007\pm 0.007$ deg \\
$x_0^{\prime}$ & N/A & $-1.114\pm 0.100$~mm\\
$y_0^{\prime}$ & N/A & $-0.846\pm 0.100$~mm\\
$z_{\circ}^{\prime}$ & 0.000$\,\pm\,$0.025~mm & 0.025$\,\pm\,$0.013~mm\\
$\phi_{\circ}^{\prime}$ & N/A & $-0.003\pm 0.023$ deg \\
$g$ & 0.010$\,\pm\,$0.010~mm & 0.023$\,\pm\,0.001$~mm\\
$\varphi$ &  $0.00 \pm 0.01$ deg & $0.0095 \pm 0.0006$ deg 
\end{tabular}
\end{ruledtabular}
\label{tab:param1}
\end{table}
The fit for Newtonian gravity alone was excellent with $\chi^2\,=\, 122$, $\nu\,=\,156$ and $Q(\chi^2,~\nu)\,=\,0.97$, where $Q(\chi^2,\nu)$ is the probability that a $\chi^2$ as poor as that observed could arise by chance. The best fit values of the experimental parameters were all within about 2 standard deviations of their measured values. Table~\ref{tab:param1} displays the measured and fitted values of the parameters. Figure~\ref{fig:cent1} shows the Newtonian fit to the centering data at an average vertical separation of 234~$\mu$m. The top panel of Fig.~\ref{fig:oncent1} shows the data and Newtonian fit for all of the on-center runs; the bottom panel shows the residuals.
\subsubsection{\label{subsubsec:constII}Experiment II}
\begin{figure}
\hfil\scalebox{.54}{\includegraphics*[0.5in,0.4in][6.5in,7.3in]{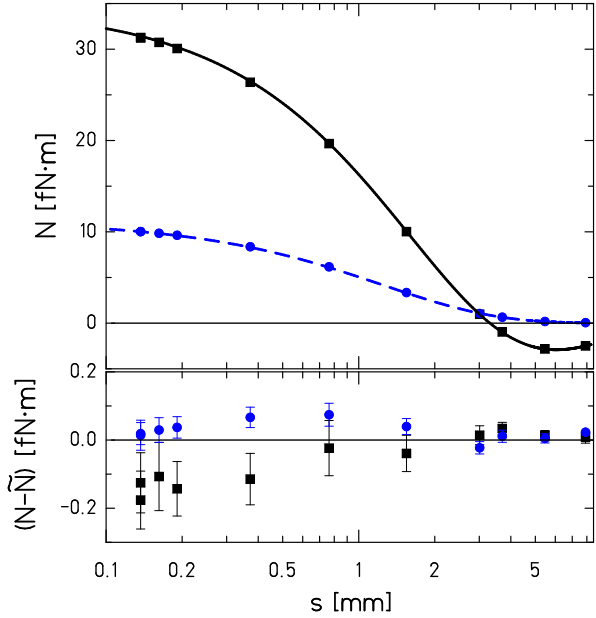}}\hfil
\caption{[color online] Top: Experiment~II on-center data and Newtonian fit; solid square and round points show $N_{10}$ and $N_{20}$ torques, respectively. Bottom: residuals of the Newtonian fit. 
The uncertainties are larger at small separations (large signals) because the error contributions from the measured free torsion period and in $\partial {\tilde N} / \partial s$ have a bigger effects at small $s$.} 
\label{fig:oncent2}
\end{figure}

\begin{table}[b]
\caption{Comparison of the measured and fitted experimental parameters for Experiment~II. The uncertainty in $M_{\rm p}$ 
reflects the uncertainty in the overall torque-scale factor as well as in the hole masses. $\varphi^{\prime}$ is the angular offset between the in-phase and out-of-phase holes in the upper attractor.}
\begin{ruledtabular}
\begin{tabular}{ccc} 
Parameter & Measured Value & Fitted Value \\
\hline
$M_{\rm p}$ & $2.6618\,\pm\,0.0140$ g & $2.6423\,\pm\,0.0057$ g\\
$M_{\rm a}^{\rm up,in}$ & $8.6228\,\pm\,0.0004$ g & $8.6227\,\pm\,0.0004$ g\\
$M_{\rm a}^{\rm up,out}$ & $19.2592\,\pm\,0.0005$ g & $19.2592\,\pm\,0.0005$ g\\
$M_{\rm a}^{\rm lo,in}$ & $125.0984\,\pm\,0.0006$ g & $125.0984\,\pm\,0.0006$ g\\
$x_0$ & N/A & $0.803\,\pm\,0.003$~mm \\
$y_0$ & N/A & $-0.189\,\pm\,0.003$~mm \\
$z_{\circ}$ & $0.000\,\pm\,0.005$~mm & $-0.001\pm 0.003$~mm \\
$\phi_{\circ}$ & N/A & $-0.001\pm 0.012$ deg \\
$g$ & 0.0014$\,\pm\,$0.0010~mm & $0.0008\pm 0.0010$~mm \\
$\varphi$ & $0.13 \pm 0.01$ deg  & $0.118\pm 0.006$~deg \\ 
$\varphi^{\prime}$ & $0.00 \pm 0.01$ deg  & $0.003 \pm 0.009$ deg \\ 

\end{tabular}
\end{ruledtabular}
\label{tab:par2}
\end{table}
The Newtonian fit for the Experiment II data was good, with $\chi^2=40.5$ for $\nu=41$, with 
$Q(\chi^2,~\nu)=0.40$. Again all fitted values of the parameters were within about 2 standard deviations of their measured values. Table~\ref{tab:par2} displays the measured and fitted values of these parameters.  
Figure~\ref{fig:oncent2} shows the data and Newtonian fit and residuals for all~11 on-center runs.
\subsubsection{\label{subsubsec:combconst}Combined Results}
We fitted both data sets simultaneously with a function consisting of the sum of the Experiment~I and Experiment~II functions. The combined fit assuming Newtonian gravity had $\chi^2=163.4$ with $\nu=197$ and $Q(\chi^2,~\nu)=0.95$. The values of the fitted and measured parameters for the combined data set were essentially unchanged.
\subsection{\label{subsec:const}Constraints on Short-Range Yukawa Interactions}
We generated constraints on Yukawa ISL violations using a fitting function that included the Newtonian and Yukawa terms. We analyzed each experiment individually as well as the combined data set. As the results for Experiment I and Experiment II were consistent, we quote only the constraints extracted from the combined data sets.  
Data were fitted for 24 fixed values of $\lambda$ between 0.020~mm and 10.0~mm, and for each case we extracted the central value $\alpha_{\lambda}$ and its 1-$\sigma$ uncertainty, $\delta\alpha_{\lambda}$. We then used these values to establish 95$\%$ confidence intervals ($\approx$2$\sigma$ limits)  on the allowed values of $+\alpha$, $-\alpha$, and $|\alpha|$. 
The constraints on $|\alpha|$ were found by integrating a normal distribution,  $N(\alpha_{\lambda},~\delta\alpha_{\lambda})$ with mean $\alpha_{\lambda}$ and width $\delta\alpha_{\lambda}$, and finding the value of $|\alpha_{\lambda}|$ that satisfied
\begin{equation}
\int_{-\infty}^{-|\alpha_{\lambda}|}\!N(\alpha_{\lambda}, \delta\alpha_{\lambda}) d\alpha_{\lambda} \,+\,\int^{\infty}_{|\alpha_{\lambda}|}\!N(\alpha_{\lambda}, \delta\alpha_{\lambda}) d \alpha_{\lambda}\,=\,0.05.
\end{equation}  

\begin{figure}
\hfil\scalebox{.58}{\includegraphics*[0.8in,0.5in][6.6in,5.7in]{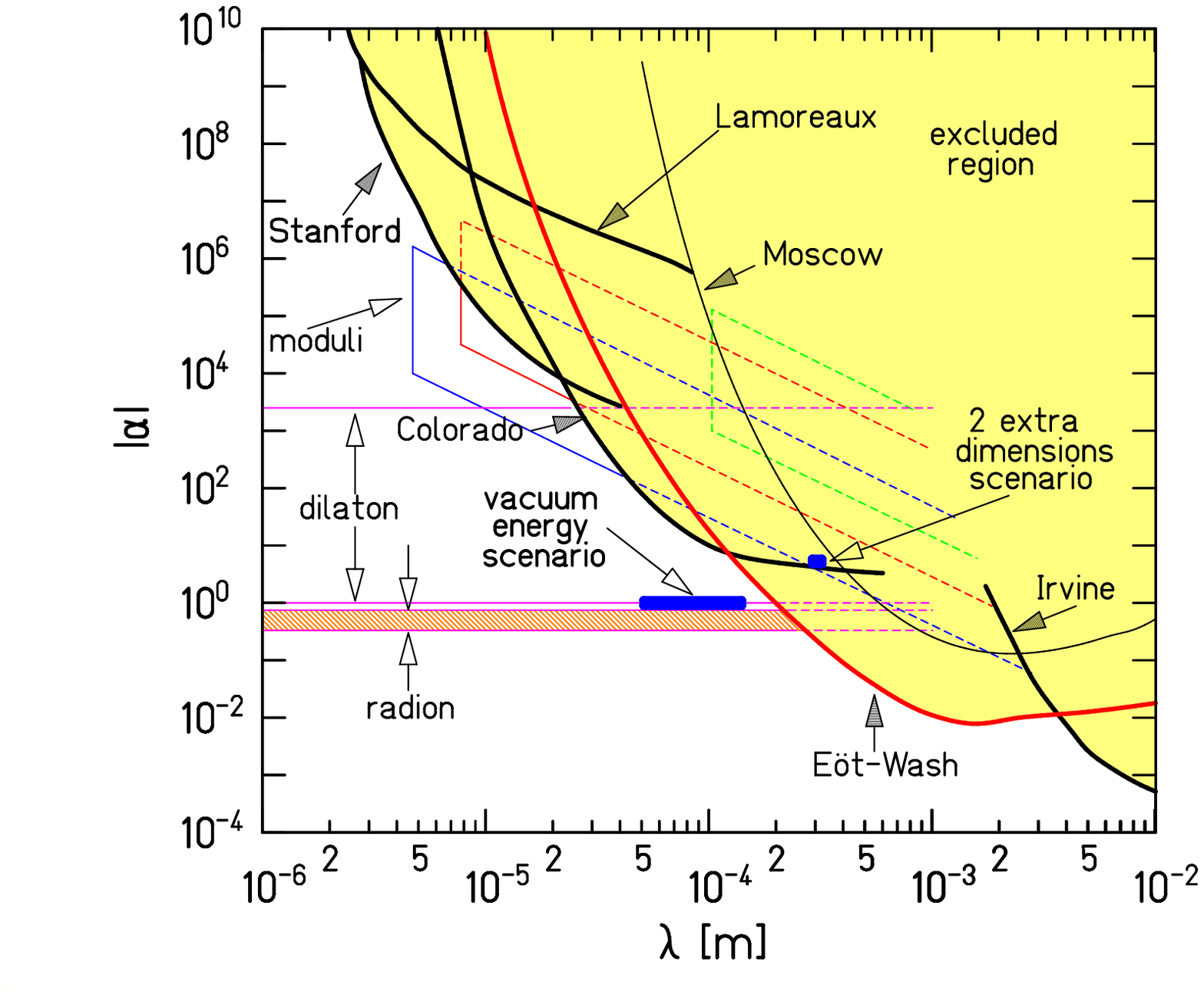}}\hfil
\caption{[color online] Yukawa constraints from our combined data set as well as from other work\cite{ho:85,mi:88,la:97,la:98,lo:03,ch:03}. The area above the heavy curves is excluded at the $95\%$ confidence level. Predicted ISL violating effects from ``extra dimensions''\cite{ar:98},  from dilaton\cite{ka:00}, moduli\cite{di:96} and radion\cite{an:98} exchange, and from a conjectured solution\cite{su:99} of the cosmological constant (vacuum energy) problem are shown as fainter lines.}
\label{alamtot}
\end{figure}

Table~\ref{tab:consttot} shows the $95\%$ confidence intervals on positive and negative $\alpha$ as well as the limit on $|\alpha|$
for the~24 different values of $\lambda$, and Figure~\ref{alamtot} displays the $|\alpha|$ constraint. It is easy to understand the general form of our constraints. In the limit where $\lambda$ is small compared to $d$, the characteristic dimension of the holes (in our case their thickness), the holes essentially behave like infinite parallel plates for which the Yukawa potential energy per unit area is
\begin{equation}
\frac{V}{A}=2 \pi \alpha \lambda^3 G \rho_1 \rho_2 \; {\rm e}^{-s/\lambda}~,
\end{equation}
where $s$ is the pendulum-to-attractor separation
The torque is proportional to the derivative $\partial V/\partial \phi$ so that for a given torque sensitivity and smallest separation $s_{\rm min}$,
we expect the constraints to have the asymptotic form 
\begin{equation}
\log \alpha(\lambda) \:\propto\: -3 \log \lambda +s_{\rm min}/\lambda~;
\end{equation}
and an asymptotic slope on a log($\alpha$) vs. log($\lambda$) plot of
\begin{equation}
\frac{d \log \alpha}{d \log \lambda}= -3 -\frac{s_{\rm min}}{\lambda}
\end{equation}
which is satisfied by our constraints shown in Fig.~\ref{alamtot}. 
On the other hand, for $\lambda >> d$, the cancelation from the lower plate reduces the
signal from the Yukawa force and the constraints eventually weaken with increasing $\lambda$.
 
Our constraints improved upon the previous results shown in Fig.~\ref{fig:old constraints} by a factor of up to $\sim10^4$, and on more recent results\cite{lo:03,ch:03} shown in Fig.~\ref{alamtot} by up to a factor of almost $10^2$. Our experiments are the only tests to date that reach gravitational sensitivity for length scales less than 500~$\mu$m. In particular, Yukawa interactions with $|\alpha| \geq 1$ are excluded at 95\% confidence for $\lambda \geq 197~\mu$m.
\begin{table}
\caption{$95\%$ confidence level constraints on Yukawa interactions from the combined data set.}
\begin{ruledtabular}
\begin{tabular}{crr}
$\lambda$~(mm) & \multicolumn{1}{c}{$\alpha$} & \multicolumn{1}{c}{$|\alpha|$} \\ 
\hline
0.010 & $(-4.7 \pm 6.4)\times10^9$ & $1.0\times10^{10}$  \\
0.025 & $(-7.4 \pm 10.4)\times10^4$ & $1.6 \times 10^5$ \\
0.050 & $(-2.2 \pm 7.9)\times10^2$ & $8.8 \times10^2$  \\
0.10  & $(0.2 \pm 1.8)\times10^1$ & $1.8 \times 10^1$  \\
0.25  & $(0.9 \pm 4.0)\times10^{-1}$ & $4.3 \times 10^{-1}$  \\
0.50  & $(1.0 \pm 4.5)\times10^{-2}$ & $4.8 \times 10^{-2}$  \\
1.00  & $(0.1 \pm 1.1)\times 10^{-2}$ & $1.1 \times 10^{-2}$  \\
1.50  & $(-1.8 \pm 7.2)\times 10^{-3}$ & $7.9 \times 10^{-3}$  \\
2.50  & $(-5.1 \pm 5.9)\times 10^{-3}$ & $1.0 \times 10^{-2}$ \\
5.00  & $(-7.3 \pm 6.7)\times 10^{-3}$ & $1.3 \times 10^{-2}$ \\
10.0  & $(-0.7 \pm 19)\times 10^{-3}$ & $1.8\times 10^{-2}$ \\
\end{tabular}
\end{ruledtabular}
\label{tab:consttot}
\end{table}
\subsection{Constraints on Power-Law Interactions}
We constrained power-law violations of the ISL by fitting our combined data set with a function that contained the
Newtonian term and a single power-law term. This procedure was carried out for power-law potentials with $k=2$, 3, 4, and 5. The results are listed in Table~\ref{tab:power-law constraints} together with constraints from previous ISL tests given in Ref.~\cite{fi:99}. 
\begin{table}
\caption{68\% confidence constraints on power-law potentials of the form given in Eq.~\protect\ref{eq:power law definition} from this work and from previous work tabulated in Ref.~\cite{fi:99}.}
\begin{ruledtabular}
\begin{tabular}{lcc}
$k$ &   $|\beta_k|$(this work) &  $|\beta_k|$(previous work)  \\
\hline
2   &   $3.6 \times 10^{-3}$   &  $1.3 \times 10^{-3}$\cite{ho:85}  \\  
3   &   $2.8 \times 10^{-3}$   &  $1.3 \times 10^{-2}$\cite{ho:85}  \\ 
4   &   $2.9 \times 10^{-3}$   &  $1.3 \times 10^{-1}$\cite{mi:88}  \\ 
5   &   $2.3 \times 10^{-3}$   &  $2.1 \times 10^{-1}$\cite{mi:88}  \\ 
\end{tabular}
\end{ruledtabular}
\label{tab:power-law constraints}
\end{table}
\subsection{Constraints on Couplings of Massive Pseudoscalars}
Second-order exchange of massive pseudoscalars was constrained by fitting the combined data set with a function containing the only the Newtonian and massive pseudoscalar terms. The results are listed in Table~\ref{tab:gamma}.
\begin{table}
\caption{95\% confidence upper bounds on $|\gamma(\lambda)|$ where $\lambda =\hbar/(mc)$ and $m$ is the pseudoscalar mass.}
\begin{ruledtabular}
\begin{tabular}{ccl}
$\lambda$ [mm] & $mc^2$ [meV] & $|\gamma(\lambda)|$ \\
\hline
0.02  &  9.85 & $6.2 \times 10^{8}$ \\
0.05  &  3.94 & $2.6 \times 10^{3}$ \\
0.10  &  1.97 &  $2.8 \times 10^{1}$ \\
0.20  &  0.985 & $1.24 $ \\
0.50  &  0.394 &  $8.7 \times 10^{-2}$ \\
1.0  &   0.197 &  $2.5 \times 10^{-2}$\\
2.0  &   0.0985 & $1.1 \times 10^{-2}$\\
3.0  &   0.0657 & $8.0 \times 10^{-3}$ \\
5.0  &   0.0394 & $6.4 \times 10^{-3}$ \\
10.0 &  0.00197 & $5.5 \times 10^{-3}$ 
\end{tabular}
\end{ruledtabular}
\label{tab:gamma}
\end{table}
%
%
\section{Some Implications of the Results}
\subsection{Extra-Dimension Scenarios}
\label{subsec:extrad}
The most basic constraint from this work is an upper limit on the maximum size of an extra dimension. If we assume that one extra dimension is much larger than all the others, and compactification on a torus, ISL violating effects will first show up as a Yukawa interaction with $\alpha=8/3$ (see Eq.~\ref{eq:R and alpha}). Our  95\% confidence level constraints require
that 
\begin{equation}
R_{\ast} \leq 160~\mu{\rm m}~.
\end{equation}
The scenario of two equal-size large new dimensions, whose radius $R_{\ast}=0.38$~mm (see Eq.~\ref{eq:true M_P}) is appropriate to give $M_{\ast}=1$~TeV,
is ruled out by our results. According to Eq.~\ref{eq:true M_P}, our 95\% confidence level constraint, $\lambda(\alpha=16/3) \leq 130~\mu$m  implies
\begin{equation}
M_{\ast} \geq 1.7~{\rm TeV/c}^2~.
\end{equation}
\subsection{New Boson-Exchange Forces}
\label{subsec:newpart}
\subsubsection{Yukawa Interactions: Radion and Dilaton Constraints}
We provide significant constraints of the radion-mediated force predicted by Eqs.~\ref{eq:radion} and \ref{eq:radion range}. For $n=1$ we find $\lambda(\alpha=1/3) \leq 270~\mu$m while for $n=6$ we have $\lambda(\alpha=3/4) \leq 215~\mu$m 
which gives the following 95\% confidence level constraints on the unification mass
\begin{eqnarray}
M_{\ast}(n=1) &\geq& 3.0~{\rm TeV/c}^2 \nonumber \\
M_{\ast}(n=6) &\geq& 4.6~{\rm TeV/c}^2~. 
\end{eqnarray} 
Note that the radion constraint given in \cite{lo:03} is not valid as it was based on an earlier, erroneous prediction for $\alpha$.

Our 95\% confidence level constraint on the dilaton mass, based on the couplings predicted in Ref.~\cite{ka:00}, is
\begin{equation}
m \geq 1.0~~~{\rm meV/c}^2~.
\end{equation}
\subsubsection{Yukawa Interaction from Axion Exchange}
The predicted Yukawa interaction from axion exchange (Eq.~\ref{eq:axion_alpha}) is too small to have been detected in this work. Even if the axion exists and has a mass in the upper portion of the allowed range (greater than 0.1 meV/$c^2$), its Yukawa interaction is unlikely to be detected in the near future. The axion's monopole-dipole and dipole-dipole interactions (discussed in Ref.~\cite{mo:84}) seem more promising avenues for detecting axions in this mass range.
\subsubsection{Power-law Interactions}
\begin{figure}[!]
\hfil\scalebox{.5}{\includegraphics*[0.7in,0.45in][7.5in,5.9in]{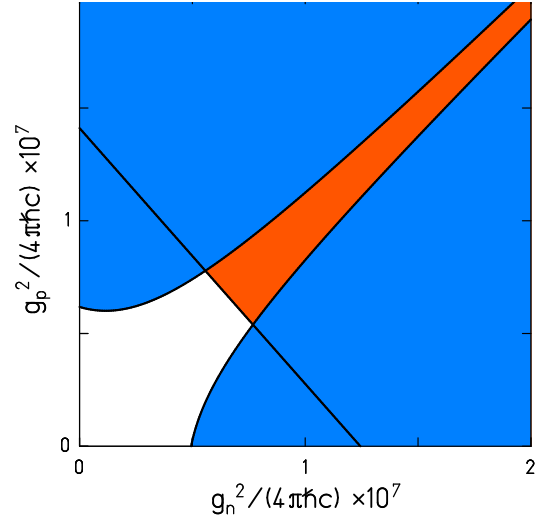}}\hfil
\caption{[color online] Constraints on the couplings of massless pseudoscalar particles to protons and neutrons. The shaded region is excluded at 68\% confidence. The down-sloping line is from this ISL test; the two curved lines are extract from a previous Equivalence-Principle test\cite{sm:00}.}
\label{fig:m=0 PS}
\end{figure}

The $\gamma_5$ couplings of massless pseudoscalar bosons to neutrons and protons, given by Eq.~\ref{eq:massless}, are constrained by our limits on $\beta_3$ from Table~\ref{tab:power-law constraints}. The contribution from
coupling to electrons can be ignored because of the very small upper limit on such couplings deduced from an electron-spin-dependent experiment (see Ref.~\cite{fi:99}). Consider the dimensionless parameter
\begin{eqnarray}
\Gamma &\equiv& \left[ \frac{Z}{\mu}\right]_{\! 1} \left[ \frac{Z}{\mu}\right]_{\! 2} \frac{g_p^4}{(\hbar c)^2} + \left[ \frac{N}{\mu}\right]_{\! 1} \left[ \frac{N}{\mu}\right]_{\! 2} \frac{g_n^4}{(\hbar c)^2} \\
&+&\Bigg( \left[ \frac{Z}{\mu}\right]_{\! 1} \left[ \frac{N}{\mu}\right]_{\! 2} + \left[ \frac{N}{\mu}\right]_{\! 1} \left[ \frac{Z}{\mu}\right]_{\! 2} \Bigg) \frac{g_p^2}{\hbar c}\frac{g_n^2}{\hbar c} \nonumber
\end{eqnarray}
where $[Z/\mu]$ and $[N/\mu]$ are defined in Ref.~\cite{sm:00} and the subscripts 1 and 2 refer to the pendulum and attractor  materials, aluminum and copper, respectively, with\begin{eqnarray}
\left[ \frac{Z}{\mu} \right]_{\rm Al}\!\!&=&0.48181~~~~~~~ \left[ \frac{N}{\mu} \right]_{\rm Al}\!\!=0.51887~,  \\
\left[ \frac{Z}{\mu} \right]_{\rm Cu}\!\!&=&0.45636~~~~~~~ \left[ \frac{N}{\mu} \right]_{\rm Cu}\!\!=0.54475~.  
\end{eqnarray}
Our experimental limit on $\Gamma$,
\begin{equation}
\Gamma = \beta_3 \frac{c G}{\hbar^3} 64 \pi^3 u^4 (1~{\rm mm})^2 = 2.56 \times 10^{-10} \beta_3~,
\end{equation} 
leads to the constraints on $g_p^2$ and $g_n^2$ shown in Fig.~\ref{fig:m=0 PS}.
Figure~\ref{fig:m=0 PS} also shows constraints we extracted from a previous Equivalence Principle test\cite{sm:00} using Eq. 32 of that work\cite{error} and assuming $\Delta a = 2.8 \times 10^{-13}$~cm/s$^2$. 
\subsubsection{Couplings of Massive Pseudoscalars}
\begin{figure}
\hfil\scalebox{.58}{\includegraphics*[0.75in,0.5in][6.5in,4.9in]{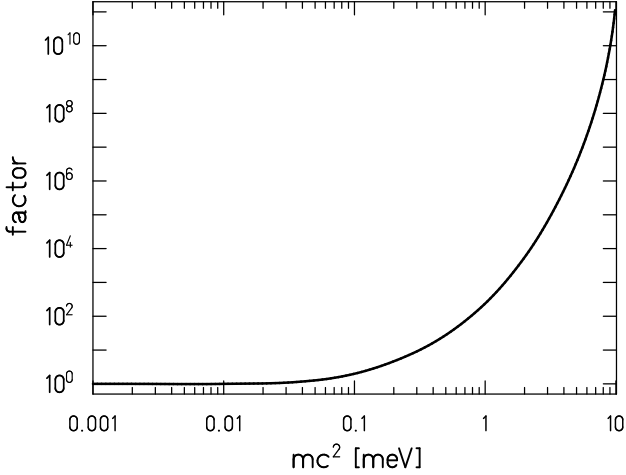}}\hfil
\caption{Scaling factor by which the 68\% confidence upper bounds on $g_p^2$ and $g_n^2$ from this experiment
(shown in Fig.~\protect\ref{fig:m=0 PS})
should be multiplied to account for a non-zero pseudoscalar mass $m$.}
\label{fig:massive PS}
\end{figure}

Our ISL constraints on $g_p^2$ and $g_n^2$, shown in
Fig.~\ref{fig:m=0 PS}, apply to massless pseudoscalars. The corresponding constraints on the couplings of massive pseudoscalars (see Eq.~\ref{eq:massive}) are obtained using
\begin{equation}
\Gamma(\lambda)=5.12 \times 10^{-10} \gamma(\lambda)~,
\end{equation}
and can be obtained using Table~\ref{tab:gamma} or Fig.~\ref{fig:massive PS}.
\subsection{\label{subsec2:cosmo}Vacuum-Energy Scenarios}
The constraints on these initial experiments were not sensitive enough to check the suggestion\cite{su:99,su:03} that gravity essentially ``shuts off'' (i.e. $\alpha \approx -1$) at $\lambda \approx 100~\mu$m.
\subsection{\label{sec:future}Future Prospects}
We are now running a second-generation instrument based on the same general
principles but with a larger number of smaller-sized holes in the pendulum
and attractor. This device has several other technical improvements that increase
the signal and reduce the noise. We expect this new instrument will have the sensitivity to check the vacuum-energy scenario.
The results will be reported in a later publication.
\begin{acknowledgments}
We are grateful to Professors Ann Nelson and David Kaplan for helpful
discussions and to the NSF (Grant PHY-9970987) and the DOE for
primary and secondary support, respectively. We thank Tom Butler for the ANSYS 
calculations of the pendulum capacitance. Rogan Carr helped with the force calculations
and Nathan Collins, Angela Kopp and Deb Spain assisted with parts of the 
experiment. CDH and EGA thank the Universit\'{a} di Trento for partial support while parts of this paper were being written.
\end{acknowledgments}
\appendix
%
%
\section{\label{sec:noise}Thermal Noise}
Thermal noise in any oscillator sets a fundamental limit on the
achievable statistical error on its amplitude.
A single-mode torsion oscillator subject to both velocity and internal
damping obeys the equation
\begin{equation}
N=I\ddot{\theta} + b \dot{\theta} + \kappa (1+i\phi) \theta~,
\end{equation}
where $N$ is the applied torque, $I$ the rotational inertia, $\theta$
the angular deflection of the oscillator, and $\kappa$ the torsional spring
constant of the suspension fiber. The velocity-damping coefficient $b$ accounts
for any losses due to viscous drag, eddy currents etc., while
the loss angle $\phi$ accounts for internal friction\cite{sa:90} of the suspension fiber.

In the familiar
case of pure velocity damping ($b > 0$, $\phi = 0$), the spectral
density of torque noise is independent of frequency
\begin{equation}
\langle N^2_{{\rm th}}(\omega)\rangle= 4k_B T \frac{I \Omega}{Q}~,
\end{equation}
where $\Omega=\sqrt{\kappa/I}$ is the free resonance frequency and
$Q=I \Omega /b$ is the quality factor of the oscillator. 
This torque noise implies a spectral density in the angular deflection
noise of
\begin{equation}
\langle\theta^2_{{\rm th}}(\omega)\rangle=\frac{4 k_B T}{QI}\frac{\Omega}
{(\Omega^2-\omega^2)^2+
(\Omega\omega/Q)^2}~.
\label{eq:theta noise for velocity damping}
\end{equation}
In this case there is no advantage, from the point of view of thermal noise, for
placing the signal at any particular frequency.

On the other hand, for pure internal damping ($b=0$, $\phi >
0$) the
spectral density of thermal noise has a $1/f$ character
\begin{equation}
\langle N^2_{{\rm th}}(\omega) \rangle= 4k_B T \frac{I \Omega^2}{\omega Q}~,
\label{eq:torque noise}
\end{equation}
where now $Q=1/\phi$. 
The corresponding spectral density of thermal noise in the twist
is
\begin{equation}\langle \theta^2_{{\rm th}(\omega)}\rangle    =\frac{4k_b T}{Q \omega I}
\frac{\Omega^2}{(\Omega^2-\omega^2)^2+(\Omega^2/Q)^2}~,
\label{eq:thermal}
\end{equation}
so that it is advantageous to boost the signal frequency above $\Omega$
until the noise contribution from
the angular deflection readout system, $\langle \theta^2_{{\rm ro}}\rangle$, becomes comparable to $\langle \theta^2_{{\rm th}}\rangle$.

\begin{figure}
\hfil\scalebox{.57}{\includegraphics*[0.7in,0.5in][6.4in,4.7in]{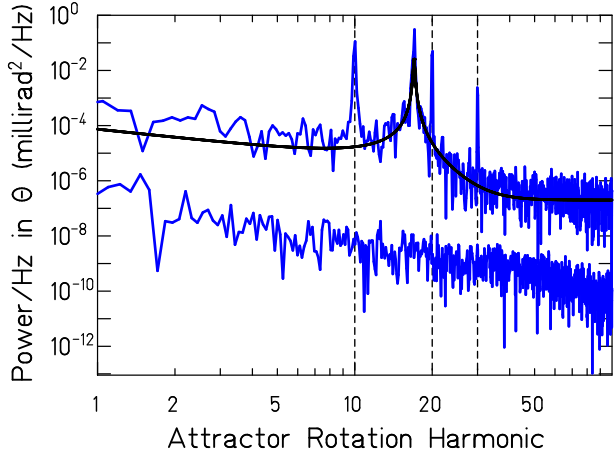}}\hfil
\caption{Fourier spectrum of the pendulum twist from an Experiment~I data run. The horizontal axis is displayed in units of attractor rotation harmonic. The 10$\omega$, 20$\omega$, and 30$\omega$ signals are easily recognizable, along with the free torsional motion at 17$\omega$.
The smooth curve shows the predicted thermal noise plus a white noise floor that arose from ADC ``bit noise''. The bottom curve, which shows the readout noise in the twist signal, is the spectral density recorded with the pendulum held stationary. The ``bit noise'' in this spectrum is negligible because a higher gain was used.} 
\label{fft}
\end{figure}

These points are illustrated in Fig.~\ref{fft}.
The top curve is the Fourier spectrum of the pendulum twist from an Experiment~I data run and shows the driven harmonic peaks as well as the free
torsional oscillation (these data were not passed through the torsion filter). The effect of the pendulum as a low-pass filter is seen by the drop in the noise floor above the resonance at~17$\omega$. The smooth curve is the predicted thermal noise from damping internal to the torsion fiber. The spectrum follows the curve well, indicating that we were operating near the thermal limit. It is interesting to note that velocity damping (from residual gas, for example) is negligible as it would produce a constant, as opposed to $1/f$, background at low frequencies. The bottom curve shows that the angle readout noise was well below the 
``torque noise'' associated with actual movement of the pendulum. 
%
%
\section{\label{sec:calc}Calculating Newtonian, Yukawa, Power-Law and Massive Pseudoscalar Forces between Cylinders}
\subsection{\label{subsec:calcapp}Approach}
\begin{figure}
\hfil\scalebox{.65}{\includegraphics*[0.4in,0.9in][4.4in,5.1in]{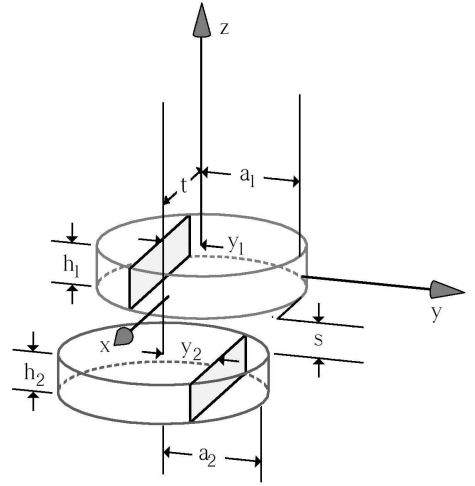}}\hfil
\caption{Coordinate system used in the force calculations.}
\label{cylinders}
\end{figure}

We assume the test bodies are perfect cylinders with symmetry axes along $\vec{z}$, radii and thicknesses $a_1$, $h_1$ and $a_2$, $h_2$, and uniform negative densities $\rho_1$ and $\rho_2$. Without loss of generality, we can take one cylinder to be centered at the origin and the other at 
coordinates
$(t,0,s+[h_1+h_2]/2)$ as shown in Fig.~\ref{cylinders}. 

The Newtonian, Yukawa, power-law and massive pseudo-scalar force calculations (see Eq.~\ref{eq:force defs}
each required evaluating 6 integrals. We evaluated analytically as many of the integrals as possible, and did the remaining ones numerically. 
We expedited the calculations by making databases of the various forces evaluated on a grid of horizontal and vertical separations $t$
and $s$ (for Yukawa and massive pseudo-scalar forces the ranges $\lambda$ were an additional parameter). 
The force between a pendulum and upper-attractor hole pair were calculated 
for 500 (100 for the non-Newtonian forces) values of $t$ spaced quadratically in the region
0.0~mm~$\leq t \leq 85$~mm (this placed more points where $F_{\rm t}$ is highly peaked) for each of 100 $s$ values 
in the relevant regime. The force from a lower-attractor hole was calculated for $0.05~{\rm mm}+h \leq s \leq 11~{\rm mm} +h$ where $h$ is the thickness of the upper disk).
The horizontal forces at arbitrary $t$ and $s$ were found by cubic spline interpolation of the database values as described in Ref.~\cite{stat}. 

In all cases, the $x$ and $z$ integrals were evaluated first to give the force per unit thickness between two infinitely thin rectangular sheets lying parallel
to the $x-z$ plane, as illustrated in Fig.~\ref{cylinders}. Because the force depends only upon the coordinate differences ($x_1-x_2$, etc.), we transformed coordinates to $x=x_1-x_2$, $X=x_1+x_2$, $z=z_1-z_2$, and $Z=z_1+z_2$. Then the $x$ integrations gave
\begin{equation}
I_x=\int \frac {\partial V} {\partial x} dx_1 dx_2 = \int_{t_3}^{t_2} V dx -\int_{t_0}^{t_1} V dx
\label{eq:xlimits}
\end{equation}
where $V$ is one of the potentials in the square brackets of Eqns.~\ref{eq:force defs} and 
\begin{eqnarray}
t_0&=&t-(a_2^2-y_2^2)^{1/2}-(a_1^2-y_1^2)^{1/2}  \\
t_1&=&t-(a_2^2-y_2^2)^{1/2}+ (a_1^2-y_1^2)^{1/2} \nonumber \\
t_2&=&t+ (a_2^2-y_2^2)^{1/2}+ (a_1^2-y_1^2)^{1/2} \nonumber \\
t_3&=&t+ (a_2^2-y_2^2)^{1/2}- (a_1^2-y_1^2)^{1/2} \nonumber
\end{eqnarray} 
Similarly, the $z$ integrations gave
\begin{eqnarray}
\label{eq:zlimits}
I_{xz}&\!\!=\!\!& \int_{}^{} dz_1 dz_2 I_x  \\
   &\!\!=\!\!&  \int_{s_0}^{s_1}\!\!\!(z \!-\!s_0)I_x dz \!+\!\int_{s_1}^{s_3}\!\!\!(s_1\!-\!s_0)I_x dz\!+\!\int_{s_3}^{s_2}\!\!\!(s_2\!-\!z)I_x dz \nonumber
\end{eqnarray}
where $s_0=s$, $s_1=s+h_1$, $s_2=s+h_1+h_2$, and $s_3=s+h_2$.
Finally, the $y$ integrals were evaluated numerically using a standard Romberg integration scheme\cite{stat}. Reflection symmetry
about the $x-z$ plane allowed us to integrate one half of one cylinder and double the result, giving
\begin{equation}
\frac {F_t(t,s)}{G \rho_1 \rho_2}\,=\,2a_1 a_2\int_{-\pi/2}^{\pi/2}\int_0^{\pi/2}\! I_{xz}\,\cos \theta_1 \,\cos \theta_2 \,d\theta_1\,d\theta_2.
\label{eq:y integrals}
\end{equation}
where we made the substitution $y_i \rightarrow a_i \sin{\theta_i}$.
\subsection{\label{subsec:calcnewt}Newtonian Force Calculation}
All four Newtonian $x$ and $z$ integrals were evaluated analytically, yielding
\begin{eqnarray}
I_{xz}^G\!\!\!\!&(&\!\!\!\!t,y,s)= -\frac {1} {2} \sum_{i=0}^3  \sum_{j=0}^3 (-1)^{i+j}  \\
\!\!& \times &\!\! \Biggl[ \;t_i\,r_{ij}+\frac{t_i}{\vert t_i\vert}(y^2-s_j^2)\ln{\left(\frac{\vert t_i\vert+r_{ij}}{\sqrt{y^2+s_j^2}}\right)} \nonumber \\ 
&~&~-2t_i s_j\ln{\left(1+\frac{r_{ij}}{s_j}\right)}+2y s_j\tan^{-1}{\left(\frac{t_i s_j}{r_{ij} y}\right)}\; \Biggr] \nonumber
\end{eqnarray}
where $y=y_1-y_2$ and $r_{ij}^2= t_i^2 + y^2 + s_j^2$.
\subsection{\label{subsec:calcyuk}Yukawa Force Calculation}
We integrated Eq.~\ref{eq:zlimits} for the Yukawa interaction by parts and made the substitution $p=1/z$. This left the following $x$ and $z$ integrals, along with the $y$ integrals, to be evaluated
numerically
\begin{eqnarray}
I_{xz}^Y\!\!\!\!&(&\!\!\!\!t,y,s,\lambda)= \lambda \Biggl[\int_{t_3}^{t_2} dx - \int_{t_0}^{t_1} dx \Biggr]  \\
\!\!& \times &\!\! \Biggl[ s_0 \int_{1/s_1}^{1/s_0}\!\! \!\! H dp +(s_1 \!-\!s_0)\int_{1/s_1}^{1/s_3}\!\! \!\! H dp + s_2\int_{1/s_3}^{1/s_2} \! \!\!\! H dp \Biggr]~,\nonumber 
\end{eqnarray}
where 
\begin{equation}
H= \exp{\frac{-\sqrt{x^2 + y^2 + 1/p^2}}{\lambda}}~.
\end{equation}
\subsection{\label{subsec:powerlaw}Power-law Force Calculation}
The power law potentials, $V \propto 1/r^k$ for $k=2-5$, required three numerical integrations. In this case, $I_{xz}$ took the form
\begin{eqnarray}
I_{xz}^P\!\!&(&\!\!\!t,y,s,k)\!=\! r_0^{k-1}\sum_{i=0}^3 (-1)^i \Bigg[ \sum_{j=0}^3 (-1)^j F_{ij}^{(k)}  \\
   &\!\!+\!\!& s_0 \!\int_{s_1}^{s_0}\!\!\!G_i^{(k)} dz + (s_1\!-\!s_0)\!\int_{s_1}^{s_3}\!\!\!G_i^{(k)} dz + s_2\!\int_{s_3}^{s_2} \!\!\!\!G_i^{(k)}dz \Bigg]~.  \nonumber
\end{eqnarray}

Letting $y=y_1-y_2$, $r_{ij}^2 = t_i^2 + y^2 + s_j^2$ $\rho_j^2 = y^2 + s_j^2$, and $\rho^2=y^2+z^2$,  we have
\begin{eqnarray}
F_{ij}^{(2)}&=& -\frac {t_i}{2} \ln \left( \frac{r_{ij}}{r_0} \right) + \rho_j \tan^{-1}{\left( \frac {t_i}{\rho_j} \right)}  \\
G_i^{(2)}&=& \frac {1}{\rho} \tan^{-1}{\left( \frac {t_i}{\rho} \right)} \nonumber \\
F_{ij}^{(3)}&=& \ln \left( \frac{t_i + r_{ij}}{r_0} \right) \nonumber \\
G_i^{(3)}&=& \frac {t_i}{\rho^2 \sqrt{(t_i^2 +\rho^2)}} \nonumber  \\
F_{ij}^{(4)}&=&\frac {1}{2\rho_j} \tan^{-1}{\left( \frac {t_i}{\rho_j} \right)} \nonumber \\
G_i^{(4)}&=& \frac {t_i}{2\rho^2 (t_i^2+\rho^2)} + \frac {1}{2\rho^3}\tan^{-1}{
\left( \frac {t_i}{\rho} \right)} \nonumber \\
F_{ij}^{(5)}&=& \frac {t_i}{3\rho_j^2 r_{ij}}  \nonumber \\
G_i^{(5)}&=& \frac {t_i}{\rho^4 \sqrt{t_i^2+\rho^2}}\left(1-\frac {t_i^2}{3(t_i^2+\rho^2)}\right)~. \nonumber
\end{eqnarray}
\subsection{Massive Pseudoscalar Exchange Calculation}
In this case, only the integrations leading to Eqs.~\ref{eq:xlimits} and \ref{eq:zlimits} could be done analytically. The 4 integrations needed to evaluate
Eqs. \ref{eq:xlimits}, \ref{eq:zlimits} and \ref{eq:y integrals} were done numerically. 


\begin{thebibliography}{}
\bibitem{remark1}Note that this correct expression for $\alpha$ differs from values quoted in some previous works.
\bibitem{error}The axis labels in Fig. 22 of Ref.~\cite{sm:00} should be interchanged.
\end{thebibliography}
\end{document}